\documentclass[12pt,preprint]{aastex}
\usepackage{epsfig}
\begin{document}
\bibliographystyle{plainnat}
\voffset-1cm
\newcommand{\gsim}{\hbox{\rlap{$^>$}$_\sim$}}
\newcommand{\lsim}{\hbox{\rlap{$^<$}$_\sim$}}

\title{Flares In Long And Short Gamma Ray Bursts}

\author{Shlomo Dado\altaffilmark{1} and Arnon Dar\altaffilmark{2}}

\altaffiltext{1}{dado@phep3.technion.ac.il\\
Physics Department, Technion, Haifa 32000,
Israel}
\altaffiltext{2}{arnon@physics.technion.ac.il\\
Physics Department, Technion, Haifa 32000, Israel}

\begin{abstract}

The many similarities between the prompt emission pulses in gamma ray 
bursts (GRBs) and X-ray flares during the fast decay and afterglow phases 
of GRBs suggest a common origin.  In the cannonball (CB) model of GRBs, 
this common origin is mass accretion episodes of fall-back matter on a 
newly born compact object. The prompt emission pulses are produced by 
a bipolar jet of highly relativistic plasmoids (CBs) ejected in the 
early, major episodes of mass accretion. As the accretion material is 
consumed, one may expect the engine's activity to  weaken. X-ray 
flares ending  the prompt emission and during the afterglow phase 
are produced in such delayed episodes of mass accretion. The common 
engine, environment and radiation mechanisms (inverse Compton scattering 
and synchrotron radiation) produce their observed similarities. Flares in 
both long GRBs and short hard gamma ray bursts (SHBs)  can also be 
produced by bipolar ejections of CBs following a phase transition in 
compact objects due to loss of angular momentum and/or cooling. Optical 
flares, however, are mostly produced in collisions of CBs with massive 
stellar winds/ejecta or with density bumps along their path. In this paper 
we show that the master formulae of the CB model of GRBs and SHBs, which 
reproduce very well their prompt emission pulses and their smooth 
afterglows, seem to reproduce also very well the lightcurves and spectral 
evolution of the prominent X-ray and optical flares that are well sampled.

\end{abstract}


\section{Introduction}

A flaring activity during the afterglow (AG) phase of a gamma ray burst 
(GRB)  was first observed in the late-time AG of GRB 970508 with the 
Narrow Field Instrument (NFI) aboard the BeppoSAX satellite in the X-ray 
band \citep{Piro1998}, and with ground based telescopes in the optical 
band \citep{Pian1998,Galama1998a}.
It was interpreted in the framework of the fireball (FB) model of GRBs as a 
delayed burst from the central GRB engine \citep{Piro1998}. 
Alternatively, in the cannonball (CB) model of GRBs it was interpreted as 
a synchrotron radiation (SR) flare from an encounter of the highly 
relativistic jetted ejecta from an underlying supernova (SN) explosion 
with a density jump in the interstellar medium \citep{DDD2002,DDD2004b}. 
Late-time flares were later discovered in the broadband AG of several 
other GRBs that were localized by the BeppoSAX 
satellite, most notably in GRB 000301C at $t\!\sim\! 4$ days after burst 
\citep{Berger2000, Sagar2000} where the flare was attributed to 
gravitational microlensing \citep{Garnavich2000} and in 
GRB 030329 \citep{Lipkin2004} where the flare was interpreted in the 
framework of 
the FB model as due to `refreshed shocks'  generated by a late activity of 
the central engine \citep{Granot2003}. In the CB model, however, these 
flares were well reproduced by the emission of SR from encounters of the 
jetted ejecta from an underlying SN explosion in a star formation region 
with density jumps within or at the border of a super bubble created by 
the star formation region \citep{DDD2002, DDD2004b}.

Early-time X-ray flares ending the prompt emission phase were also 
detected by the wide field camera (WFC) aboard BeppoSAX in a few GRBs such 
as GRB 011121. In the FB model they were attributed to the 
onset of the external shock in the circumburst material. In the CB model 
they were interpreted as being due to the last episodes of bipolar CB 
ejections from a shutting off central engine. Shortly after the launch of 
the Swift satellite in November 2004, data collected with its X-ray 
telescope (XRT) showed that X-ray flares are quite common in all the 
phases of the emission from GRBs. In more than 50\% of the GRBs observed 
with the Swift X-ray telescope (XRT), flares were observed at the end of 
the prompt emission and/or the early AG phase \citep{Burrows2005, 
Burrows2007, Falcone2007}. In some cases X-ray 
flares were observed also at very late times, of the order of several days 
after the prompt emission. Although the information on flares is much more 
sketchy compared to that on the prompt gamma ray pulses, their spectral 
and temporal behaviours show clearly that the X-ray flares during the 
prompt $\gamma$-ray emission follow the pattern of the $\gamma$-ray 
pulses, suggesting they are the low energy part of these pulses.  The 
X-ray flares ending the prompt emission and those superimposed on the 
early-time afterglow have a fast spectral evolution. Their peak 
intensities decrease with time, and their spectral and temporal behaviours 
are similar to those of the prompt X/$\gamma$-ray pulses, except that they 
are progressively softer and last longer. In most cases their $\gamma$-ray 
emission probably is below the detection sensitivity of the Swift broad 
alert telescope (BAT). In some cases their fluence in the XRT band 
exceeded that of the prompt emission in the BAT 15-150 keV band 
\citep{Chincarini2008a,Chincarini2008b}.

Late-time ($t\gsim 10^4$ s) X-ray flares, however, seem to exhibit 
temporal and spectral 
behaviours that are different from those of most of the early-time 
flares. Their power-law decline is more moderate, they are "achromatic"  
with a power-law spectrum almost identical to that of the 
late-time AG, and they show very little spectral evolution
(see, e.g., the late time broad band flares in   
GRBs 060614 \citep{Mangano2007} and 081028 \citep{Margutti2009},
and the hardness ratio during late time flares in Swift GRBs reported 
in the Swift-XRT lightcurve repository, \citet{Evans2009}).

Flares in the X-ray lightcurve of Swift GRBs were studied 
phenomenologically by various observer groups \citep{Burrows2005, 
Burrows2007, Kocevski2007, BK2007, Falcone2007, 
Chincarini2007a,Chincarini2007b,Chincarini2008a, 
Chincarini2008b}. Modifications of previously suggested models and 
new theoretical models were proposed and discussed by several authors 
\citep{Proga2005, King2005, Dai2006, Fan2006, Perna2006}, but none of 
these proposed models was shown to actually derive the observed 
lightcurves and spectral evolution of either early-time or 
late-time flares from underlying physical assumptions.

\begin{figure}[]
\centering
\epsfig{file=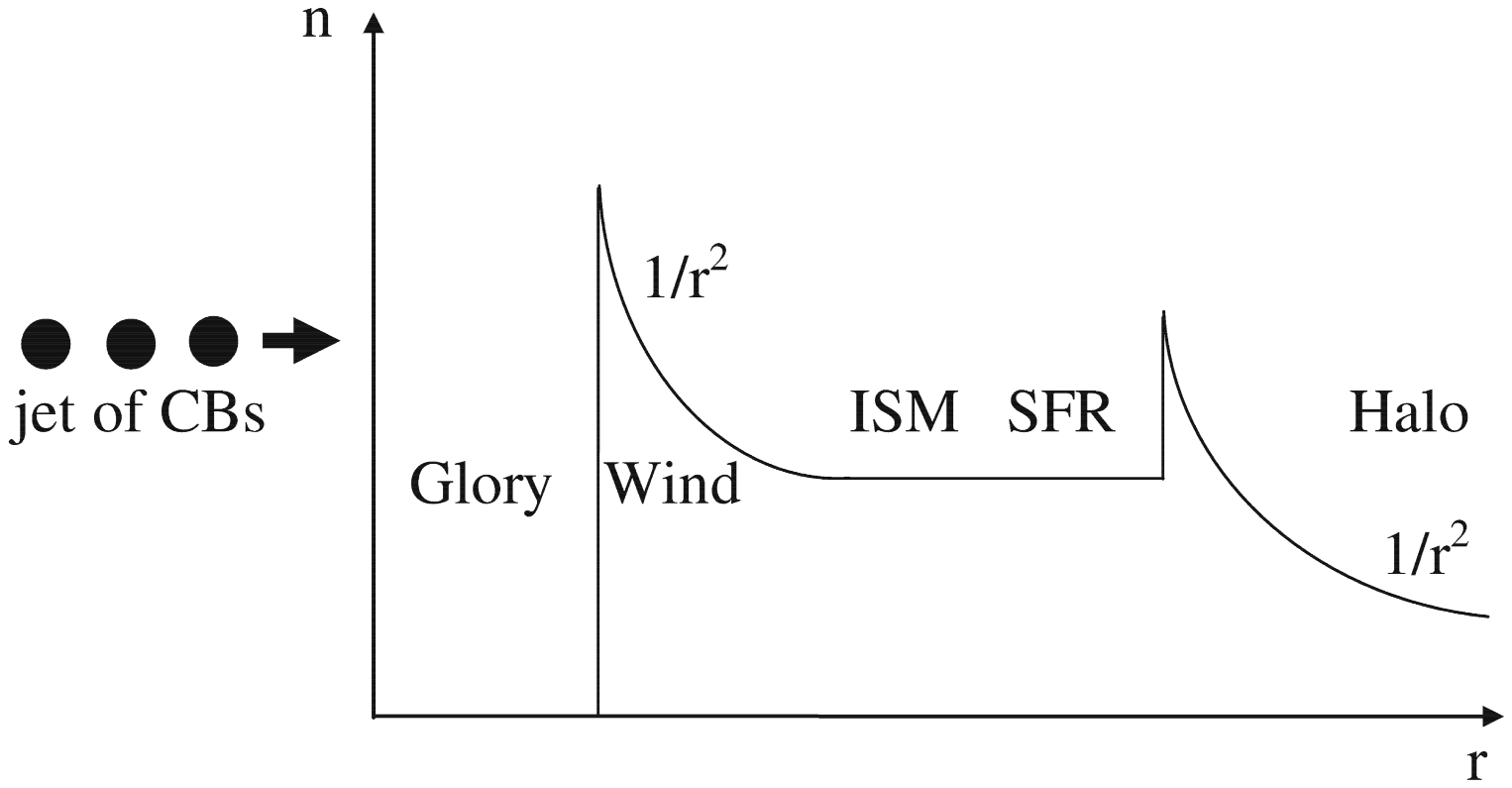,width=18cm}
\caption{Schematic illustration, not in scale, of the typical environment
encountered by
a highly relativistic jet ejected in core collapse SN that escapes
from the star formation region into the galactic halo.}
\label{fig1}
\end{figure}

Flares are a natural consequence of the cannonball (CB) model of GRBs, 
which was motivated by a GRB-microquasar analogy \citep{DDD2002,DD2004}.
In the CB model, {\it long-duration} GRBs and their AGs are 
produced by bipolar jets of highly relativistic plasmoids of ordinary 
matter ejected in accretion episodes on the newly formed compact stellar 
object \citep{Shaviv1995, Dar1998} in core-collapse supernova 
(SN) explosions \citep{Dar1992, DP1999}. It is hypothesized that 
an accretion disk or a torus is produced around the newly formed compact 
object, either by stellar material originally close to the surface of the 
imploding core and left behind by the explosion-generating outgoing shock, 
or by more distant stellar matter falling back after its passage 
\citep{Rujula1987}. As observed in microquasars \citep{Mirabel1999},
each time part of the 
accretion disk falls abruptly onto the compact object, two jets of 
cannonballs (CBs) made of {\it ordinary-matter plasma} are emitted with 
large bulk-motion Lorentz factors in opposite directions along the 
rotation axis, wherefrom matter has already fallen back onto the compact 
object due to lack of rotational support. The entire radiation emitted 
from a GRB is produced by the interaction of the jet of CBs with the 
environmemt along its path, as illustrated in Fig.~\ref{fig1}. The prompt 
$\gamma$-ray and X-ray emission is dominated by inverse Compton scattering 
(ICS) of photons of the SN light filling the cavity produced by the 
pre-supernova wind/ejecta blown from the progenitor star long before the 
GRB. The CBs' electrons Compton up-scatter this `glory' light into a 
narrow conical beam of $\gamma$ rays along the CBs' direction of motion. 
Each CB produces a single GRB pulse. An X-ray `flare' coincident in time 
with a prompt $\gamma$-ray pulse is simply its low-energy part. The 
natural explanation of flares ending the prompt emission and during the 
early time afterglow is the same: ICS of glory photons by the electrons of 
CBs ejected in late accretion episodes of fall-back matter on the newly 
formed central object. Early-time X-ray flares without an accompanying 
detectable $\gamma$-ray emission are usually IC flares (ICFs) produced by 
CBs with relatively smaller Lorentz factors, due to weakening activity of 
the engine: As the accretion material is consumed, one may expect the 
`engine' to have a few progressively-weakening dying pangs. Like the 
lightcurves of the prompt GRB pulses, the lightcurves of ICFs exhibit a 
rapid softening during their fast decline phase 
(see, e.g., the XRT hardness ratio reported for Swift GRBs  
in the Swift lightcurve repository, \citet{Evans2007,Evans2009}).

In the CB model, each IC flare (ICF) is followed by the emission of SR 
flare (SRF) from the encounter of the CB with the wind/ejecta, which  was 
blown from the progenitor star long before the GRB (see Fig.~\ref{fig1}). 
Because of time-aberration in the observer frame, these SRFs 
appear to have only short lag-times relative to the corresponding ICFs. 
Below their peak energy, the spectral behaviour of the ICS 
pulses/flares is roughly $F_\nu\!\sim\!\nu^0 $, while the 
SR emission from fast cooling electrons has typically
$F_\nu\!\sim\!\nu^{-1.1} $. 
Thus, while the prompt keV-MeV pulses/flares are dominated by ICS 
of glory light, the `prompt' optical emission  
is dominated by SR. As the glory extends into the wind, often 
the SR emission begins before the ICs pulse/flare has ended.

The initial expansion of the CBs and the slow-down of the leading ones by 
the circumburst matter may merge most of them into a single leading 
CB \citep{DD2004, DDD2009a} during the afterglow phase. Its 
collimated beam of the prompt gamma rays ionizes the matter in front of 
it. The ions continuously impinging on a CB with a relative Lorentz 
factor 
$\gamma(t)$, where $\gamma(t)$ is the bulk motion Lorentz factor of the 
CB, generate within it an equipartition turbulent magnetic field. The 
intercepted electrons are isotropized and Fermi accelerated by these 
fields and emit isotropic synchrotron radiation in the CB's rest frame, 
which is Doppler boosted and beamed relativistically into a narrow cone 
with a typical opening angle $\!\sim\!1/\gamma(t)$ in the observer's rest 
frame. Late SRFs are produced 
mainly when the CBs encounter winds or density bumps along their path 
first from the progenitor star and later in the interstellar medium (ISM). 
The lightcurve of these flares depends on the unknown density profile of 
the encountered wind/density bump that cannot be predicted a-priori. 
But, 
both the early-time and the late-time SRFs have a typical SR spectrum and 
a weak spectral evolution that are quite different from those of the 
accretion induced ICFs and can be used to identify their origin -- late 
ejection episodes from the central engine or density bumps.

In the CB model, short hard bursts (SHBs) are also produced by bipolar
jets of plasmoids ejected in mergers of compact objects in close binary 
systems such as neutron stars merger \citep{Goodman1987} or in mass 
accretion episodes on compact objects in close binary systems, or in phase 
transitions in compact stars (neutron stars, hyper stars and strange quark 
stars) \citep{DDD2009b}. Bipolar ejections in late 
accretion episodes or phase transitions after cooling and loss of angular 
momentum probably produce the observed ICFs in SHBs, and SR radiation from 
encounters with winds/density bumps produces SRFs.

Flares were routinely included in the CB model description of the 
afterglows of GRBs and SHBs \citep{DDD2002, DDD2004b, DDD2009a, DDD2009b}. 
They were calculated from the master formulae of the model, which describe 
well the prompt ICS emission and the emission of SR at all times. It was 
shown that ICS explains successfully both the prompt keV-MeV pulses and 
the X-ray flares ending the prompt emission, while the SR emitted in the 
collision of the jet of CBs with the wind/ejecta from the progenitor star 
explains well the prompt optical flares measured with robotic telescopes 
in very bright GRBs \citep{DDD2009a, DD2008}. However, attention was 
focused there on the general properties of the prompt emission and the 
smooth afterglow, rather than on flares during the afterglow phase. 
Moreover, in many GRB afterglows, flares are weak, or are blended, or are 
not well sampled, and their properties could neither be inferred reliably 
from the AG lightcurve nor used reliably to test theoretical models. The 
situation concerning prominent X-ray flares observed by the Swift XRT is 
different. They are well resolved and their spectral properties are much 
better measured. In this paper we focus our attention on GRB X-ray and 
optical lightcurves with prominent flares.  In particular, we compare the 
CB model lightcurves and their spectral evolution with representatives set 
of X-ray and optical lightcurves measured with the Swift XRT, and with 
ground-based robotic telescopes and Swift UVO in space, respectively,
which have prominent flares that are well sampled. Such a comparison 
provides stringent tests of both the CB model and its interpretation of 
the origin of prompt emission pulses in GRBs, their afterglows and the 
early and late time flares in their lightcurves. We show that the CB model 
correctly predicts their main observed properties, and reproduces well 
their entire lightcurves and spectral properties. For completeness, we 
first summarize the relevant master formula of the CB model and their 
simplified forms that we later use in our analysis of the X-ray 
and optical lightcurve of GRBs.

\section{ICS flares}

In this section we summarize the master formula of the CB model
for the pulse shape and spectral evolution of ICS pulses/flares
(see \citet{DDD2009a} and references therein 
for their derivation).
Let  $t$ denote the time in the observer frame 
after the beginning of a flare.  
The light-curve of a flare, produced  
by the electrons in the CBs by ICS of  
thin bremsstrahlung photons filling the cavity formed by a wind 
blown by the progenitor star long before the 
GRB, is generally well approximated by 
\citep{DDD2009a}:
\begin{equation}
E\, {d^2N_\gamma\over dt\,dE}(E,t)\approx
A\, {t^2/\Delta t^2  \over(1+t^2/\Delta t^2)^2}\,
e^{-E/ E_p(t)}
\propto  e^{-E /E_p(0)}\, F(E\,t^2),
\label{ICSlc}
\end{equation}
where $A$ is a constant that depends on the CB's baryon number,
Lorentz and Doppler factors, and on the density
of the glory light and on the redshift and distance of the GRB,
and $E_p(t)$, the peak energy of $E\, d^2N_\gamma/ dE\, dt$ 
at time $t$, is given roughly by: 
\begin{equation}
E_p(t)\approx  E_p(0)\, {t_p^2 \over t^2+t_p^2}\,,
\label{PeakE}
\end{equation}
with $t_p$ being the time (after the beginning of the flare) when the ICS 
photon count-rate reaches its peak value.
For $E\!\ll\!E_p$, it satisfies 
$E_p(t_p)\!=\!E_p$ where $E_p$ is the 
peak energy of the time-integrated spectrum of the flare. 
Thus, in the CB model, each ICS pulse in the GRB lightcurve
is described by four parameters, $A,$
$\Delta t(E),$  $E_p(0)$ and the 
beginning time of the pulse when $t$ is taken to be 0.

\begin{figure}[]
\centering
\vspace{-1cm}
\vbox{
\hbox{
\epsfig{file=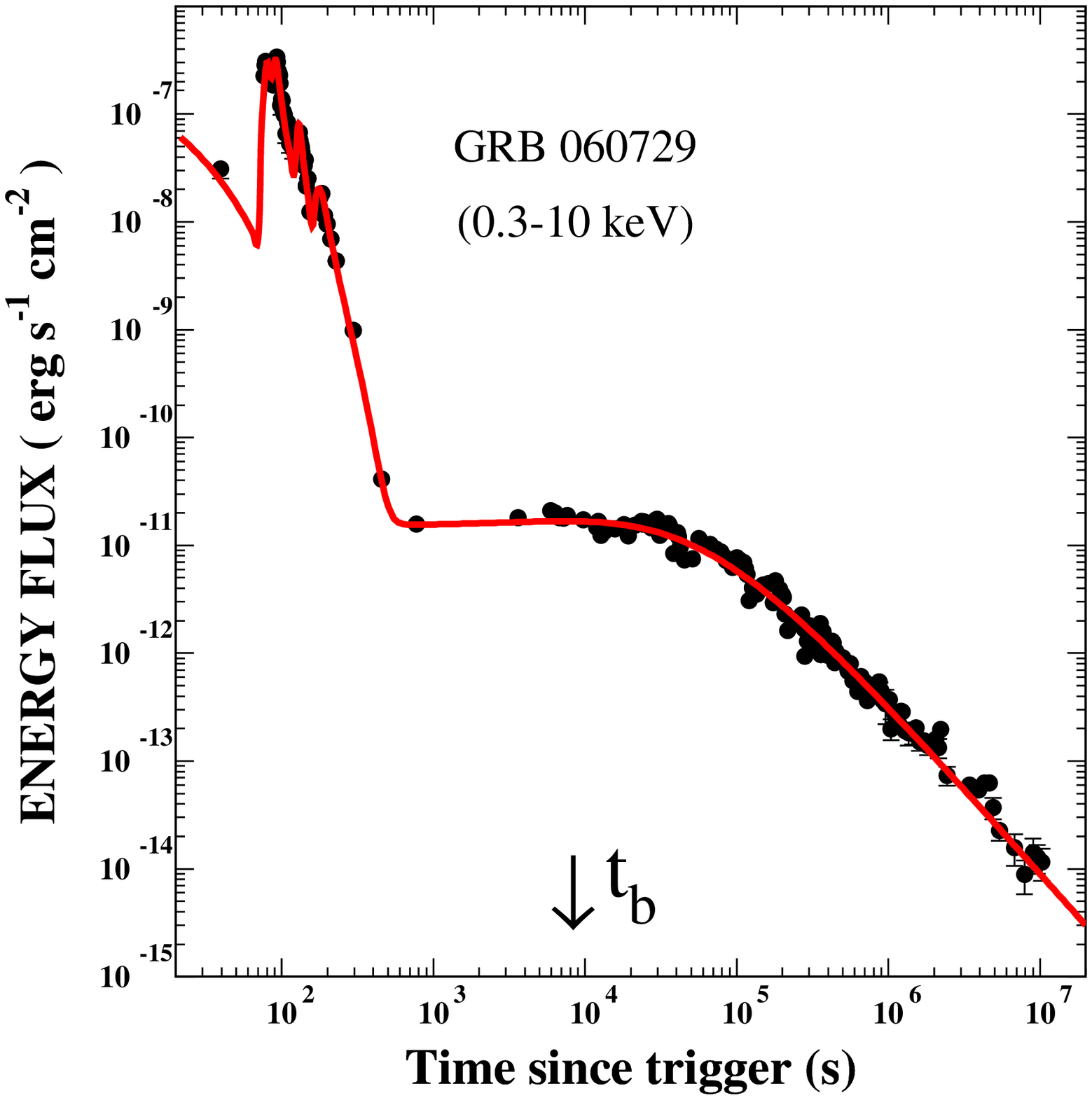,width=7.0cm,height=6.0cm}
\epsfig{file=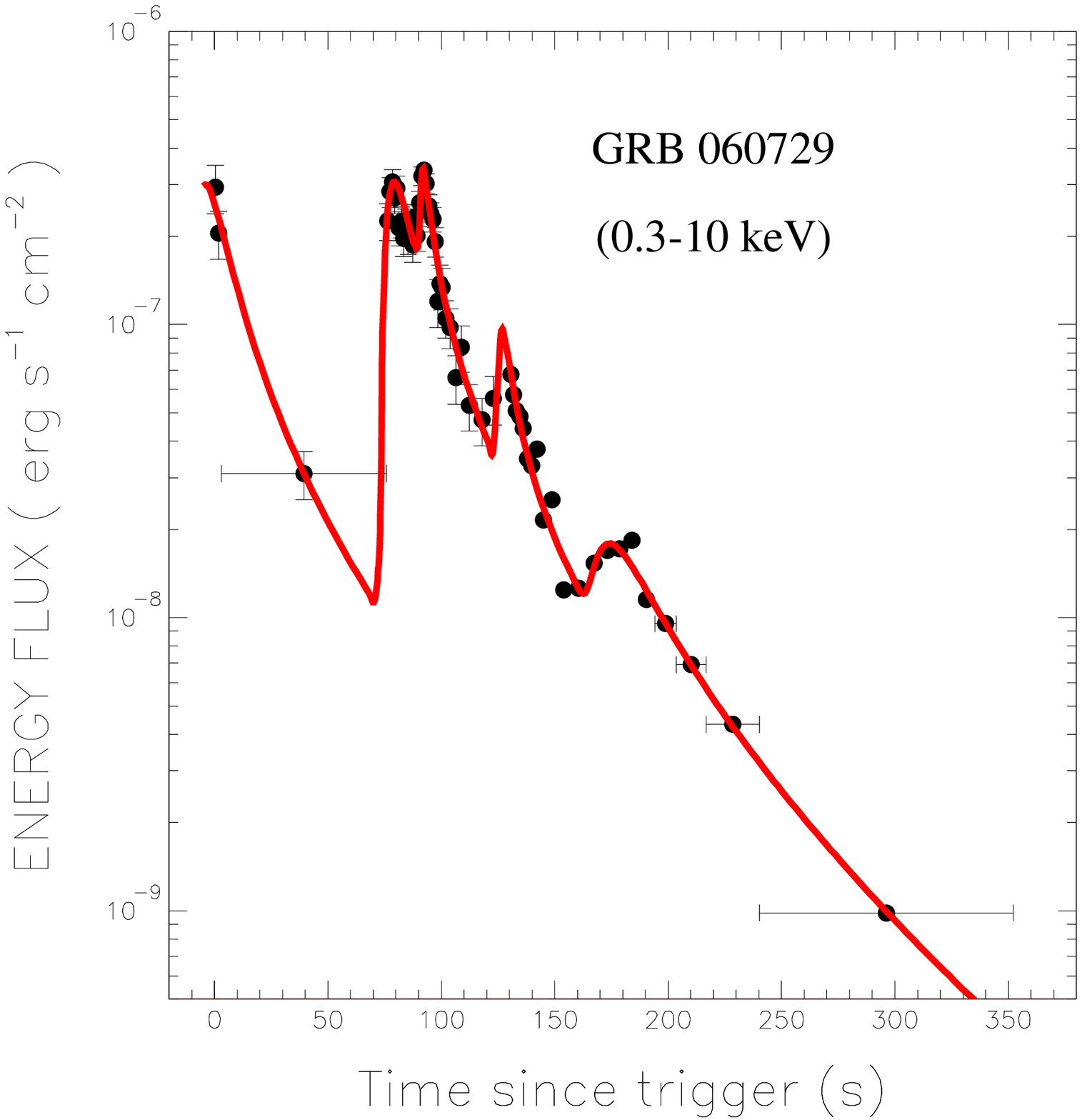,width=7.0cm,height=6.0cm}
}}
\vbox{
\hbox{
\epsfig{file=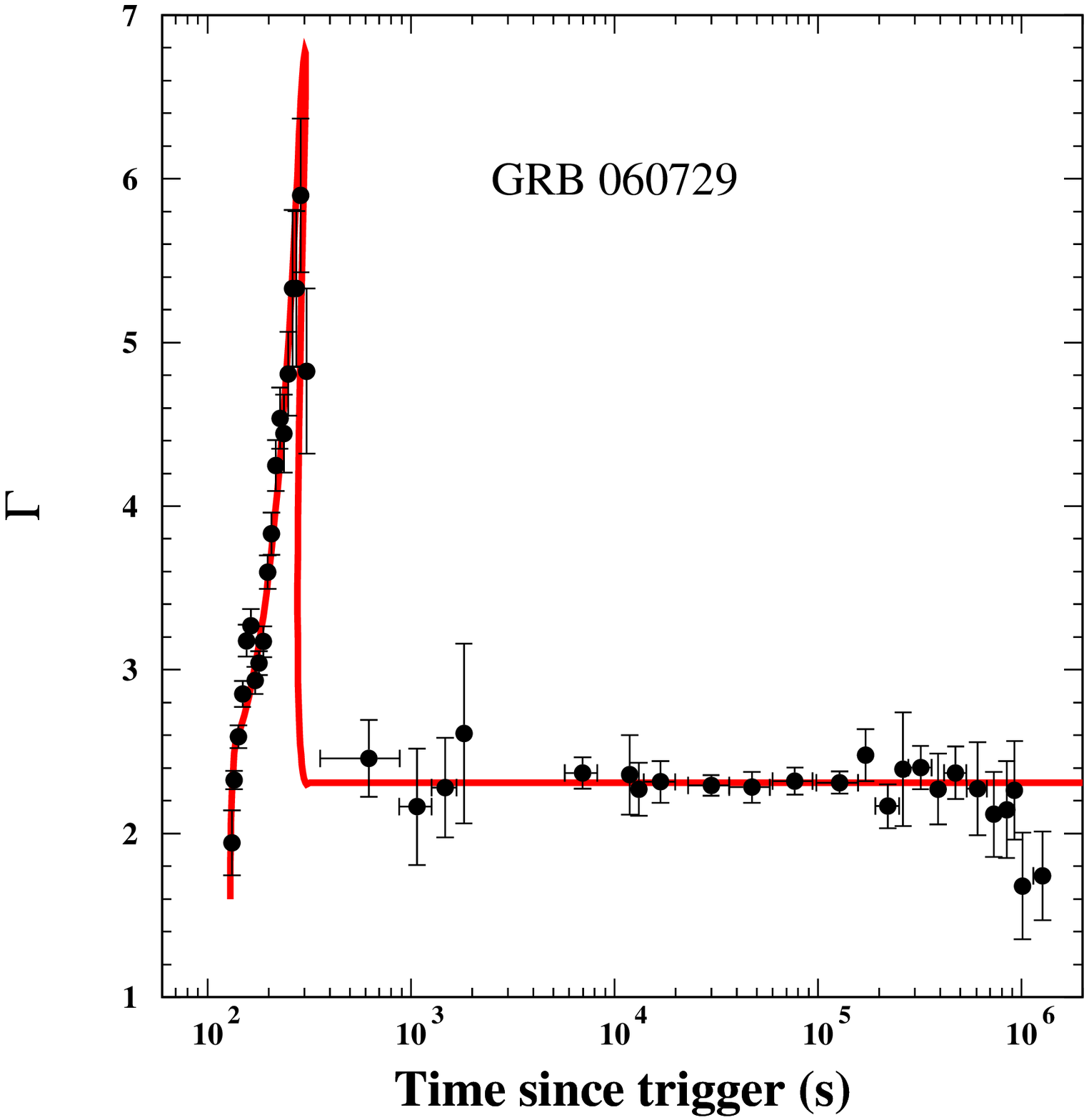,width=7.0cm,height=6.0cm}
\epsfig{file=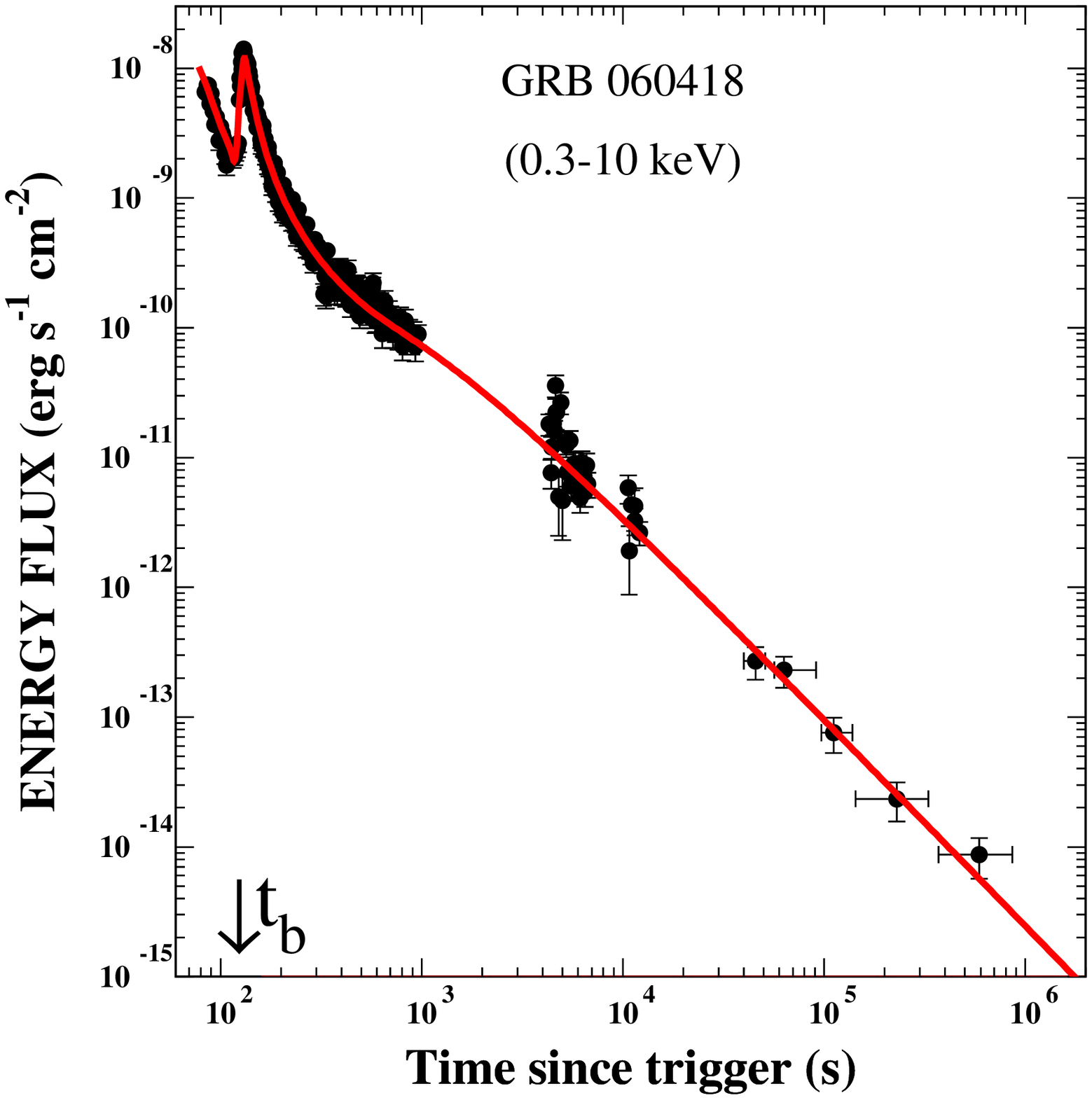,width=7.0cm,height=6.0cm}
}}
\vbox{
\hbox{
\epsfig{file=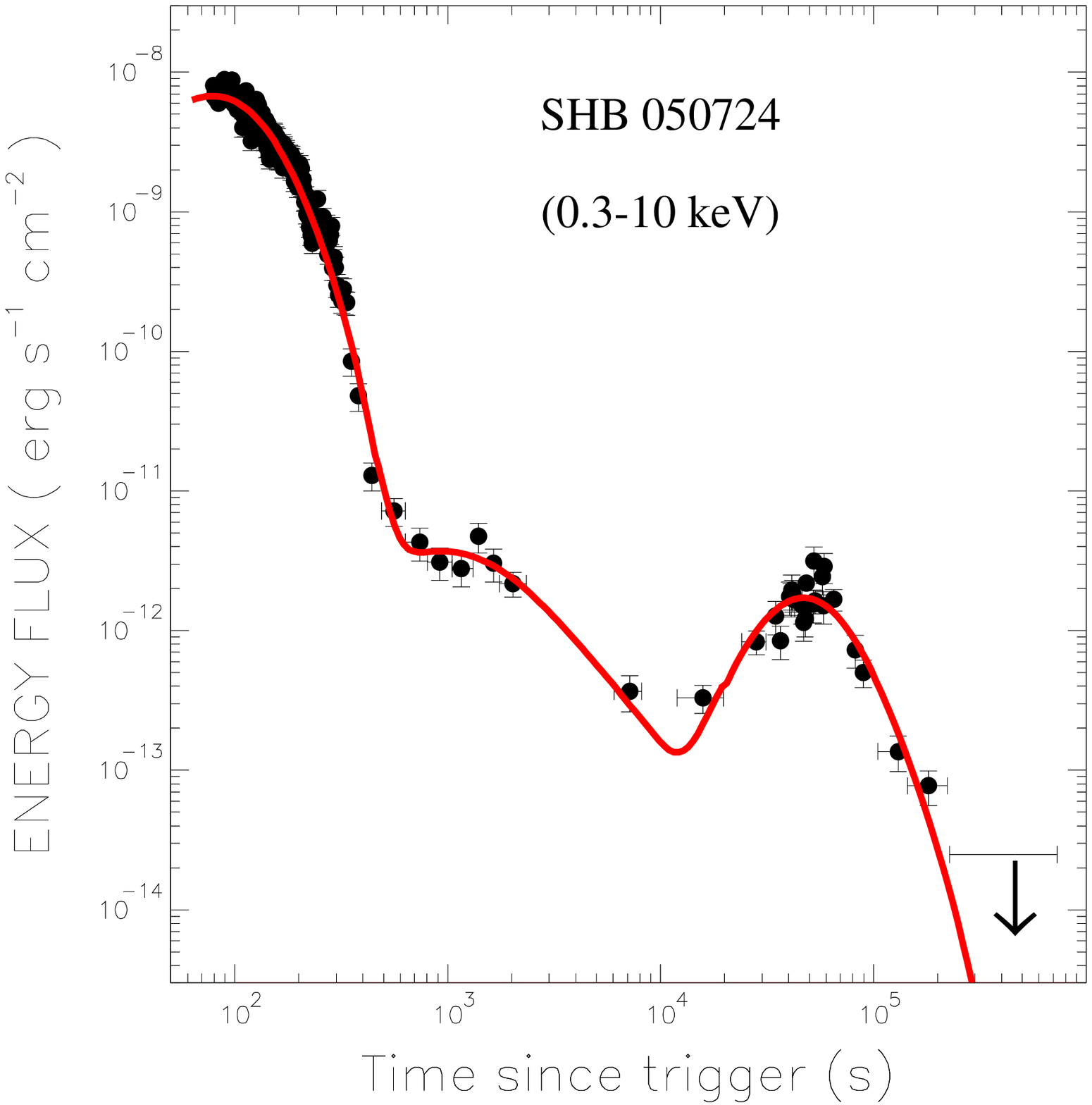,width=7.0cm,height=6cm }
\epsfig{file=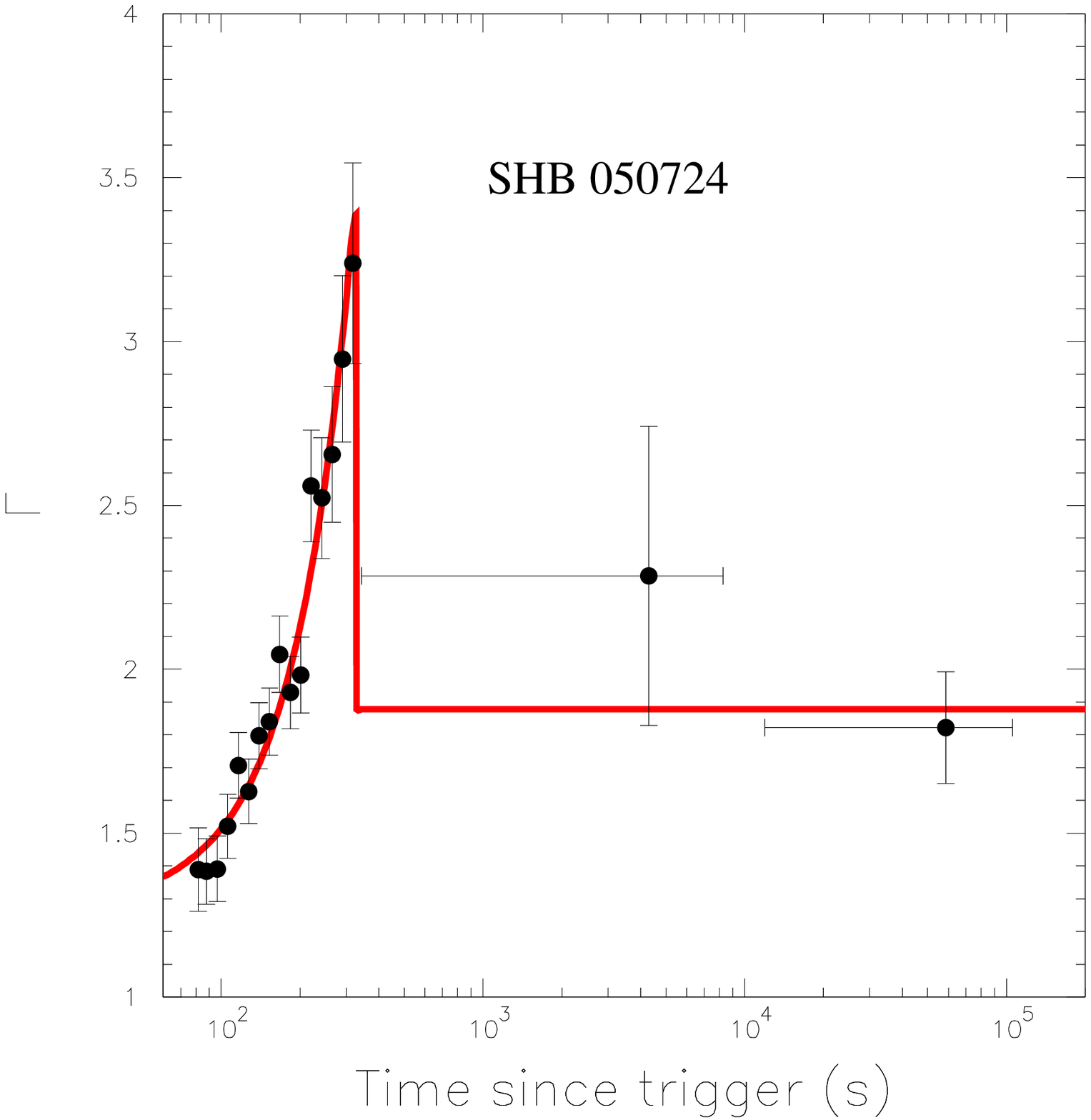,width=7.0cm,height=6cm}
}}
\caption{Comparison between 0.3-10 keV lightcurves
of Swift GRBs (Swift/XRT lightcurve repository \citep{Evans2009})
with early-time X-ray flares and their CB model 
description with the parameters listed in Table 1.
{\bf Top left (a):} Comparison between the Swift `canonical X-ray
lightcurve' \citep{Nousek2006} of GRB 060729 
and its CB model description. 
{\bf Top right (b):}  Zoom on
the prompt emission and fast decay phase in {\bf a}. 
The CB model lightcurve of the prompt emission consists of 
a sum of 5 ICFs. 
{\bf Middle left (c):} Comparison between the  evolution
of the effective photon spectral index of GRB 060729
in the 0.3-10 keV X-ray band as inferred from the Swift 
XRT observations by \citet{Zhang2007}  
and that inferred from the CB model lightcurve.
{\bf Middle right (d):} 
Comparison between the non-canonical X-ray lightcurve 
of GRB 060418 and its CB model description. 
{\bf Bottom left (e):}
Comparison between the XRT and CB model lightcurves of SHB 050724.
{\bf Bottom right (f):} Comparison between the evolution
of the effective photon spectral index 
in the 0.3-10 keV X-ray band 
as inferred  by \citet{Zhang2007} from the observations 
of SHB 050724 with the Swift XRT 
and that inferred from the CB model lightcurve.
} 
\label{fig2}
\end{figure}

\begin{figure}[]
\centering
\vspace{-1cm}
\vbox{
\hbox{
\epsfig{file=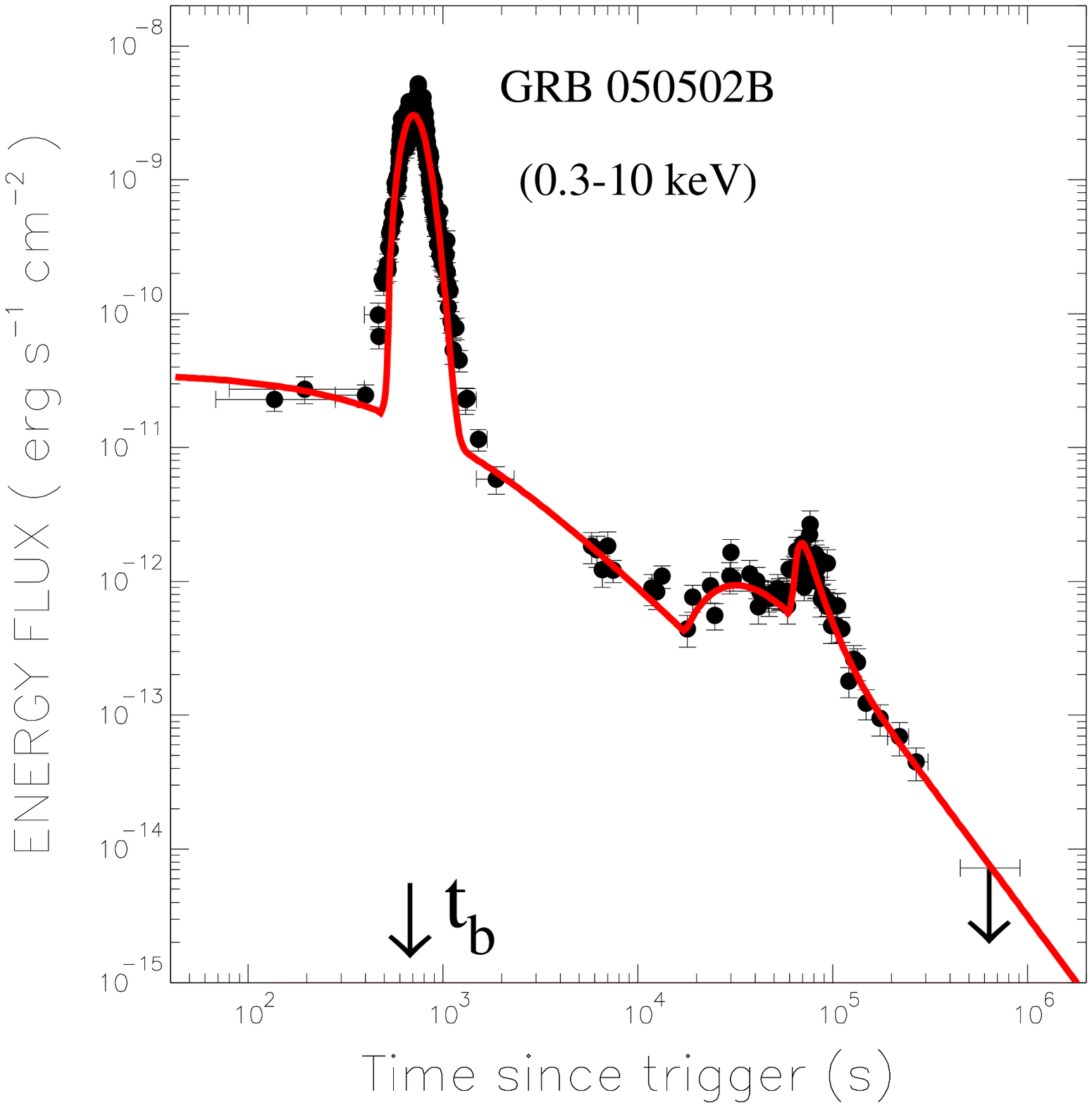,width=7.0cm,height=6.0cm}
\epsfig{file=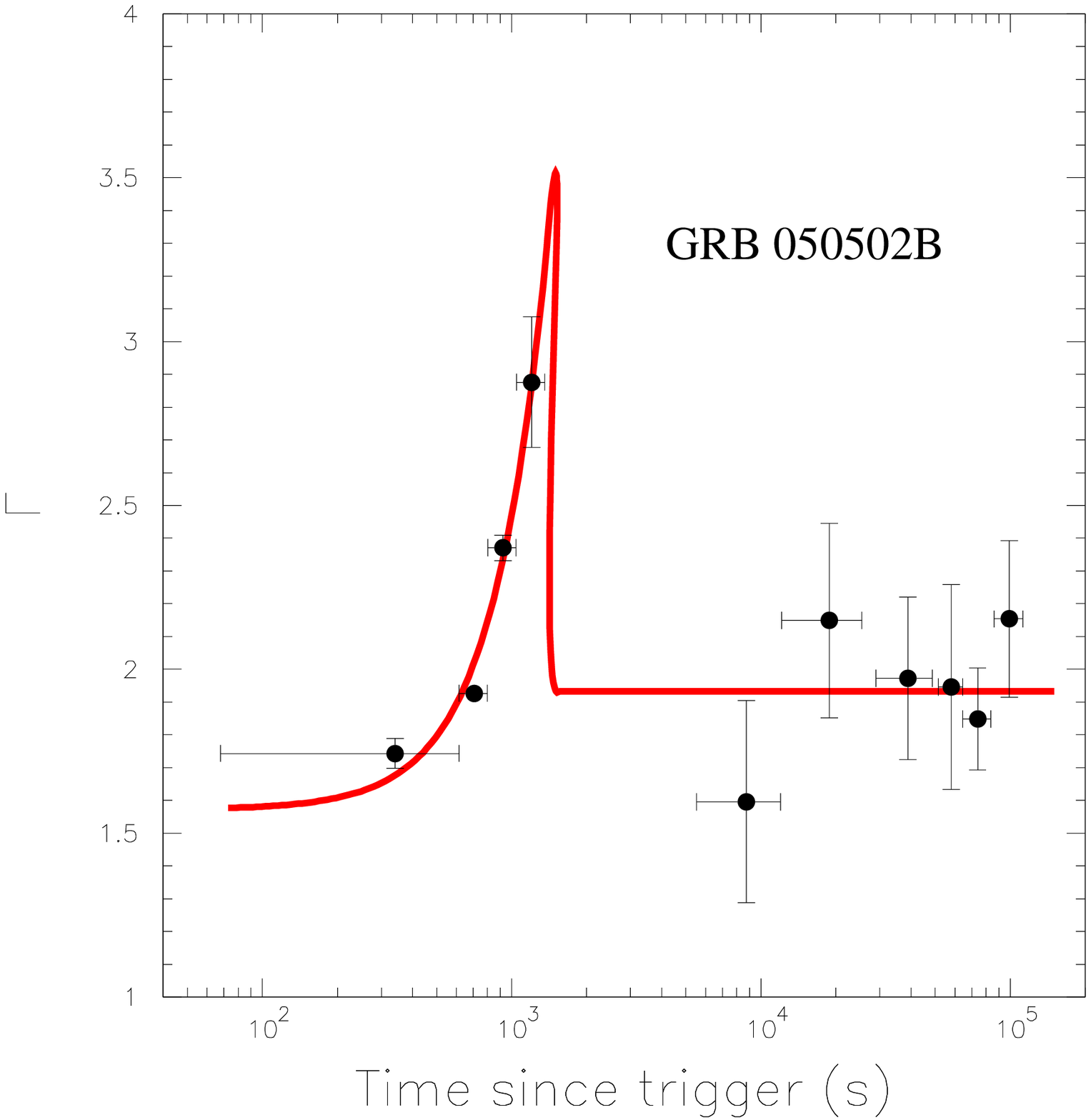,width=7.0cm,height=6.0cm}
}}
\vbox{
\hbox{
\epsfig{file=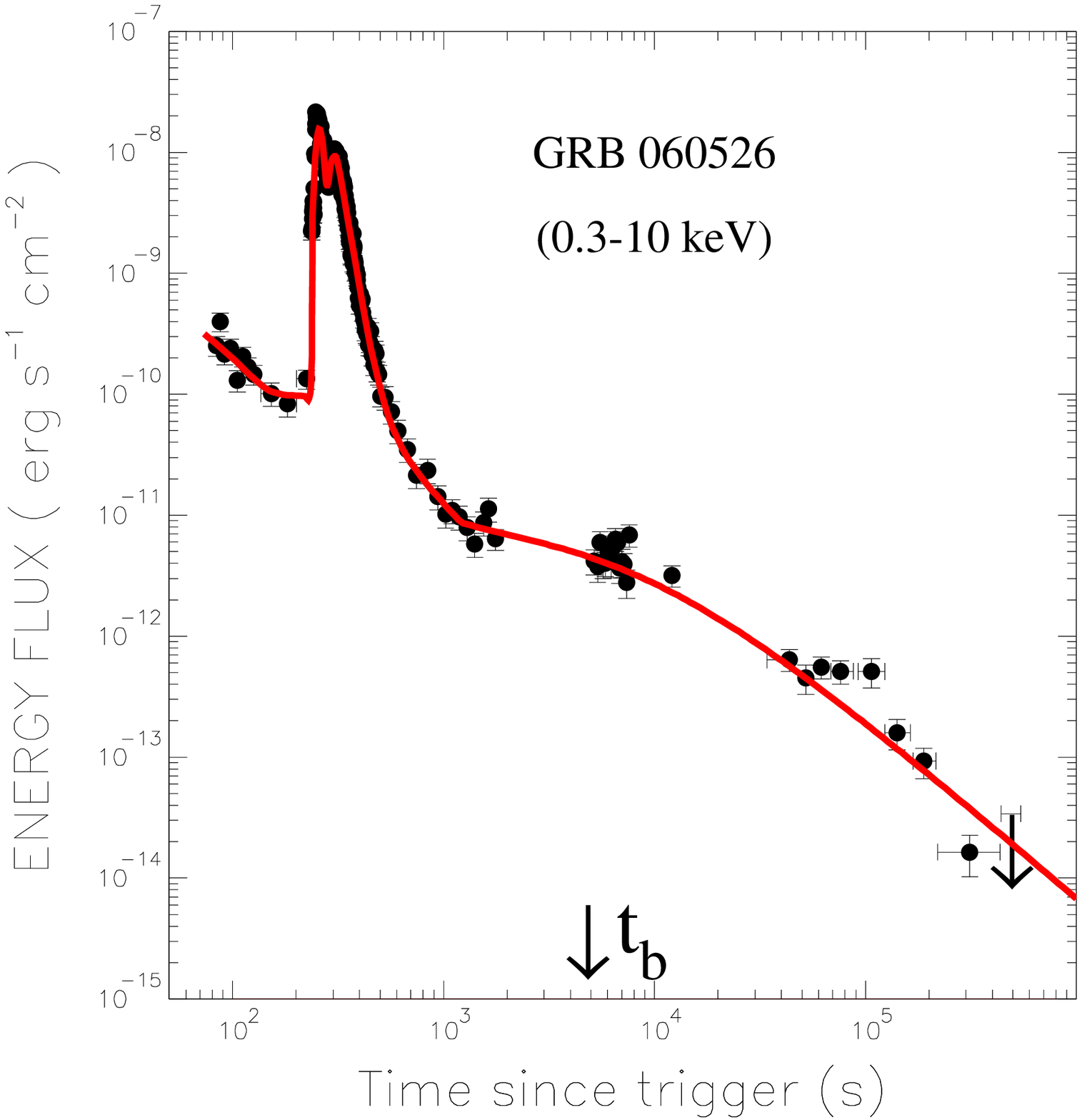,width=7.0cm,height=6.0cm}
\epsfig{file=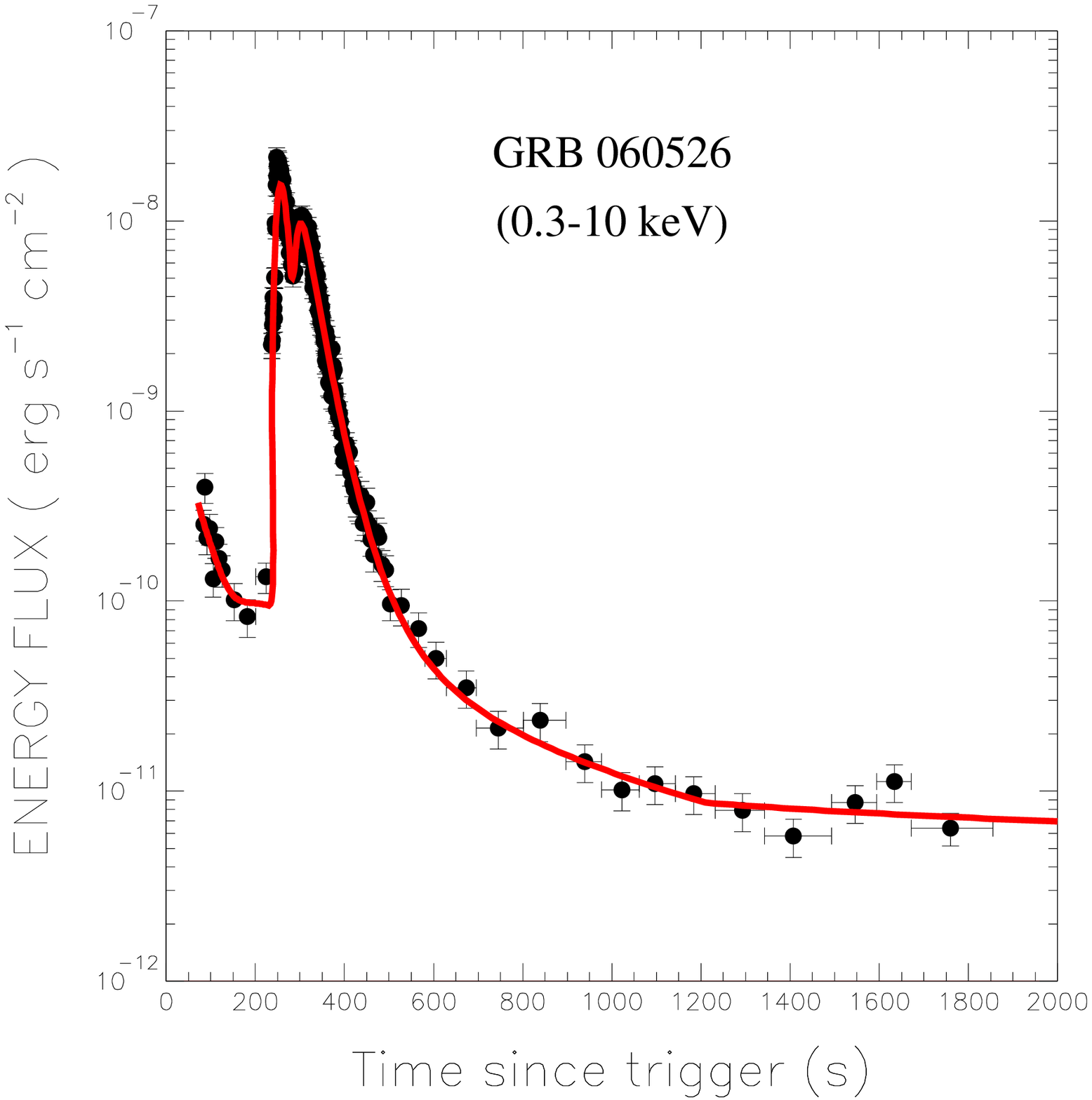,width=7.0cm,height=6.0cm}
}}
\vbox{
\hbox{
\epsfig{file=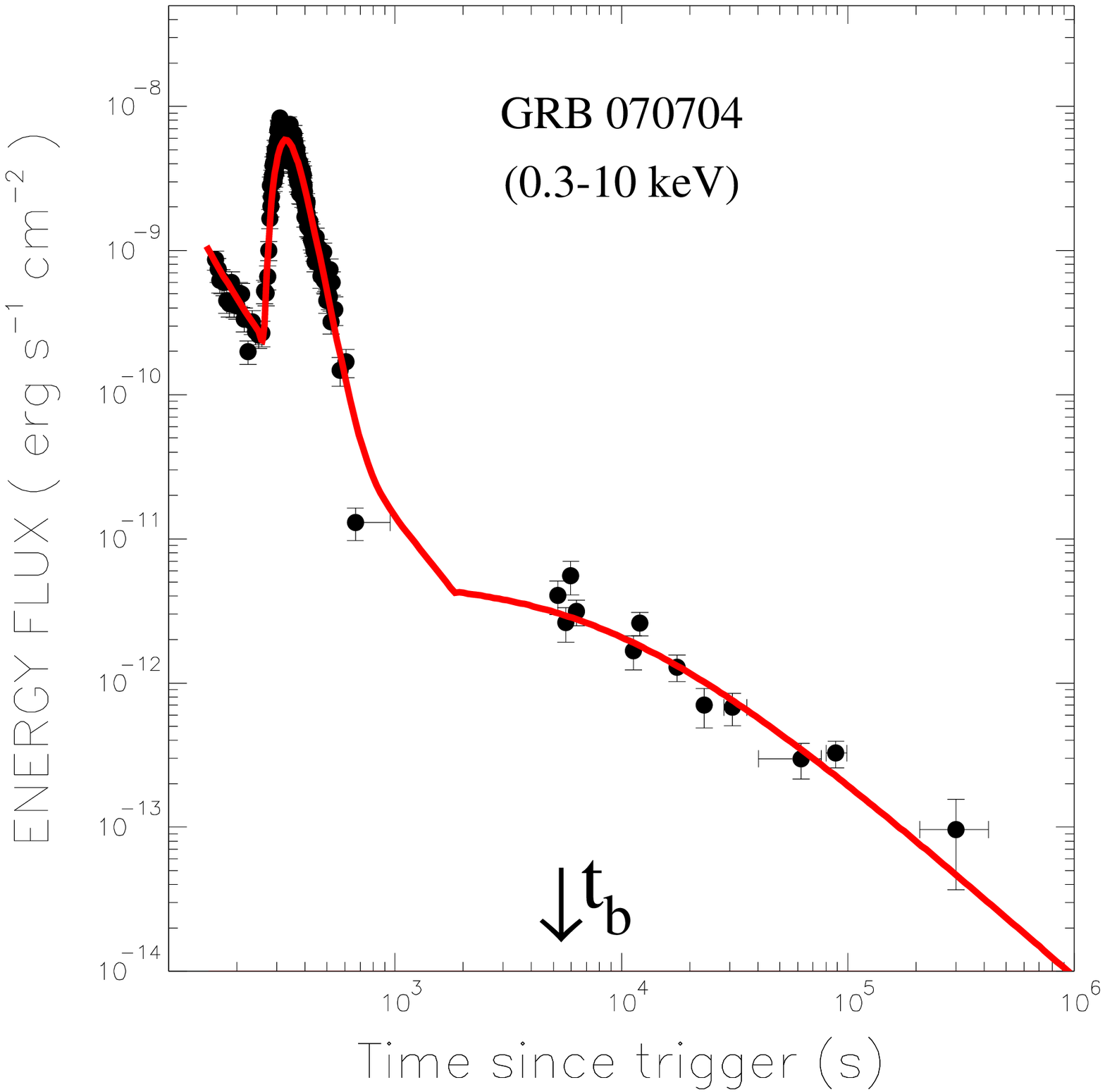,width=7.0cm,height=6cm }
\epsfig{file=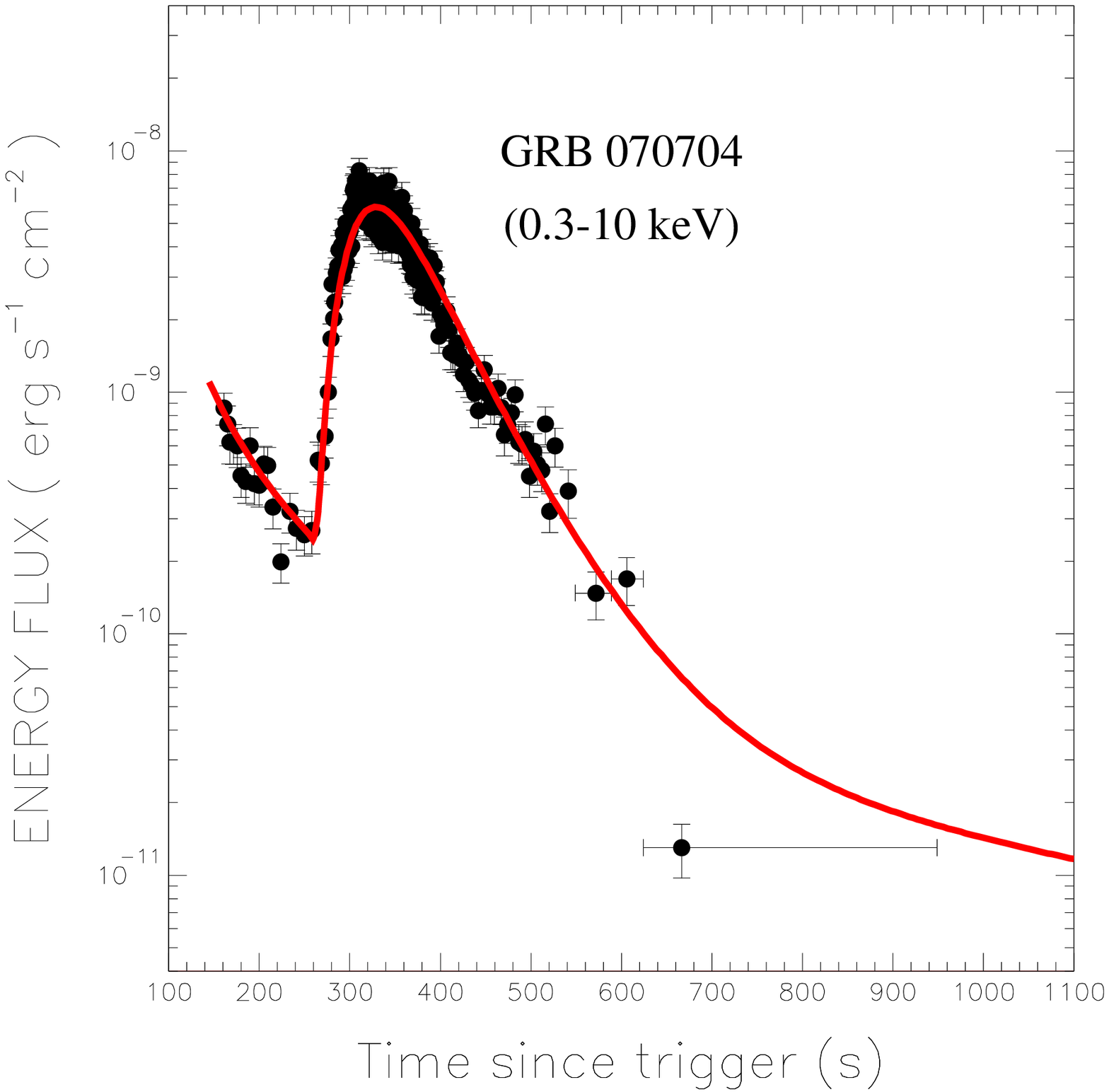,width=7.0cm,height=6cm}
}}
\caption{Comparison between the 0.3-10 KeV X-ray lightcurves
of GRBs with prominent flares in their X-ray afterglow that were 
measured by the Swift XRT and reported in the Swift/XRT lightcurve
repository {\it http://www.swift.ac.uk/xrt$_{-}$curves/}  
\citep{Evans2009} and their CB model description with the parameters 
listed in Table 1. 
{\bf Top left (a):}   GRB 050502B. 
{\bf Top right (b):}  Comparison between the  evolution
                      of the effective photon spectral index of GRB 050502B
                      in the 0.3-10 keV X-ray band as inferred 
                       by \citep{Zhang2007} from the 
                      Swift XRT observations 
                      and that inferred from the CB model lightcurve.
                      Note the similarity between this Fig. and 
                      Figs.~2c,f.
{\bf Middle left (c):} GRB 060526.
{\bf Middle right (d):} Zoom on
                     the prominent X-ray flare
                        in the early-time AG of GRB 060526.
{\bf Bottom left (e):} GRB 070704.
{\bf Bottom right (f):} Zoom on
                       the prominent X-ray flare in the 
                       early-time AG of GRB 070704.
} 
\label{fig3}
\end{figure}

Eq.~(\ref{ICSlc}) with $E_p$ given by Eq.~(\ref{PeakE}) describes well
the shape and the spectral evolution of GRB pulses and of early-time
X-ray flares. In particular, it correctly describes their rapid spectral
softening during their fast decline as demonstrated in 
Figs.~\ref{fig2}c,f, \ref{fig3}b and \ref{fig7}d, and 
in \citet{DDD2008b} for many more cases.

If absorption in the CB is dominated by free-free transitions, then 
$\Delta t(E)\!\propto\! E^{-0.5}$, and for $E\!\ll\!E_p$ the lightcurve 
of an ICF is approximately a function of $E\,t^2$ (the `$Et^2$' 
law'), with a peak at $t\!=\!\Delta t$, a full width at half maximum, FWHM 
$\!\approx\!2\,\Delta t$ and a rise time from half peak value to peak 
value, ${\rm RT\!\approx\! 0.30\,FWHM}$ independent of E.  Note that the 
approximate $E\,t^2$ law makes the fast decline sensitive only to the 
product $E_p  t_p^2$ and not to their individual values. This 
degeneracy in the pulse shape can be removed by inferring $E_p$ from the 
broad band spectrum of the ICF.

The late-time decay of the energy flux of the prompt emission pulses
and ICFs in  an 
energy band $[E1,E2]$, which follows from Eq.~(\ref{ICSlc}), is given 
approximately by:
\begin{equation}
\int_{E1}^{E2} E\, {d^2N_\gamma\over dt\,dE}(E,t)\, dE\approx
A\, {E_p(t)\,\Delta t^2\over t^2}\,
[e^{-E1/E_p(t)}-e^{-E2/ E_p(t)}].
\label{InICSlc}
\end{equation}
Thus, for the Swift XRT lightcurves where $E1\!=\!0.3$ keV and 
$E2\!=\!10$ keV, as long as $E_p(t) \!\gg\! E2\!\geq\!E1$, the energy flux 
in an ICS pulse/flare decays like $t^{-2}$ until it is taken over by the 
SR afterglow. When $E_p(t)\!\gg\! E1$ but $E_p(t)\lsim E2$ 
the energy flux 
decays like $t^{-4},$ and when $E_p(t)\lsim E1$ the energy flux decays 
like $t^{\!-\!4}\, e^{\!-\!E\,t^2/2\, E_p\, t_p^2}$.

For a single ICF beginning at $t$=0, which is superimposed on a
smooth SR afterglow, the effective photon spectral index 
$\Gamma(E,t)\equiv dlog(dN/dE)/dlogE$ is given approximately by 
\citep{DDD2008b}:
\begin{equation}
\Gamma(E,t)\approx [1-\beta_g + E/E_p(t)] 
\Theta(t_{AG}-t)\, \Theta(t)
                   + \Gamma_{SR}\, \Theta(t-t_{AG})\,,
\label{Geff}
\end{equation}   
where $t_{AG}$ is the time when the SR afterglow takes over the ICS 
emission, $\beta_g\!\approx\! 0$ for a glory with thin thermal 
bremsstrahlung spectrum,
and $\Gamma_{SR}$  is the best fit photon spectral index 
of the smooth X-ray AG.

All the above properties are clear fingerprints of ICFs produced 
by highly relativistic CBs fired  in mass accretion episodes 
on the newly formed compact central object. Moreover, the observations
of X-ray flares indicate that their widths are proportional to their 
emission time $t_f$ after the GRB trigger. In the CB model, this 
implies that their peak energy, equivalent isotropic gamma ray energy 
and peak luminosioty decrease like
$E_p\!\propto\!t_f^{-1}\,, $ $E_{iso}\!\propto\! [(1+z)/t_f]^2$ 
and $L_p\!\propto\! [(1+z)/t_f]^3\,.$

\section{The SR lightcurve} 

In the CB model, the lightcurve in the 
observer frame of the SR emitted by a CB is given by 
the master formula \citep{DDD2002, DDD2009a}: 
\begin{equation}
F_\nu[t] \propto A(\nu,t)\,{n(t)\, R(t)^2\ \gamma(t)^2\, \delta(t)^4
    \over \nu_b(t)}\, \left[{\nu\over \nu_b(t)}\right]^{-1/2}\,
         \left[1+{\nu\over \nu_b(t)}\right]^{(1-p)/2}\,,
\label{FnuSR} 
\end{equation}
where $t$ is the time after the ejection of the CB, $R(t)$ is its radius, 
$n$ is the density along its trajectory, $A(\nu,t)$ is the attenuation of 
radiation along the line of sight to it, $\nu_b(t)$ is the typical 
frequency in the observer frame of the SR emitted by the electrons that 
are swept into it at time $t$ with a relative Lorentz factor 
$\gamma(t)$,
\begin{equation}
\nu_b(t)\propto {n^{1/2}\, \gamma(t)^3\, \delta(t)\over 1+z }\,,
\label{FnuSRapp} 
\end{equation}
and $p\!\sim\! 2.2$ is the spectral index of the Fermi accelerated 
electrons in it.

\subsection{The early time SRFs}

The SR radiation that is emitted from the encounter of a CB with the 
wind/ejecta of the progenitor star, with a density profile 
$n(r)\!\propto\!e^{\!-\!a/(r\!-\!r_w)}/ 
     (r\!-\!r_w)^2$ for $r\!>\!r_w$ and $n(r)\!=\!0$ for $r\!<\!r_w$, 
follows from Eq.~(\ref{FnuSR}) and 
is given approximately by \cite{DDD2009a}: 
\begin{equation} 
F_\nu \propto  {e^{-a/t}\, t^{1-\beta} \over t^2+t_{exp}^2}\, 
\nu^{-\beta}\,,
\label{SRP} 
\end{equation} 
where $t\!=\!T\!-\!T_w$ with $T$ being the time after trigger and 
$T_w$ 
the time at the CB-wind encounter, $t_{exp}$ is the typical slow-down 
time of the fast CB expansion, $\beta\!=\!\Gamma\!-\!1$, and the 
exponent describes the decreasing attenuation of the emitted radiation 
when the CB penetrates the wind and/or the initial rise 
in the wind density due to an exponential cutoff of the wind ejection.  

Note that for $t^2\!\gg\!t_{exp}^2 $ 
the asymptotic decline of an SRF is a simple power law
\citep{DDD2003},
\begin{equation}
F_\nu[t] \propto   t^{-\Gamma}\, \nu^{-\Gamma+1}\, ,
\label{SRtail}
\end{equation}
while that of an ICF is, 
\begin{equation}
F_\nu[t]\propto  t^{-2}\, E^{-\beta_g}\, e^{-E/E_p(t)}
    \sim t^{-2}\,  E^{-\beta_g}\, e^{-E\,(t^2+t_p^2)/2\, E_p\, t_p^2}\,.  
\label{ICStail}
\end{equation}
Thus, ICFs and SRFs may be distinguished by their different 
tempo-spectral evolution.

In the X-ray band, early-time ICFs are usually much brighter than their 
following SRFs. But, due to their rapid late-time decay,  occasionally 
the $\!\sim\! t^{-\Gamma}$ tail (Eq.~\ref{SRtail}) of the SRF, 
which follows an ICF, can be seen before the plateau/shallow decay  phase 
of the early-time X-ray AG takes over (see, e.g., Fig.~3 in 
\citet{DDD2009a}).

The early time SR has usually $\beta_{OX}\!\sim\!0.75$ and $F_\nu$ that
decreases strongly between the optical and the X-ray band. Consequently, 
although ICS dominates the prompt X-ray emission, in the optical band SRFs 
are usually much brighter than their preceding ICFs, which typically 
have $\beta\!\sim\!0$ around their peak time. Consequently, optical flares 
are usually SRFs. The evolution of the effective spectral index during an 
optical SRF that  follows from Eq.~(\ref{FnuSR}) has the simple 
form,
\begin{equation}
\beta \equiv {dlog\, F_\nu[t]\over dlog\, \nu}=
                    1/2+(\beta_X-1/2)\, {\nu\over \nu +\nu_b}\,.
\label{betaO}
\end{equation}
In early-time optical SRFs, $\nu_b$ that is initially below the optical 
band can cross above it as $n$ of the wind/ejecta increases and then cross 
back as the density decreases. Such a change in $\beta_O$, from 
$\beta_O\!\sim\!\beta_X$ towards $\beta_O\!\sim\!1/2$ and back to 
$\beta_O\!\sim\!\beta_X$, has been observed in some early time optical 
flares, e.g., in GRB 071031 \citep{Kruhler2009}.

\subsection{The late-time SR afterglow} 

When the merged CBs coast through the constant-density ISM,
their SR in the X-ray band is well above the cooling frequency 
of the Fermi accelerated electrons and their unabsorbed AG
as given by Eq.~(\ref{FnuSR})
reduces to \citep{DDD2009a}: 
\begin{equation}
F_{ISM}[\nu,t] \propto 
           \gamma(t)^{3-\beta_X}\, \delta(t)^{3\,\beta_X+1}\, 
\nu^{-\beta_X}
\label{Fnu}
\end{equation}
where $\delta\! =\! 1/\gamma\, (1\!-\!\beta\, cos\theta)$ is
their Doppler factor with $\theta$ being the angle between the line of 
sight
to the CBs and their direction of motion. For $\gamma^2 \gg 1$ and 
$\theta^2
\ll 1$, $\delta \approx 2\, \gamma/(1\!+\!\gamma^2\, \theta^2)$ to an
excellent approximation. In the CB model 
the canonical value of the spectral index well above 
the bend frequency $\nu_b$ has the value $\beta_X\!\approx\!1.1\,.$  
For a CB of a baryon number 
$N_{_B}$,  a radius $R$ and an initial Lorentz factor $\gamma_0$, 
relativistic energy-momentum
conservation yields  the deceleration law of the CB in 
an ISM with a constant density $n$ \citep{DDD2009a}:
\begin{equation}
\gamma(t) = {\gamma_0\over [\sqrt{(1+\theta^2\,\gamma_0^2)^2 +t/t_0}
          - \theta^2\,\gamma_0^2]^{1/2}}\,,
\label{goft}
\end{equation}
with $t_0={(1\!+\!z)\, N_{_{\rm B}}/ 8\,c\, n\,\pi\, R^2\,\gamma_0^3}\,.$
As can be seen from Eq.~(\ref{goft}), $\gamma$  and $\delta$
change little as long as $t\!\ll\! t_b\!=\![1\!+\gamma_0^2\,
\theta^2]^2\,t_0\,, $ and Eq.~(\ref{Fnu}) yields the {\it `plateau'}
phase of `canonical AGs' \citep{Nousek2006}.
For $t\!\gg\!t_b$, $\gamma$  and $\delta$ decrease like $t^{-1/4}\,.$
The transition $\gamma_0\!\rightarrow\!
\gamma_0\,(t/t_0)^{-1/4}$
around $t_b$
induces a bend, the so called `jet  break',
in the synchrotron AG
from a plateau to an asymptotic power-law decay,
\begin{equation}
F_{ISM}[\nu,t] \propto t^{-\beta_X-1/2}\,\nu^{-\beta_X}\, .
\label{Asymptotic}
\end{equation}
Thus, the shape of the entire lightcurve of the SR afterglow 
from a CB that enters the constant density 
ISM depends only on three parameters, the  product $\gamma_0\, \theta$,
the deceleration parameter $t_0$ (or the break time $t_b$)
and the spectral index $\beta_X$. The post break decline is 
given by the  simple power-law (Eq. \ref{Asymptotic}) independent of 
the values of $\gamma_0\, \theta$ and $t_b$.
In cases where $t_b$ is earlier
than the beginning of the XRT observations or is hidden under 
the prompt emission, the entire observed lightcurve of the AG
has this asymptotic power-law behaviour \citep{DDD2008a}.

\subsection{Late-time SRFs}

The lightcurve of late-time SRFs strongly depend on the
density profile of the density bumps. For a wind-like 
density jump $n\!\propto\! e^{-a_w/(r\!-\!r_w)}/(r\!-\!r_w)^2$ beyond 
$r\!=\!r_w$, the lightcurve is given by Eq.~(\ref{SRP}).
Such a density profile is expected for the boundaries of star formation 
regions where GRBs usually take place. At late times both the optical 
and the X-ray bands are above the bend frequency, 
$\beta_O\!\approx\!\beta_X$, and the asymptotic decline 
of the corresponding late-time SRF is given by:
\begin{equation}
F_\nu[t] \propto t^{-\Gamma}\,\nu^{-\Gamma-1}\, ,
\label{Asymptoticrm2}
\end{equation}
where $t\!=\!T\!-\!T_w$ is the observer-time after the CB has reached 
$r_w$. 
At late times, the bend frequency is well below the optical band and 
remains so during the crossing of density bumps. Thus, no 
detectable spectral variation is expected in late-time SRFs.

\subsection{Correlations}
Because of the large bulk motion Lorentz factors of the jets of CBs, 
Doppler boosting, relativistic beaming and time aberration yield
strong dependence of their observed radiations 
on $\gamma$ and $\delta$. This dependence
dominates the flare observables and can be used to
correlate triplets of independent flare observables without knowing the
exact values of $\gamma$ and $\delta$. Moreover, many independent
observables depend on the same combinations of $\gamma$ and $\delta$,  
which results in pair correlations. Finally, due to selection effects,
various observables depend strongly only on the Lorentz factor or on the
Doppler factor, which also yields pair-correlations. These 
correlations between various radiations, between flare observables
and between flare and  GRB observables
are discussed in detail in \citet{DD2010}.

\section{Case Studies}

Although in the CB model early-time optical SRFs follow the X-ray ICFs, 
they have a different origin and a different tempo-spectral evolution. 
Thus, we shall compare separately the observational data on X-ray flares 
and optical flares and the CB model predictions.

\subsection{X-ray flares} 

In the CB model, the X-ray emission in GRBs, XRFs and SHBs during the 
prompt emission phase is a sum of X-ray flares that are part of their 
prompt gamma ray pulses. That has been shown repeatedly in CB model 
publications (see, e.g., \citet{DDD2009a,DDD2009b}) and independently by 
many other authors. We have also shown that the fast decline phase of 
their prompt emission with its rapid spectral softening is just the tail 
of the last prompt emission pulses, because the exponential factor in 
Eq.~(\ref{ICSlc}) suppresses very fast the relative contribution of the 
earlier pulses by the time the data sample the later pulses or flares. 
This is demonstrated in Figs.~2a,b,c where we compare between the 
`canonical X-ray lightcurve' \citep{Nousek2006} of GRB 060729 and the 
evolution of its spectral index as inferred by \citet{Zhang2007} and the 
CB model descriptions of these lightcurves. This has already been 
demonstrated in detail in \citet{DDD2008b,DDD2009a} 
for many other GRBs and XRFs. It is also the case for SHBs as shown
in Figs.~2e,2f for SHB 050724 and in \citet{DDD2009b} for many other 
SHBs. GRBs with a non-canonical X-ray AG, are 
simply GRBs where the emission of SR begins to dominate the 
X-ray lightcurve before the fast decline phase of the prompt emission.
Such a case is shown in Fig.~3d where we compare
the XRT lightcurve of GRB 060418 and its CB model description. 
Many other cases are shown in \citet{DDD2008b,DDD2009a,DDD2009b}.

In order to have a stringent test of the CB model interpretation of X-ray 
flares during the afterglow phase of GRBs, we have selected a sample of 14 
GRBs with X-ray lightcurves that are reported in the Swift/XRT GRB 
lightcurve repository \citep{Evans2007,Evans2009} and have well sampled 
prominent flares during their AG phase. This sample includes GRBs 050502B, 
050916, 060526, 060929, 070704, 080506, 080607, 080810, 080906, 081102, 
090417B, 090621A, 090709A, and the SHB 050724. For all these GRBs we have 
fitted the entire XRT lightcurve with the master formulae of the CB 
model. In order to minimize the number of adjustable parameters in the 
theoretical lightcurves, we adopted the standard simplifying CB model 
assumptions \citep{DDD2009a}: The burst environment is a cavity full of 
thin bremsstrahlung optical photons (glory light) enclosed within a 
wind/ejecta that have a density profile $n\!\propto 
e^{\!-\!a_w/(r\!-\!r_w)}/(r\!-\!r_w)^2$ beyond $r_w$ until the density is 
taken over by the constant density of the ISM. The CBs were taken to be 
well separated during the prompt emission phase and to be well represented 
by a single effective CB during the afterglow phase.  The latest one or 
two observed pulses/flares in the prompt emission were assumed to dominate 
its fast decay. This fast ICF decay is overtaken by the tail of the SR 
emission from the encounter of the CBs with the wind/ejecta as given by 
Eq.~(\ref{SRP}), or by the plateau phase of the SR afterglow emitted from 
the decelerating CB in the constant-density ISM. The smooth afterglow was 
calculated from Eqs.~(\ref{Fnu}), (\ref{goft}), using best fit values of 
the normalization, $\gamma_0\,\theta$, $t_0$, and $p/2$=$\beta_X$ with 
$\beta_X$ within the error range reported for $\Gamma\!=\!\beta_X$+1 in 
the Swift X-ray repository \citep{Evans2009}.

Prominent X-ray flares during the afterglow with a rapid spectral 
softening during their decline phase were assumed to be ICS pulses/flares. 
Such flares were superimposed on the CB model smooth SR afterglow. Flares 
with a constant hardness ratio similar to that of the smooth SR afterglow 
were assumed to be SRFs. They were generated by introducing density bumps 
with a windy profile into the master formula of the SR lightcurve 
\citep{DDD2009a}.

The fitted parameters of the CB model descriptions of the 14 X-ray light 
curves of the above GRBs are listed in Tables \ref{t1},\ref{t2}. When 
only 
the tail of the SRF was visible we used its parameter-free asymptotic 
form, Eq.~(\ref{SRtail}).  Because of the use of simplifying assumptions, 
the values of the parameters may be  effective 
values and not true values.  
Therefore, we refer to the CB model fits as `descriptions' rather than 
as best fit predictions.  
In order to avoid 
repetitions and an excessively long section, we limit our detailed 
discussion to three representative cases, GRB 050502B representing long 
soft GRBs, SHB 050724 representing short hard bursts, and the recent 
peculiar GRB 090709A with a suspected  8 s periodicity in its prompt 
emission.

\noindent
{\bf GRB 050502B} was studied in detail by \citet{Falcone2006}. The XRT 
began taking data 63 s after the BAT trigger and 
followed its X-ray lightcurve until 10.6 days after burst. The measured 
lightcurve in the 0.3-10 keV band is shown in Figure 2a. Following an 
initial low-flux level, the XRT detected a giant X-ray flare, which 
began 
at $345\!\pm 30$ s, reached a peak value around 770 s, with intensity more 
than 500 times larger than that of the underlying afterglow. The fluence 
of the flare, $(1.0\!\pm\! 0.05)\times 10^6\, {\rm erg\, cm^{-2}}$ in the 
0.2-10.0 keV energy band, exceeded that of the prompt emission measured 
with the Swift BAT in the 15-350 keV energy band. After several hours, two 
weaker flares in the X-ray emission occurred consecutively beyond which 
the decay of the X-ray lightcurve became steeper. Except for the giant 
flare, the spectrum of the afterglow and the two late flares was well fit 
by a power-law with a photon spectral index $\Gamma$=1.945 (+0.077, 
-0.100) \citep{Evans2007, Evans2009}. The spectrum of the flare was well 
fit with a power-law with an exponential cutoff; however, due to its non 
detection by the BAT, its value could not be well determined from the 
spectrum measured only by the XRT. The photon spectral index of the giant 
flare before peak-time was $\Gamma\!\approx\!1$, much harder than that 
measured before and after the flare. During the fast decline phase of the 
flare, a rapid spectral softening took place and the spectral index 
increased rapidly to a value well above the constant value of the smooth 
afterglow.

\noindent 
CB model fits to the X-ray lightcurve of GRB 050502B and the 
evolution  of its spectral index are shown in Figs.~3a,b. They show 
that the observed lightcurve and spectral evolution of the giant X-ray 
flare are well described by the master formula (Eq.~(\ref{ICSlc})) of the 
CB model for an ICF. As can be seen in Figs.~\ref{fig3}a,b, the fast decay 
and spectral softening stopped simultaneously when the AG was taken over 
by the smooth SR from the decelerating CB in an ISM of a constant density. 
The late-time afterglow shows two flares superimposed on the smooth AG, 
similar to those observed in many long GRBs. Their spectrum, which is 
similar to the smooth AG, suggests that they were produced by enhancement 
of the emitted SR when the CB encountered density bumps in its voyage 
through the host's ISM. Such late flares with a typical SR spectrum and 
little spectral evolution produced by density bumps along the CB 
trajectory in the ISM add no specific information on the origin of GRBs.

\noindent
{\bf SHB 050724} at redshift z = 0.257 was studied in detail by 
\citet{Campana2006}, citet{Grupe2006} and \citet{Malesani2007}. 
The Swift BAT triggered on the burst at 12:34:09 UT on July 24, 2005. The 
burst had $T_{90}$=3.0$\!\pm 1.0\!$ s, but most of the energy of the 
initial SHB was released in a hard spike with a duration of 0.25 s. The 
bulk of the burst energy was not emitted in the short initial spike but in 
an extended soft emission component that lasted $\!\sim\!150$ s. Swift 
XRT began observing the afterglow 74 s after the BAT trigger. The Chandra 
X-ray observatory performed two observations, two days and about three 
weeks after the burst. The complete X-ray lightcurve is shown in Fig.~2e. 
It has a rapid decay with a fast spectral softening ending with a sharp 
transition to a shallower decay with a much harder power-law spectrum, 
$\Gamma\!=\!1.79\!\pm\!0.12$ (Swift repository, \citet{Evans2009}).  The 
AG steepens gradually into a late power-law decay. A large flare 
superimposed on the canonical lightcurve occurred around 50 ks after burst 
with a fluence of $\sim$7\% of that of the prompt burst. The flare has 
been 
detected also in the optical and NIR bands, e.g., \citet{Malesani2007}. 
Spectral analysis of the XRT data \citep{Campana2006} showed no 
evolution during the afterglow phase, including the large late flare. 
Spectral analysis of the Chandra observations from the fading tail of this 
flare confirmed this result \citep{Grupe2006}. The burst took place 2.5 
kpc (in projection) from the center of an elliptical host galaxy 
\citep{Malesani2007}.

\noindent
CB model fits to the X-ray lightcurve of SHB 
050724 and the evolution of its spectral index are shown in 
Figs.~2e,f. The 
initial fast decay with a rapid spectral softening is described 
in the CB model by the 
tail of an ICF. As can be seen in Figs.~2b,2c, the fast decay and 
spectral softening stopped simultaneously when the AG was taken over by 
the SR from the decelerating CB in an ISM of a constant density. As 
expected, the late-time SR afterglow is 
similar 
in shape to the SR afterglow of long GRBs. Also the late-time flare 
superimposed on the canonical lightcurve is similar to those observed in 
many long GRBs. Its spectrum, which is similar to the smooth AG, suggests 
that it was produced by enhancement of the emitted SR when the CB 
encountered a density bump in its voyage through the host's ISM. The 
Chandra data show that the canonical AG continued to decay after the flare 
with the same slope and the same spectral index, $\beta_X\!=\!0.79\!\pm\! 
0.15$, as that of the AG before the SRFs. As shown in Fig.~2e, the 
complete XRT lightcurve is well described by the CB model. Moreover, the 
CB model relation for the asymptotic decline of the smooth AG is well 
satisfied:  The temporal behaviour of the smooth AG was best fit with 
p=1.56, implying an unabsorbed spectral index, $\beta_X$=p/2=0.78, in 
agreement with that inferred from the XRT and Chandra observations. The 
observed elliptical host galaxy of SHB 050724 was argued to provide strong 
support for a neutron star merger origin of this SHB. But, it was pointed 
out that neutron star mergers do not produce the late accretion episodes 
needed to power a late central activity that 
could produce the large flare around 50 ks after burst \citep{Grupe2006}. 
In the CB model, a late flare with a typical SR spectrum and 
little spectral evolution is produced by density bumps along the CB 
trajectory in the ISM. Such flares neither rule out nor support any 
specific origin of the SHB.

\noindent

{\bf GRB 090709A} triggered the Swift BAT on July 9, 2009 at 07:38:34 UT 
\citep{Morris2009}. Its prompt emission lightcurve within 100 seconds 
after trigger appeared to show a quasi periodic variation with a period of 
$\!\sim\!8$ seconds \citep{Markwardt2009}, which may have also seen in  
independent measurements of its lightcurve with Konus-Wind 
\citep{Golenetskii2009}, INTEGRAL \citep{Gotz2009}, and Suzaku 
\citep{Ohno2009}. 
The Swift XRT began its follow-up observations 68 seconds after the BAT 
trigger \citep{Morris2009}. However, analysis by 
\citet{Mirabal2009} of the XRT data during 79-469 seconds after the BAT 
trigger did not 
reveal any significant periodicity. From deep optical observations 
that did not detect its host galaxy, 
it was concluded that GRB 090709A took place 
either in the Milky Way or at a redshift between 8 and 10
\citep{Castro2009}. The large 
redshift and the fluence of $9.1\times 10^{-5}\, {\rm erg\, cm^{-2}}$ 
measured with Konus-Wind in the 20 keV-3 MeV energy range imply an 
isotropic equivalent gamma-ray energy between $8.6\times 10^{54}$ and 
$1.1\times 10^{55}$ ${\rm erg\,cm^{-2}}$, which makes GRB 090709A more 
luminous than any GRB with known redshift.

Figs.~\ref{fig6}e,f present a comparison between the XRT 
lightcurve of GRB 090709A 
and its CB model description assuming it was an ordinary GRB. The early 
time flare around $t$=90 seconds probably is the X-ray part of the prompt 
emission pulse with a peak around 87 seconds in the Swift BAT lightcurve 
that is slightly delayed relative to the gamma-ray peak, as expected from 
the $E\,t^2$ law of the CB model. Its fast decline and rapid softening are 
those expected from an ICF. They are taken over by SR around 150 
seconds after the BAT trigger. The next 3 peaks probably are SR peaks as 
suggested by their shape and their hardness ratio that is roughly the 
same as that of the SR afterglow. The best fit CB model lightcurve of the 
smooth AG yields $p$=2.16, which, in the CB model, implies 
$\beta_X\!=\!1.082$, in good agreement with the best fit spectral index 
reported in the Swift repository \citep{Evans2009}, 
$\beta_X\!=\!\Gamma\!-\!1\!=\!$1.081 (+0.076, -0.074). All together, GRB 
090709A looks like a normal GRB with a normal early time flaring 
activity that took place at a relatively very large redshift, and its 
X-ray lightcurve is well reproduced by the CB model.

\subsection{Optical flares}

The CB model predictions for early-time and late-time flares in the 
optical lightcurves of GRBs are compared with observations 
for a representative set of GRBs in 
Figs.~\ref{fig7}-\ref{fig11}. For all these GRBs we show both their 
measured optical and X-ray lightcurves (if available) and their CB model 
descriptions with the parameters listed in Tables~\ref{t1}-\ref{t5}.

Only in a very few bright GRBs was the prompt optical emission  
resolved into separate flares. Two such cases, GRB 080319B \citep{Wozniak2009}
and GRB 071003 \citep{Perley2008}
are shown in Fig.~\ref{fig7}. In most GRBs and XRFs, the prompt optical 
emission that appears as a single extended flare,
probably, is a sum of unresolved 
flares. Cases where the prompt optical flare was partially resolved 
into a sum of flares or where there is  
clear evidence for overlapping flares are, 
e.g., GRB 080330 \citep{Guidorzi2009},
XRF 071031 \citep{Kruhler2009}, GRB 061007 \citep{Rykoff2009},
which are shown in Figs.~\ref{fig8},\ref{fig9}. In most GRBs 
where the prompt optical emission was detected with robotic telescopes 
from the ground or from space (with the Swift/UVO), the prompt emission 
appears like a single flare, probably as a result of either strongly
overlapping flares, or insufficient temporal resolution due to low 
statistics, or because the prompt emission did consist of a single flare. 
Examples of GRBs with a `single' prompt optical flare are 
990123 \citep{Akerlof1999}, 030418 \citep{Rykoff2004}, 050820A
\citep{Vestrand2006}, 081203A \citep{Kuin2009}, 090102 \citep{Gendre2009}, 
090618 \citep{Li2009} and 091029 \citep{LaCluyze2009}, which are shown in 
Figs.~\ref{fig9}-\ref{fig11}. Examples of bright GRBs with well resolved 
late-time optical flares include GRB 030329 \citep{Lipkin2004}
and GRB 060206 \citep{Wozniak2006} shown in Fig.~\ref{fig11}f.
However, in order not to inflate this paper, only GRB 080319B, XRF 
071031, GRB 061007 and GRB 990123 are discussed here in detail.

\noindent {\bf GRB 080319B} at redshift $z\!=\!0.937$, the brightest GRB 
observed so far,was simultaneously detected by the Swift-Burst Alert 
Telescope (BAT) and the Konus gamma-ray detector aboard the Wind satellite 
\citep{Racusin2008, Golenetskii2008}. The location of GRB 080319B was 
fortuitously only 10$^o$ away from GRB 080319A, which was detected by 
Swift less than 30 minutes earlier, and allowed several wide field 
telescopes to detect the optical emission of GRB 080319B instantly. It 
started after the beginning of the prompt keV-MeV emission and it peaked 
26 s after the Swift trigger at magnitude $V\!=\!5.3$ \citep{Racusin2008, 
Wozniak2009} visible to the naked eye. The extreme brightness of the burst 
and its gamma-ray, X-ray and $UVOIR$ afterglows led to a flurry of 
follow-up observations with a variety of space- and ground-based 
telescopes, which were summarized by \citet{Bloom2008, Racusin2008, 
Wozniak2009}. Its isotropic equivalent gamma-ray energy release was 
$E_{iso}\!\approx\! 1.3\times 10^{54}$ erg, similar to that of GRB 990123. 
The fast spectral variation of its hard X-ray and gamma ray emission was 
well parametrized with an exponentially cut-off power-law with a cut-off 
energy that was strongly correlated with the peak structure of the 
lightcurve and a low-energy photon spectral index, $\Gamma\!\approx\!1$, 
which changed abruptly into $\Gamma\!\approx\!2.1$ after the fast decay 
phase of the prompt emission. The optical and gamma-ray lightcurves 
during the explosion were not correlated (see, e.g., Fig.~1 in 
\citet{Racusin2008}): The onset of the optical emission lagged behind the 
gamma ray emission by several seconds and decayed more slowly at the end 
of the prompt emission. The typical time scales of their temporal 
variability were entirely different. The extremely bright optical emission 
could not be reconciled with a single emission mechanism - extrapolating 
the gamma-ray spectrum to the optical band underestimates the optical flux 
by more than 4 orders of magnitude. Their spectra were also quite 
different. Contrary to fireball (FB) model expectations, the X-ray and 
$UVO$ afterglow lightcurves were also chromatic, with no obvious `jet 
breaks' and with spectral and temporal power-law decays that did not 
satisfy the closure relations expected in the FB model (see, however, 
\citet{Bloom2008,Racusin2008, Wozniak2009, Kumar2008} for attempts to 
reconcile the observations with the fireball model).

The prompt $\gamma$-ray and hard X-ray emission in GRB 080319B is composed
of many narrow peaks (see Fig.~1 in \citet{Racusin2008}), most of which
are not well resolved.  Its  0.3-10 keV X-ray lightcurve 
measured with the Swift XRT \citep{Racusin2008} and its CB
model description assuming a constant ISM density
until around $t\!\sim\! 4\times 10^5$ s, presumably when 
the CB escaped the star formation region into 
the halo of the host galaxy, are shown in Fig.~7a.   The best fit
parameters are listed in Table~\ref{t1}. 
The late time
temporal decay of the X-ray AG is well described by a power-law with
$\alpha_X\!=\!1.54\!\pm\! 0.04 $, except around 40 ks, where the
lightcurve is poorly sampled. This value of $\alpha_X$ satisfies
well the CB model closure relation 
$\alpha_X$=$\beta_X\!+\!1/2$=$1.53!\pm\!0.064$.
As  expected for GRBs with large measured
$E_p$, $E_{iso}$ and $L_p$ \citep{DDD2008b}, no AG break was observed 
in the XRT lightcurve before $6\times 10^5$ s. The wiggling of the 
measured lightcurve around
a power-law decay is probably due to variations in the
ISM density along the CB trajectory, which we have not tried to
parametrize. A late-time SRF with a typical late-time decay 
$\!\sim\!(t\!-\!t_f)^{\!-\!2.1}$
probably was observed around $6\times 10^5$ s.

In Fig.~7b we compare the measured $R$-band (and $V$ band renormalized to 
the $R$ band) lightcurve of GRB 080319B \citep{Racusin2008} and its CB 
model description assuming that the initially expanding 3 CBs merged into 
a single CB by the end of the prompt ICS emission of gamma-rays and hard 
X-rays around 300 s (observer time), which decelerates in roughly a 
constant density ISM. The afterglow parameters are listed in 
Table~\ref{t3}. The `missing jet break' probably is hidden under the 
prompt emission. Shown also is the contribution to the $R$-band afterglow 
from an SN akin to SN1998bw \citep{Galama1998b} displaced to the GRB site.

The early-time optical emission that was resolved into three 
prominent peaks and its CB model description in terms of 3 SR peaks, 
each one described by Eq.~(\ref{SRP}) with the parameters listed in 
Table~\ref{t3}, are compared in Fig.~7c. The decay of the prompt 
emission can also be reproduced assuming a single CB crossing 3 wind 
shells, which were ejected by the progenitor star long before its SN 
explosion, rather than 3 CBs crossing a continuous pre-supernova wind 
blown by the SN progenitor.

In Fig.~7d, the observed evolution of the spectral index in the 15-150 keV 
band during the prompt emission and the afterglow \citep{Racusin2008} are 
compared with that predicted by the CB model. The predicted sharp 
transition from the prompt emission, which is dominated by ICS of 
thin bremmstrahlung ($\Gamma\!\approx\!1$), to SR 
($\Gamma_x\!\approx\!2.1$), which 
dominates the emission once the CBs encounter the progenitor's 
wind/ejecta, is clearly observed.

\noindent 
{\bf XRF 071031} was studied in detail by Kruhler et al.~(2009). The 
Swift/XRT began follow-up observations of XRF 071031 103 s after the 
burst. 
The early XRT lightcurve (XRT lightcurve repository, \citet{Evans2009} is 
dominated by bright flares at around 120 s, 150 s, 200 s, 250 s and 450 s. 
The late X-ray data exhibit re-brightenings at 5.5 ks, 20 ks and 55 ks 
superimposed on a smooth AG. The complete XRT lightcurve was reproduced 
with the CB model, assuming 7 early-time ICFs plus two late time SRFs 
superimposed on a smooth SR afterglow. Their parameters are listed in 
Table \ref{t2}. As shown in Figs.~\ref{fig8}c,d the CB model describes 
well 
the very 
complex XRT lightcurve ($\chi^2/dof\!=\!431/426$). The values of the 
photon spectral index of the late time afterglow, $\Gamma_X$=1.86 (+0.14, 
-0.15), and the index of the late time power-law decay, 
$\alpha_X\!=\!1.4\!\pm 0.1\!$, which were inferred from the Swift/XRT 
observations (Swift XRT lightcurve repository, \citet{Evans2009}, satisfy 
well the CB model closure relation, $\alpha_X\!=\!\Gamma_X$-1/2.

The emission in the optical band was detected and followed up from the 
ground in automated observations by GROND \citep{Kruhler2009} with good 
temporal resolution, which began 225 s after burst and lasted nearly 7 h. 
The white lightcurve obtained by adding the various optical bands, in 
order to increase sensitivity, shows a broad peak with clear deviations 
from a power-law rise and decay. Fig.~\ref{fig8}e shows the CB model 
description of 
this lightcurve in terms of strongly overlapping 4 SRFs following their 
preceding ICFs in the X-ray lightcurve. The parameters used in the CB 
model 
description are listed in Table~\ref{t3}. Comparison between the 
optical 
spectral index lightcurve that was inferred by \citet{Kruhler2009} 
and the CB model prediction as given by Eq.~(\ref{betaO})  
is shown in Fig.~\ref{fig8}f.

\noindent {\bf GRB 061007} at a redshift $z$=1.26 \citep{Jakobsson2006}, 
which was detected by 
Swift, had a T90 duration of 75.3 s in the Swift BAT 15-350 keV band, and 
consisted of three large peaks with a long faint tail 
\citep{Markwardt2006, Sakamoto2008}. The Swift XRT began 
observations 80 s 
after the start of the burst. The optical observations began 27 s after 
the start of the burst by the ROTSE-IIIa robotic telescope 
\citep{Rykoff2009} 
The optical lightcurve shows a sharp rise, brightening by over 
a factor of 50 in less than 5 s, followed by two peaks and a steady 
power-law decline. The $\gamma$-ray and X-ray lightcurves have multiple 
peaks that are not contemporaneous with the optical peaks 
\citep{Rykoff2009}, followed by a steady decline in the X-rays that tracks the 
optical decline. The three overlapping optical peaks could be the three 
SRFs following the three large BAT peaks. 
In Fig.~\ref{fig9}c we compare the 0.3-10 keV Swift XRT lightcurve of
GRB 061007 (Swift XRT lightcurve repository, \citet{Evans2009}),
and its CB model description as the tail of the prompt emission
SRFs overtaken around $t$=300 s by a standard AG with an early-time break 
hidden 
under the prompt emission. The best fit parameters are listed in 
Table~\ref{t1}. Note that both the early-time and late-time 
decays satisfy well the CB model closure relations:
$\alpha\!=\!\Gamma\!=\!1.884 (+0.023, -0.023)$ for the early-time decay 
and $\alpha\!=\!\Gamma\!-\!1/2\!=\!1.518 (+0.087, -0.083)$
for the late-time decay, respectively.
In Fig.~\ref{fig9}d we compare the optical lightcurve of GRB 061007 
as measured by ROTSE-IIIa \citep{Rykoff2009} and its CB model 
description in terms of three overlapping SRFs and a standard 
AG taking over around $t$=400 s. 
Note that the late-time optical AG also satisfies well
the CB model prediction, 
$\alpha_O\!=\!\beta_O\!+\!1/2\!=\!\beta_X\!+\!1/2\!=\!1.518 (+0.087, -0.083)
$.

\noindent 
{\bf GRB 990123} at redshift $z\! =\! 1.600$ \citep{Kulkarni1999, 
Andersen1999} was for a long time the brightest known GRB. It was 
also the first GRB in which an optical emission was detected during the 
prompt $\gamma$/X-ray emission. GRB 990123 was detected and localized by 
the Burst And Transient Source Experiment (BATSE) on board the Compton 
Gamma Ray Observatory (CGRO) in the keV-MeV range and at higher 
energies by the COMPTEL, OSSE and EGRET instruments \citep{Briggs1999}. 
It was also detected and localized by the Gamma Ray Burst Monitor (GBM) 
aboard the BeppoSAX satellite \citep{Maiorano2005}.
The lightcurves of the prompt emission 
in the keV-MeV range showed a complex structure of at least 9 
pulses/flares (see Fig.~\ref{fig9}a). The prompt optical emission that 
was detected by
the Robotic Optical Transient Search Experiment (ROTSE) at Los Alamos
22 s after the onset of the burst, brightened and peaked at magnitude 
$V\!\sim\! 9$, about 50 s after
the GRB onset, and decayed rapidly with time \citep{Akerlof1999}.
The prompt optical flare was not resolved into separate flares.
It was followed in the $UVONIR$ bands with large ground based telescopes
\citep{Castro1999, Galama1999, Kulkarni1999, Fruchter1999,
Holland2000} and with
the Hubble Space Telescope until it faded to a magnitude
$V\!=\! 27.7\!\pm\!0.15$, two months after burst
\citep{Fruchter2000}. The broad band
$\gamma$-ray, X-ray, $UVO$ and $NIR$ lightcurves of GRB 990123 were
reanalyzed recently within the synchrotron fireball (FB) model by
Corsi et al.~(2005). Essentially they found that the
spectral and temporal properties of the prompt optical emission are
uncorrelated to the $\gamma$ and X-ray emission, implying different
physical origins, that the optical and X-ray afterglow lightcurves are
chromatic contrary to expectations, and that their spectral and temporal
power-law decays do not satisfy the closure relations of the FB model.

In Fig.~\ref{fig9}a we compare the BATSE multi-peak lightcurve of GRB 
990123 
in the 20-50 keV channel \citep{Briggs1999} and its CB model 
description. The count-rate in the 20-50 keV energy band was calculated 
from the integral $\int F_\nu\, dE/E$ using Eq.~(\ref{ICSlc}) with the 
best fit parameters, which are listed in Table~\ref{t4} for the 9 peaks 
suggested by the multichannel BATSE data and by the BeppoSAX data 
\citep{Maiorano2005}. As shown in Fig.~\ref{fig9}a, the shape of the 
peaks 
and the entire lightcurve are well reproduced by Eq.~(\ref{ICSlc}). The 
2-10 keV lightcurve of the X-ray afterglow of GRB 990123 that was 
measured with 
BeppoSAX \citep{Maiorano2005} for $t\!<\!2.5$ days (not shown here) was 
best fit by the CB model with $p\!=\!1.79$, implying $\beta_X\!=\!0.90$, 
consistent with $\beta_X\!=0.94\!\pm\!0.12$ that was inferred by 
\citet{Maiorano2005} from their data.  The observed temporal power-law 
decay index 
of the late-time X-ray afterglow, $\alpha_X\!=\!1.46\!\pm\!0.04$ 
\citep{Maiorano2005}, also obeys the CB model relation , 
$\alpha_X\!=\!\beta_X\!+\!1/2\!=\!1.44\!\pm\!0.13\, .$

In Fig.~\ref{fig9}b we compare the observations of the optical lightcurve 
of 
GRB 990123 from onset \citep{Akerlof1999} until late time  
\citep{Castro1999, Galama1999, Kulkarni1999, Fruchter1999,
Holland2000}, normalized to the
$V$-band, and its CB model description
as given by Eq.~(\ref{Fnu}) with
the afterglow parameters $\gamma\, \theta\!=\!0.24$, $t_0=2250$ s
and $p=1.79\,.$ Due to a gap in the
data between 500 s and 15,000 s, the expected transition from a
circumstellar density profile $\propto\! 1/r^2$ to a constant ISM density
was not well determined. However, the gradual bending (`jet break') of the
optical AG to an asymptotic power-law decay, $F_\nu\!\propto\!
t^{\!-\!\beta_O\!-\!1/2} \nu^{-\beta_O}\,,$ is well reproduced with the
expected late-time spectral index $\beta_O\!\sim\!\beta_X\!\sim\! 1.1\,.$

\section{Summary and Conclusions}

In the CB model, GRBs, XRFs and SHBs and their afterglows are produced by 
the interaction of bipolar jets of highly relativistic plasmoids 
(cannonballs) of ordinary matter, which  are ejected in mass accretion 
episodes on a newly formed compact stellar object, with the radiation and 
matter that they encounter along their path. As observed in microquasars, 
each time part of the accretion disk falls abruptly onto the compact 
object, two jets of cannonballs (CBs) made of {\it ordinary-matter plasma} 
with large bulk-motion Lorentz factors are emitted in opposite directions 
along the rotation axis, wherefrom matter has already fallen back onto 
the compact object due to lack of rotational support. The prompt 
$\gamma$-ray and X-ray emission is dominated by inverse Compton scattering 
(ICS) of photons of the glory - a quasi isotropic optical light emitted by 
the supernova and scattered by the wind/ejecta blown from the progenitor 
star long before the GRB. The CBs' electrons Compton up-scatter the glory 
photons into a narrow conical beam of $\gamma$ rays along the CBs' 
direction of motion. An X-ray `flare' coincident in time with a prompt 
$\gamma$-ray pulse is simply its low-energy part. The early-time X-ray 
flares without a detectable accompanying $\gamma$-ray emission are 
usually ICFs produced by CBs with relatively smaller Lorentz factors due 
to a weakening activity of the central engine: As the accretion material 
is consumed, the `engine' has a few progressively-weakening dying pangs. 
The lightcurves of ICFs, like those of the prompt emission pulses, exhibit 
a rapid softening during their fast decline phase.  Roughly, the 
lightcurves are a 
function of the product $E\,t^2$ and not of the individual values of the 
photon energy $E$ and the time $t$ after the beginning of the flare. The 
peak energy, isotropic equivalent energy and peak luminosity of the 
ICFs are correlated like those of the prompt GRB pulses.

In the CB model, each ICF is followed by SRF from the encounter of the CB 
with the wind/ejecta that was blown from the progenitor star long before 
the GRB. Because of time-aberration, in the observer frame, these SRFs lag 
after their preceding ICFs by a short time of the order of the ICS pulse 
duration. Optical flares are usually much wider than their corresponding 
gamma/X-ray pulses and overlap, which makes it difficult to associate the 
early time optical flares with their preceding gamma/X-ray pulses/flares 
and measure their lag-time. Only in single-pulse GRBs that are bright 
enough to be detected with robotic telescopes and/or Swift UVO and in very 
bright GRBs, such as 080319B and 071031, where the optical flares were 
partially resolved with robotic telescopes, could the predicted 
association between early time optical flares following gamma/X-ray 
pulses/flares be tested.

Often the fast decay of an X-ray ICF is taken over by SR of X-rays from 
the CB encounter with the wind enclosing the glory light before it 
disappears under the plateau/shallow decay phase of the AG (see, e.g., 
Figs. 3a,b,c in \citet{DDD2009a}).

Late-time flares are usually SRFs produced by CB encounters with the bumpy 
boundary of the star formation region, or with density bumps within this 
region where the GRB took place. Their exact profile is not known apriori 
but a wind-like profile seems to be a good working hypothesis,

Unlike the empirical parametrizations (such as Band function 
\citep{Band1993}, 
cut-off power-law, Beuermann function \citep{Beuermann1999}, broken 
power-law, segmented power-law, etc.) used in most of the published 
standard analyses of GRB lightcurves, which have never been properly 
derived from underlying physical assumptions, the master formulae of the 
CB model were derived in fair approximations from its underlying physical 
assumptions. As shown in this paper, the lightcurves and spectral 
evolution of X-ray flares and optical flares in GRBs, XRFs and SHBs are 
well described by the master formulae of the cannonball model of 
GRBs. So far, no new assumptions or modifications of these formulae were 
needed when applied to well sampled flares in the GRB lightcurves. 
Probably, in the future, when much more refined data and better sampled 
lightcurves of GRBs and their AGs will become available, the CB model with 
its current simplifying assumptions, which were introduced in order to 
avoid `over parametrization' and make it predictive and falsifyable, will 
have to be refined in order to reproduce such data with sufficient 
accuracy.

{\bf Acknowledgment.} We thank an anonymous referee for useful suggestions 
and T. Kruhler for making available to us tabulated data of the optical 
lightcurves of GRB 071031 measured with GROND.

{}

\begin{deluxetable}{llllc}
\tablewidth{0pt}
\tablecaption{The AG parameters used in the CB model
description of the smooth X-ray AG of Swift GRBs
with superimposed flares.  
The CB model spectral index obtained from the temporal shape 
of the AG and that inferred from
the measured XRT spectrum as reported in the Swift lightcurve repository 
\citep{Evans2009}  are compared in the last two columns.} 
\tablehead{\colhead{GRB/SHB} & \colhead{$t_b$ [s]}& 
\colhead{$\gamma_0\, \theta$ } & \colhead{$\beta_X\!=\!p/2$ }& 
\colhead{$\beta_X\!=\!\Gamma_{Swift}\!-\!1$ }
}       
\startdata
GRB050502B &  674  & 0.73  & 0.96  & 0.945 (+0.077, -0.100)  \\
SHB050724  &       &       &       & 0.81 (+0.14, -0.17)     \\
GRB050820A & 9545  & 1.19  & 1.10  & 0.966 (+0.051, -0.050)  \\
GRB050916  & 2178  & 0.74  & 0.90  &                         \\
GRB060206  & 9402  & 1.04  & 1.12  & 1.21 (+0.27, -0.24)     \\
GRB060418  &  123  & 1.73  & 1.05  & 0.96 (+0.15, -0.14)     \\
GRB060526  & 4828  & 1.35  & 1.02  & 0.931 (+0.086, -0.084)  \\
GRB060729  &32665  & 2.52  & 1.10  & 1.067 (+0.038, -0.037)  \\
GRB060929  & 5383  & 1.27  & 0.91  & 1.22 (+0.25, -0.25)     \\
GRB061007  &   40  & 0.15  & 1.08  & 1.018 (+0.087, -0.083)  \\
GRB070704  & 5183  & 1.27  & 0.90  & 0.98 (+0.18, -0.33)     \\
GRB071003  & 1214  & 1.94  & 1.08  & 0.984 (+0.107, -0.059)  \\
XRF071031  & 5451  & 2.56  & 0.74  & 0.86 (+0.14, -0.15)     \\
GRB080319B &   86  & 0.14  & 1.08  & 1.03 (+0.064, -0.063)   \\
XRF080330  &  541  & 4.61  & 0.89  & 0.89 (+0.13, -0.12)     \\
GRB080506  & 7801  & 1.76  & 0.91  & 0.990 (+0.122, -0.077)  \\
GRB080607  & 2904  & 0.53  & 0.91  & 1.102 (+0.098, -0.092)  \\
GRB080810  & 3452  & 0.63  & 1.25  & 1.156 (+0.099, -0.089)  \\
GRB080906  & 5790  & 0.90  & 0.96  & 1.049 (+0.069, -0.164)  \\
GRB081102  & 1891  & 0.47  & 0.92  & 0.921 (+0.079, -0.114)  \\
GRB081203A &  436  & 0.91  & 1.15  & 1.096 (+0.089, -0.081)  \\
GRB090102  &  348  & 0.80  & 0.95  & 0.858 (+0.078, -0.076)  \\
GRB090417B & 1259  & 0.63  & 1.04  & 1.09 (+0.11, -0.11)     \\
GRB090618  & 1540  & 1.10  & 1.04  & 1.008 (+0.047, -0.046)  \\
GRB090621A & 1309  & 0.83  & 0.94  & 1.01 (+0.18, -0.18)     \\
GRB090709A & 1098  & 0.51  & 1.08  & 1.081(+0.076, -0.074)   \\
GRB090812  & 598   & 1.34  & 0.92  & 0.914 (+0.138, -0.089)  \\
GRB091029  & 5753  & 2.32  & 1.01  & 1.197 (+0.077, -0.070)  \\

\enddata
\label{t1}
\end{deluxetable}

\begin{deluxetable}{lllllllc}
\tablewidth{0pt}
\tablecaption{The parameters of the ICFs and SRFs used in the CB model
description of X-ray  lightcurves of Swift GRBs. 
Values of parameters that are not well determined by a best fit are 
inserted within parentheses}
\tablehead{
\colhead{GRB/SHB} &  \colhead{flares}  
& \colhead{$t_0$ [s]  } & \colhead{$\Delta t$ [s]} 
&\colhead{$E_p[keV]/a[s]$}
& \colhead{$t_0$ [s]  } & \colhead{$\Delta t$ [s]} 
&\colhead{$E_p[keV]/a[s]$}
}
\startdata
GRB050502B & ICF,SRF,  & 485.7 & 308.3 & 0.27~keV & 13750 &28950 &6013~s\\
           & SRF       &  58950 & 18911 &  3350~s & & & \\
GRB060418  & ICF,ICF   &  60    & 12    &  0.13keV &9.9 & 118& 0.13keV \\
GRB060206  & SRF,SRF   &  1.19  & 32.2  &  235~s & 653 & 3427 & 3645~s \\
           & SRF       &  609359 & 245909 & 1980979~s   & & &\\
SHB050724  & ICF,ICF   & 1.58 & 114  & 1.08~keV& 469 & 1422 & 1.32~keV\\
           & ICF       & 9757& 49512 & 3.53~keV & & & \\
GRB050916  & SRF,ICF   &  &  &  &17836 &1335 &91~keV \\
           & ICF       &  9757& 49512 & 3.53~keV & & & \\
GRB050916  & SRF,ICF   &  &  &  &17836 &1335 &91~keV \\
GRB060526  & SRF,ICF   &   &  &  & 233 & 42.5 & 0.15~keV\\
           & ICF,SRF   & 280  & 25.7 &60.2 keV & 85 &106 &288~s\\
GRB060929  & SRF,ICF   &   &   &  &472  &45  & 117~keV \\
GRB071031  & ICF,ICF   & 84.8 &46.2 &13.2~keV& 131 & 33.8 & 0.74~keV\\
           & ICF,ICF   & 181.7 & 30.5 & 4.3~keV & 228 & 40.6& 14.63~keV\\
           & ICF,ICF   & 328 & 176 & 2.74~keV   & 572 & 173.6 &19,74~keV\\
           & ICF       & 4351& 1201& 4.5~keV& & & \\   
           & SRF,SRF   & 16743 & 5463& 4.2~s & 33241 &(53238) &$\ll a$\\  
GRB070704  & SRF,ICF   & (63) & (0.35) & (25~s) & 472 & 45.22 & 117~keV \\
GRB080506  & ICF,SRF   & 130 & 36.7 & 14.4~keV & 211& 43& 171~s\\
           & ICF,ICF   & 433  &(41) &(159~keV)& 684 s& (160)& (156~keV)\\
GRB080607  & ICS,ICF   & 65.9 & 8.6 & 145 keV & 113.9 & (15.4) &(350~keV)\\
           & SRF,SRF   & 222 & 332 & 0.6~s &652 &1018 &335~s \\
GRB081203A & SRF       & 79  & 10.8 & (0)  & & &\\
GRB080810  & ICF,SRF,  & 74& 16& 25.5 keV& 98.6   & 4.6  & 34~s\\
           & ICF,SRF   &196 & (14) & (203~keV) & 265 & 208 & (7.2~s)\\
GRB080906  & SRF,SRF,  & (49) & (1) &(47.7~s) &157  &27.5 & 10.75~s\\
           & SRF,ICF  & 490  & 8.58 & 200 s & 836& 78 & 145~keV\\
GRB081102  & SRF,ICF   & 63 & 11 & 42.7~s &914  &88 & 111~keV \\
GRB090417B & SRF,ICF   & 336.2  & 203 & 82.6~s &1227.8 &431 &1.96~keV  \\
GRB090618  & ICF,SRF   & 65.9   &12.9 &        & (0)   & 884 & (0)   \\    
GRB090621A & SRF,ICF   & 150.6  & 2.6~s & 215 & 217.9 & 50.4 & 0.95~keV \\
GRB090709A &ICF,SRF,   & 70.3 & 20.61 & 2.38~keV & 75.5 & 14.3& 116 s\\
           & SRF,ICF  & 165.8  & 110.2   & 149 &324.6 & 37 & 187 s\\
GRB091029  & ICF,ICF  & 0  & 38.8    & 9.8~keV &218.6 &102.5& 10.9~keV\\ 
           & SRF      & 209360 & 245909& & & &\\  
\enddata
\label{t2}
\end{deluxetable}

\begin{deluxetable}{llllc}
\vskip -5.cm
\tablewidth{0pt}
\tablecaption{The parameters of the CB model description 
of the early-time optical 
SRFs/peaks in GRBs 080319B, 060206, 071003, 071031 and 080330}
\tablehead {\colhead{Flare} & \colhead {$t_0\, [{\rm s}]$} &
\colhead {$a\, [{\rm s}]$} & \colhead {$ t_{exp}\, [{\rm s}]$} & 
\colhead{$\beta_O$} 
}
\startdata
\hline
GRB 080319B: & & & &  \\
\hline
SRF1 & 9.50 & 3.52 &10.24 & 0.70   \\
SRF2 & 32.37& 3.52 & 5.83 & 0.70   \\
SRF3 & 39.54& 3.52 & 6.96 & 0.70   \\
\hline
GRB 060206: & & & & \\
\hline
SRF1 & $\sim 0$ &  548 & $\ll a$ & 0.50  \\
SRF2 & 2526       & 1214 &  887     & 0.80  \\
SRF3 & 4286       & 3113 &  31      & 0.87  \\        
\hline
GRB 071003: & & & & \\
\hline
SRF1 & 4.11  & $\ll t_{exp}$ & 16.7           & 0.79\\
SRF2 & 125   & 139           & 14.6           & 1.03  \\ 
SRF3 & 441   & 348           & $\gg a$      & 0.75 \\ 
\hline
GRB 071031: & & & & \\
\hline
SRF1 & 90  & 97            & 1875 & 0.62\\
SRF2 & 540 & $\ll t_{exp}$ & 43   & 0.63  \\ 
SRF3 & 2335 & 7900          & 158  & 0.63 \\ 
SRF4 & 10318 & 13372       & 134  & 0.63  \\ 
\hline
GRB 080330: & & & & \\
\hline
SRF1 &0.29 &   32 & 946   & 0.67  \\
SRF2 &3.40 & 1716 & 235   & 0.67  \\
SRF3 &3416 & 1491 & 18337 & 0.67  \\
\enddata
\label{t3}
\end{deluxetable}

\newpage
\begin{deluxetable}{llllc}
\vskip -2.cm
\tablewidth{0pt}
\tablecaption{CB model parameters of the ICS $\gamma$ peaks in GRB 990123}
\tablehead{
\colhead{Peak} & \colhead{$t_0\, [{\rm s}]$} & \colhead{$Dt\,[{\rm
s}]$}& \colhead{$E_p\,[{\rm keV}]$} &\colhead{$A\,
[{\rm counts\,s^{-1}}]$}
}
\startdata
1 & -5.57  & 15.4  & 300 & $ 4.00\times 10^3$\\
2 & 19.42  & 4.28  & 1450& $ 3.95\times 10^4$\\
3 & 29.88  & 0.87  & 500 & $ 6.74\times 10^3$\\
4 & 32.95  & 5.43  & 800 & $ 2.75\times 10^4$\\
5 & 43.67  & 4.99  & 500 & $ 1.11\times 10^4$\\
6 & 52.77  & 4.30  & 450 & $ 1.05\times 10^4$\\
7 & 61.09  & 4.10  & 450 & $ 1.21\times 10^4$\\
8 & 70.35  & 5.63  & 600 & $ 1.69\times 10^4$\\
9 & 85.50  & 1.86  & 200 & $ 6.64\times 10^3$\\
\enddata
\label{t4}
\end{deluxetable}

\begin{deluxetable}{lllllc}
\tablewidth{0pt}
\tablecaption{The parameters used in the CB model
description of unresolved prompt emission SRF
in the optical lightcurves of Swift GRBs.}
\tablehead{ \colhead{GRB} &  \colhead{flare}
& \colhead{$t_0$ [s]} &\colhead{$a$ [s]} & 
\colhead{ $t_{exp}$ [s]} & \colhead{$\beta$} 
}
\startdata
GRB990123 & SRF &  22 &$\lsim 10$&  1.67    & 0.50 \\
GRB030418 & SRF & 200 & 4097     &  $\ll a$ & 0.56 \\
GRB050820A& SRF & 109 &  479     &  $\ll a$ &      \\
GRB061007 & SRF & 15.8&  90.37   &  $\ll a$ & 0.60 \\
GRB081203A& SRF & 0   &  316     &  368    & 0.90 \\
GRB090102 & SRF & 1.65&  0.13    &  2.16   & 0.53 \\
GRB090618 & SRF & 118 & $\sim 0$ &  61     & 0.51 \\
\enddata
\label{t5}
\end{deluxetable}

\newpage

\newpage

\newpage

\newpage
\begin{figure}[]
\centering
\vspace{-1cm}
\vbox{
\hbox{
\epsfig{file=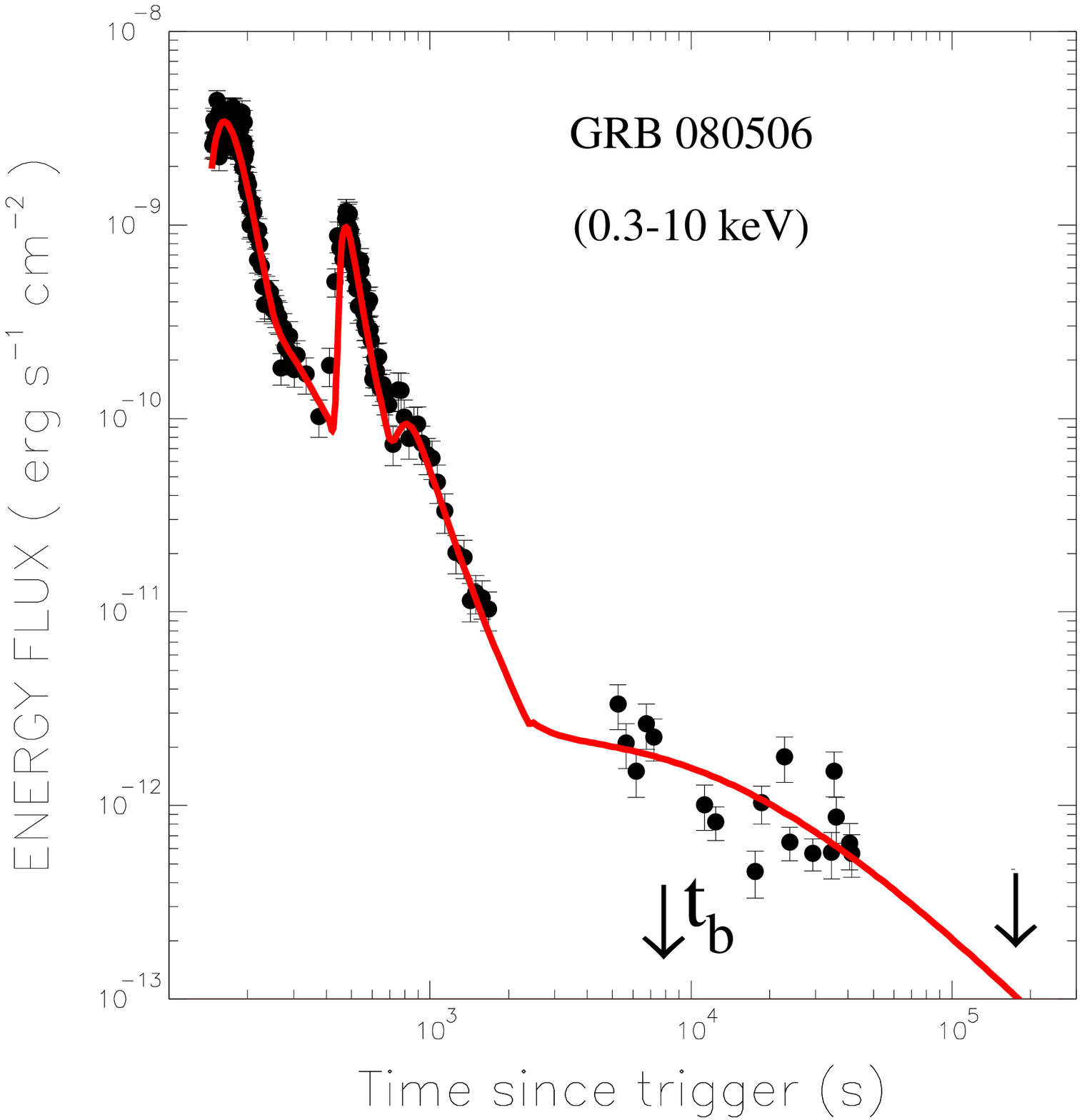,width=7.0cm,height=6.0cm}
\epsfig{file=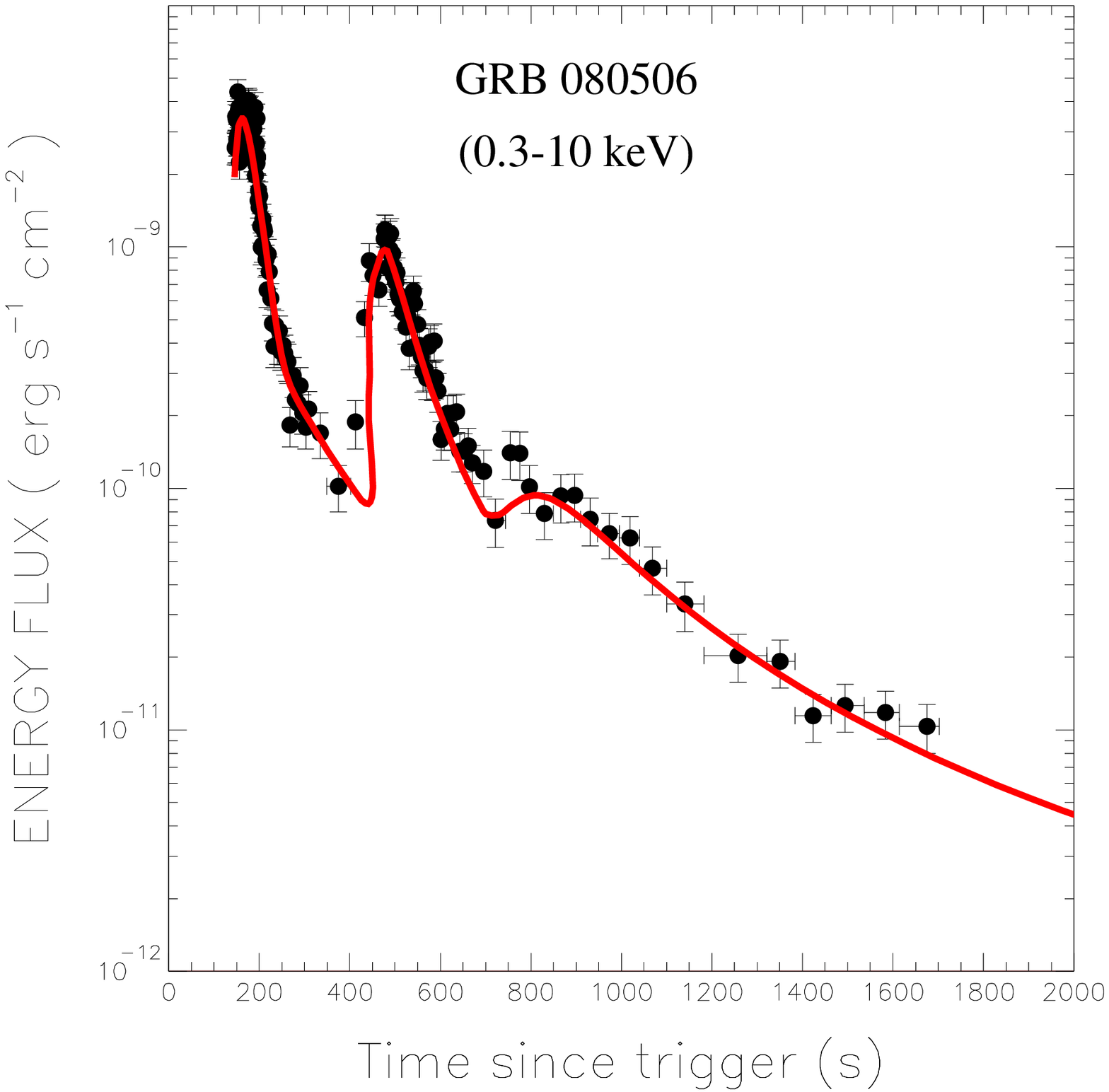,width=7.0cm,height=6.0cm}
}}
\vbox{
\hbox{
\epsfig{file=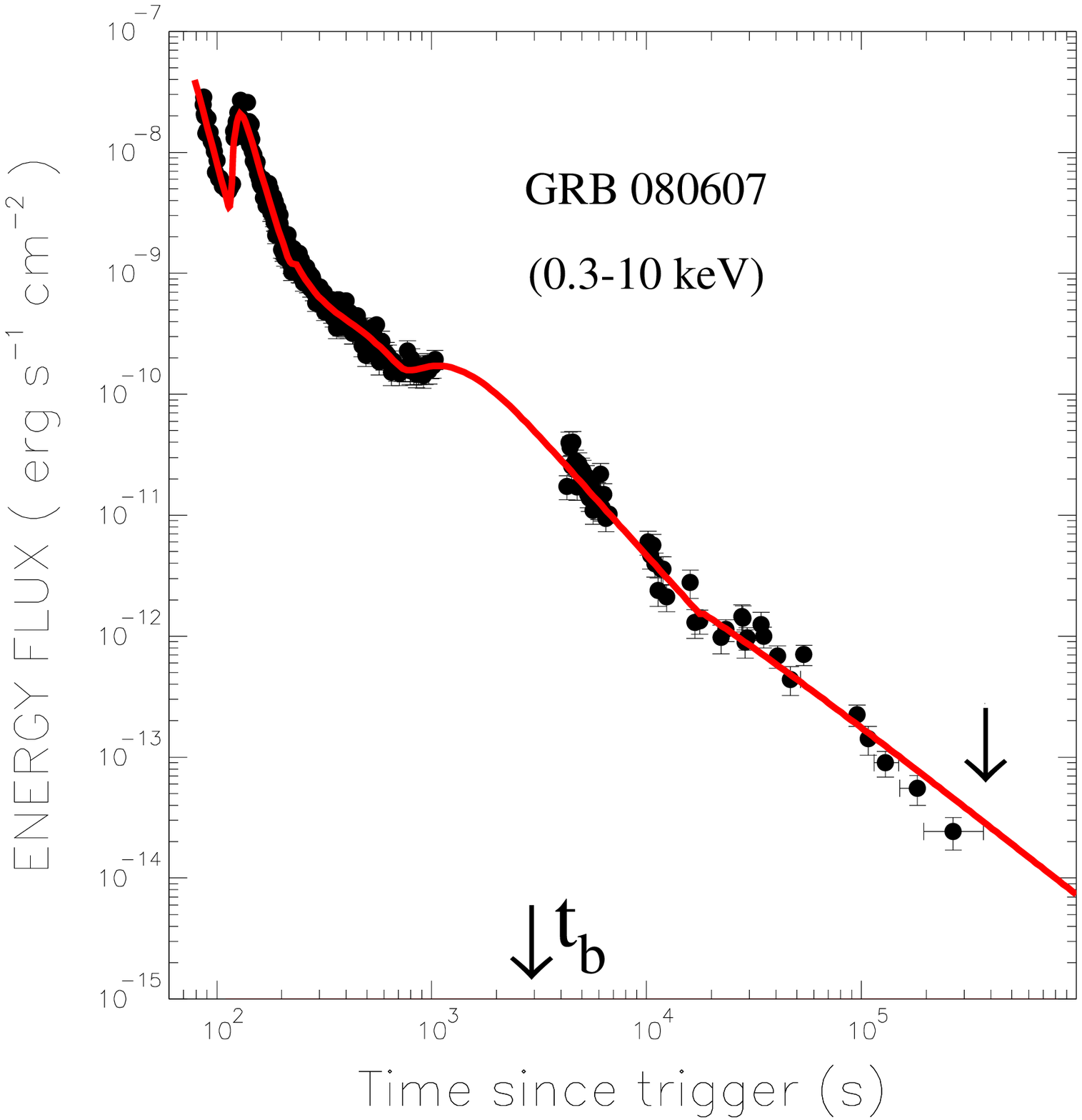,width=7.0cm,height=6.0cm}
\epsfig{file=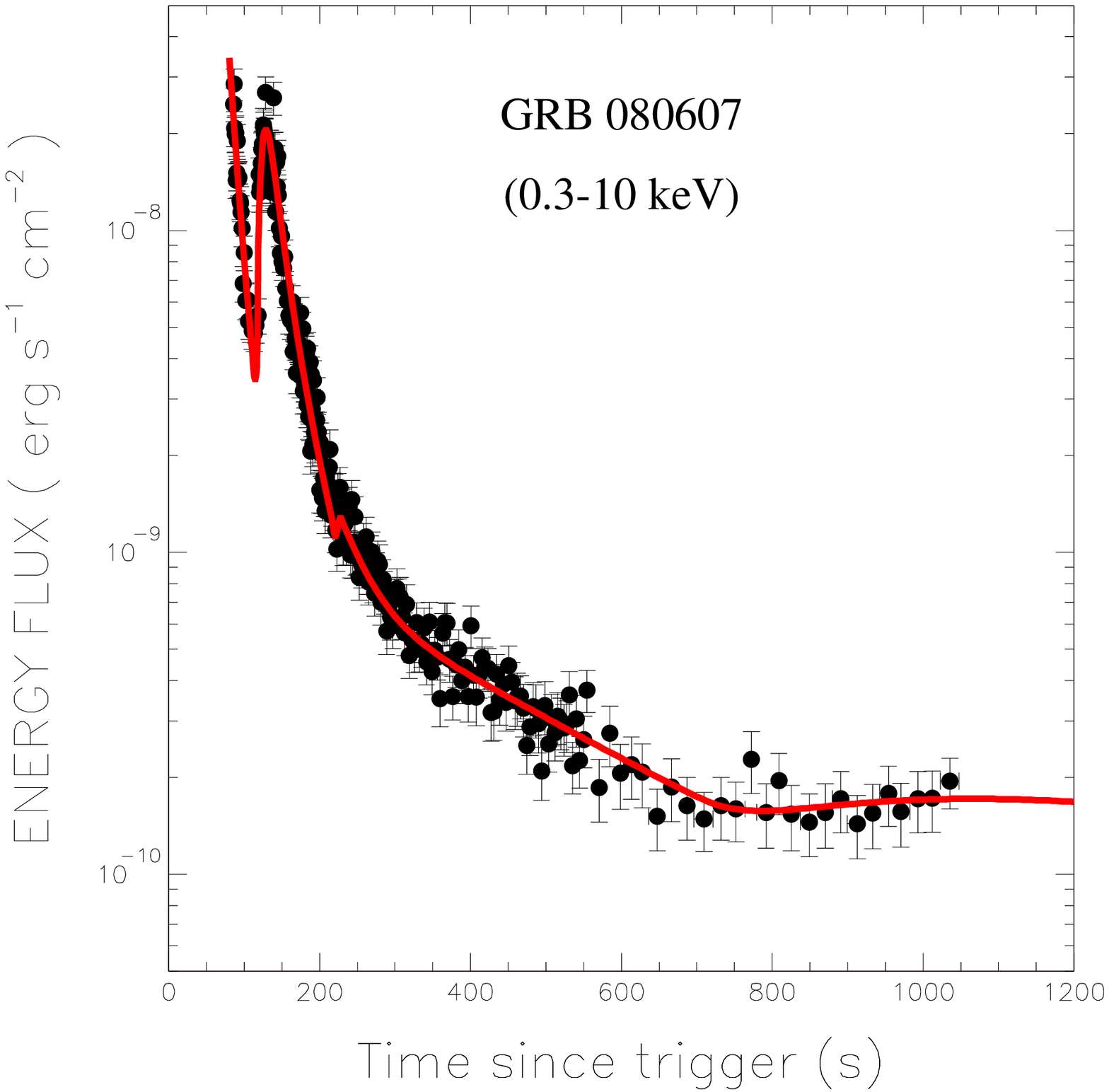,width=7.0cm,height=6.0cm}
}}
\vbox{
\hbox{
\epsfig{file=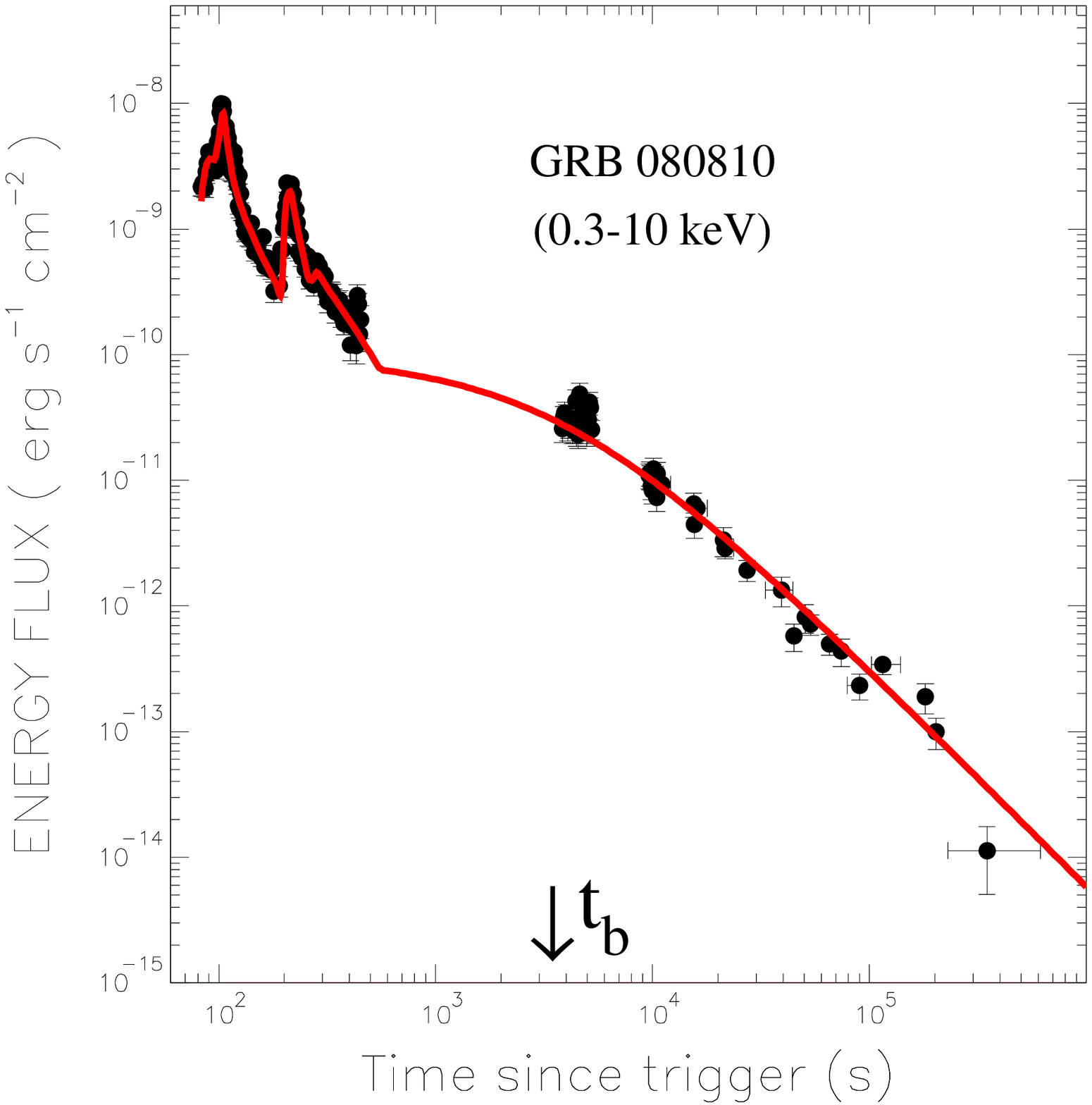,width=7.0cm,height=6cm }
\epsfig{file=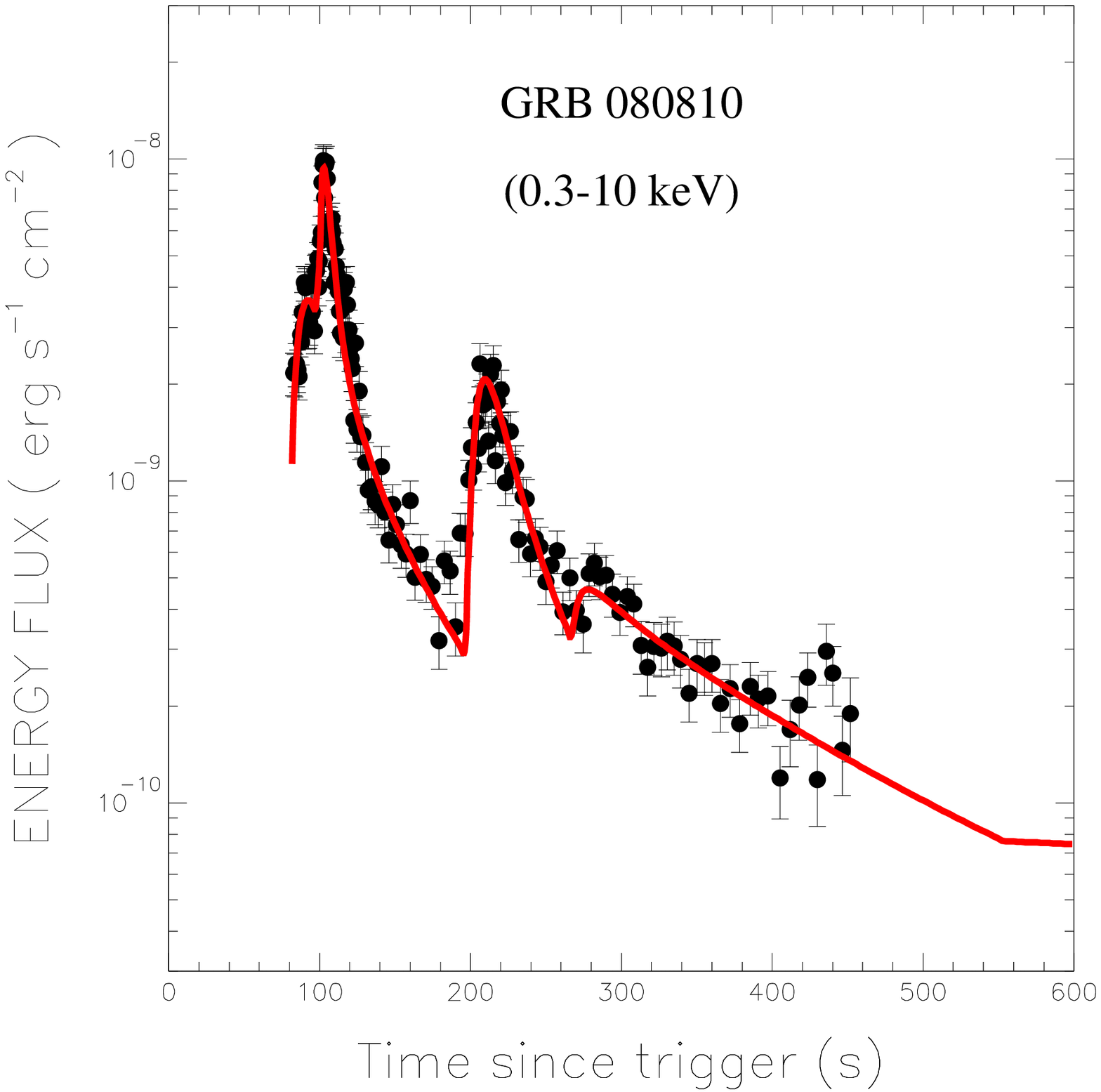,width=7.0cm,height=6cm}
}}
\caption{Comparison between the 0.3-10 KeV X-ray lightcurves
of GRBs with prominent flares in their X-ray afterglow that were
measured by the Swift XRT and reported in the Swift/XRT lightcurve
repository {\it http://www.swift.ac.uk/xrt$_{-}$curves/} 
\citep{Evans2009} and their CB model description with the parameters 
listed in Table 1.
{\bf Top left (a):}  GRB 080506.
{\bf Top right (b):}  Zoom on
                      the prominent X-ray flare
                      in the early-time AG of GRB 080506.
{\bf Middle left (c):} GRB 080607. 
{\bf Middle right (d):} Zoom on
                       the prominent X-ray flare
                        in the early-time AG of GRB 080607.
{\bf Bottom left (e):} GRB 080810.
{\bf Bottom right (f):} Zoom on the
                        prominent X-ray flare
                        in the early-time AG of GRB 080810.
}
\label{fig4}
\end{figure}

\newpage
\begin{figure}[]
\centering
\vspace{-1cm}
\vbox{
\hbox{
\epsfig{file=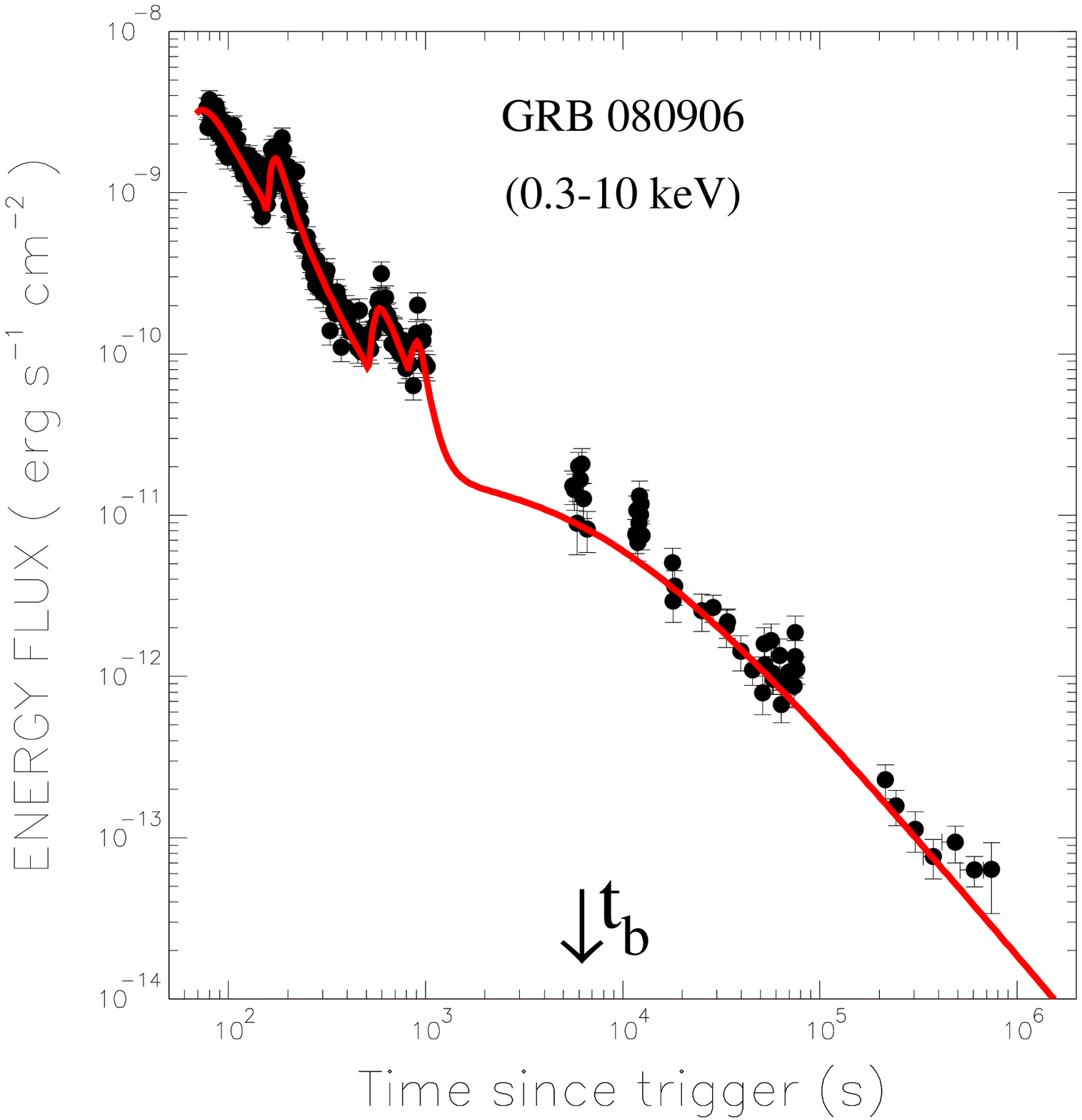,width=8.0cm,height=6.0cm}
\epsfig{file=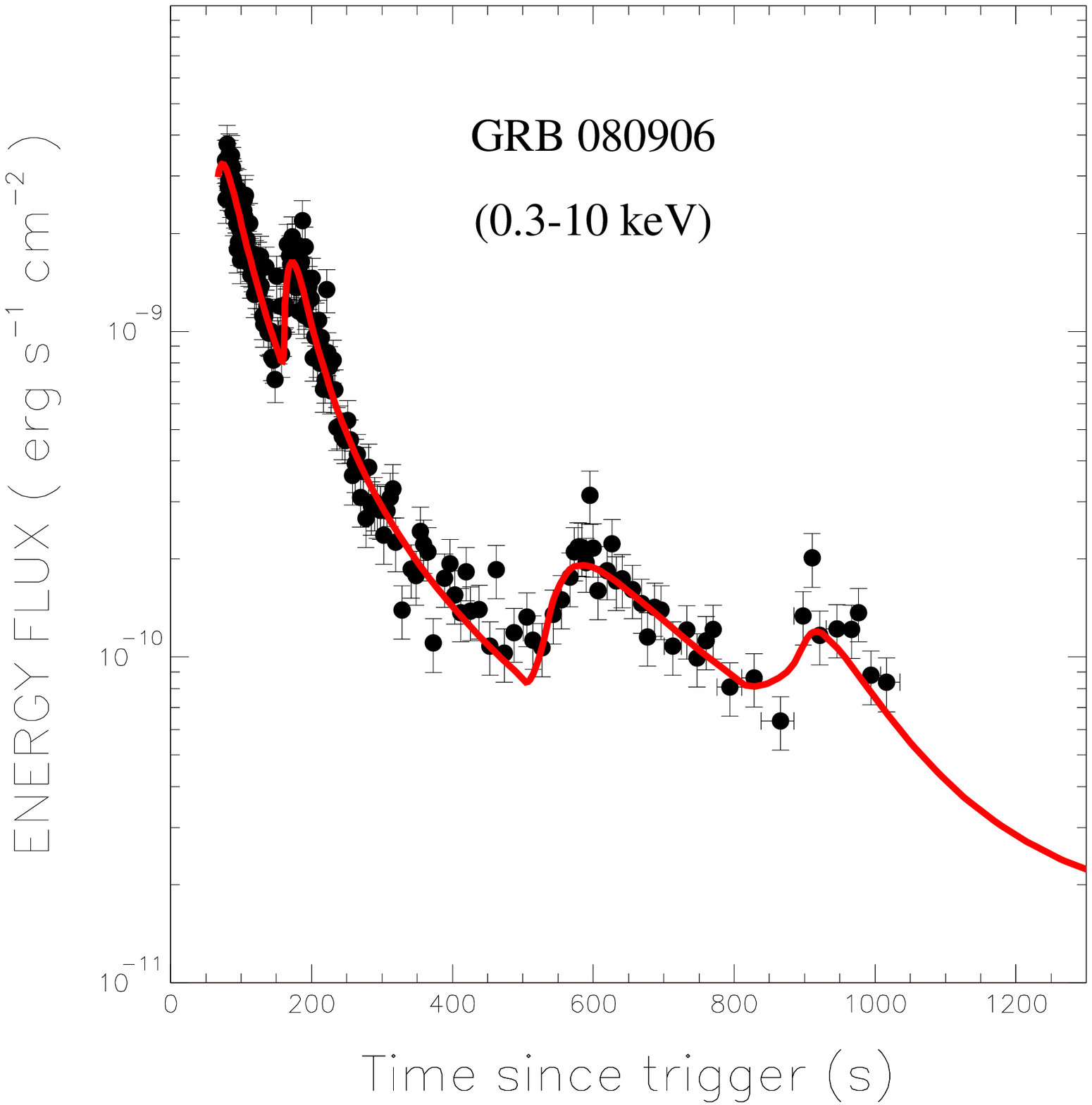,width=8.0cm,height=6.0cm}
}}
\vbox{
\hbox{
\epsfig{file=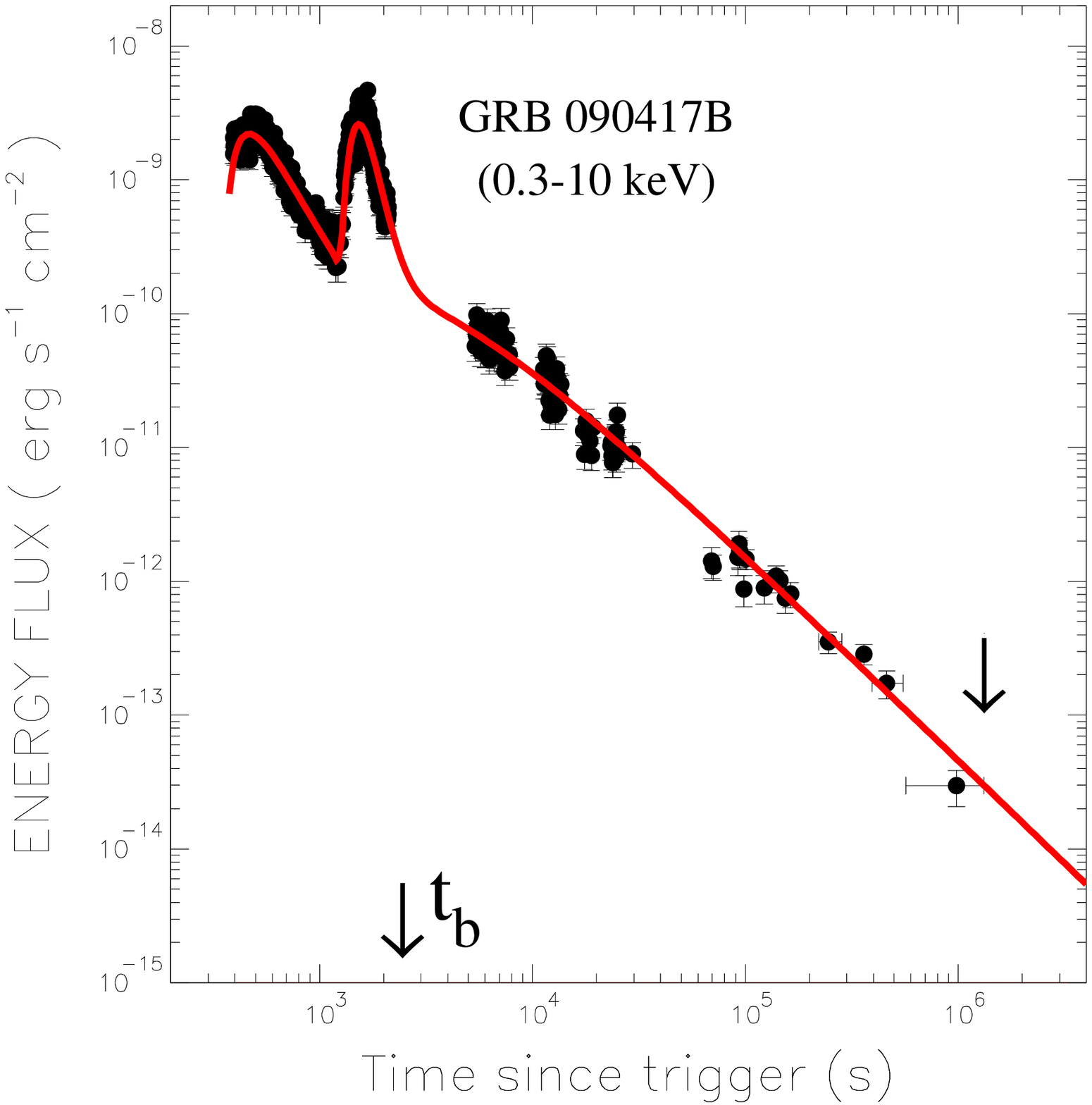,width=8.0cm,height=6.0cm}
\epsfig{file=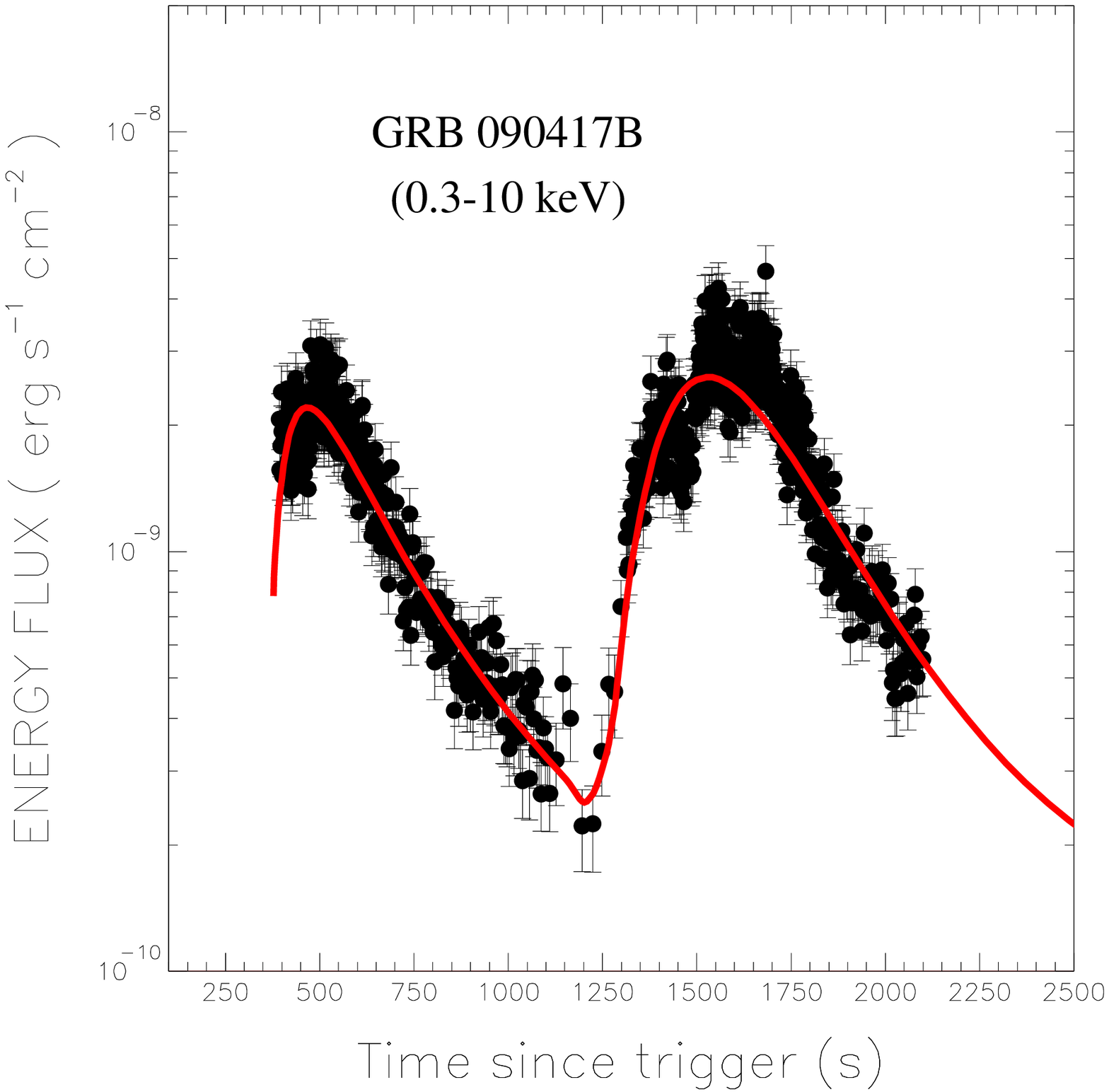,width=8.0cm,height=6.0cm}
}}
\vbox{
\hbox{
\epsfig{file=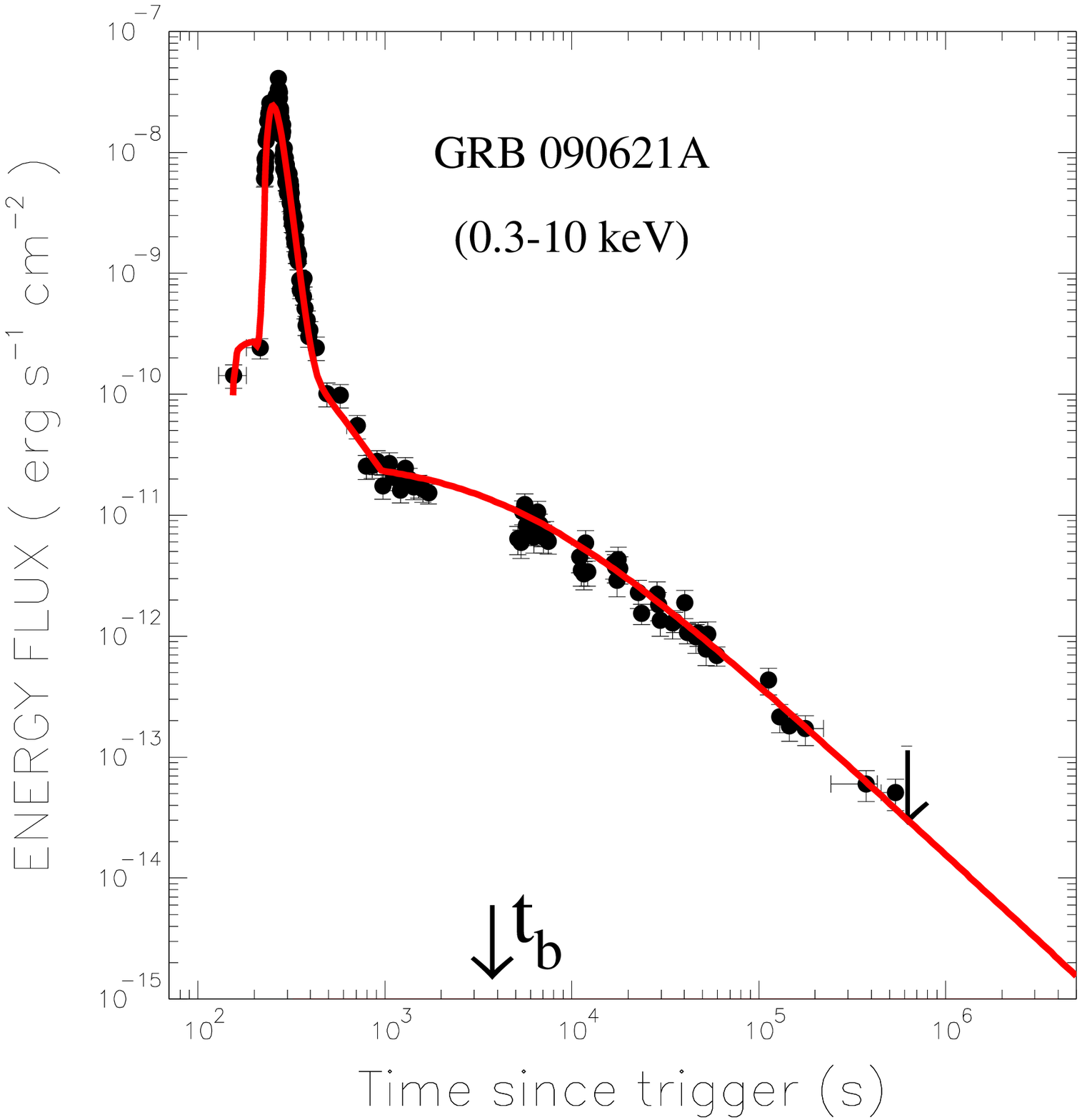,width=7.0cm,height=6cm }
\epsfig{file=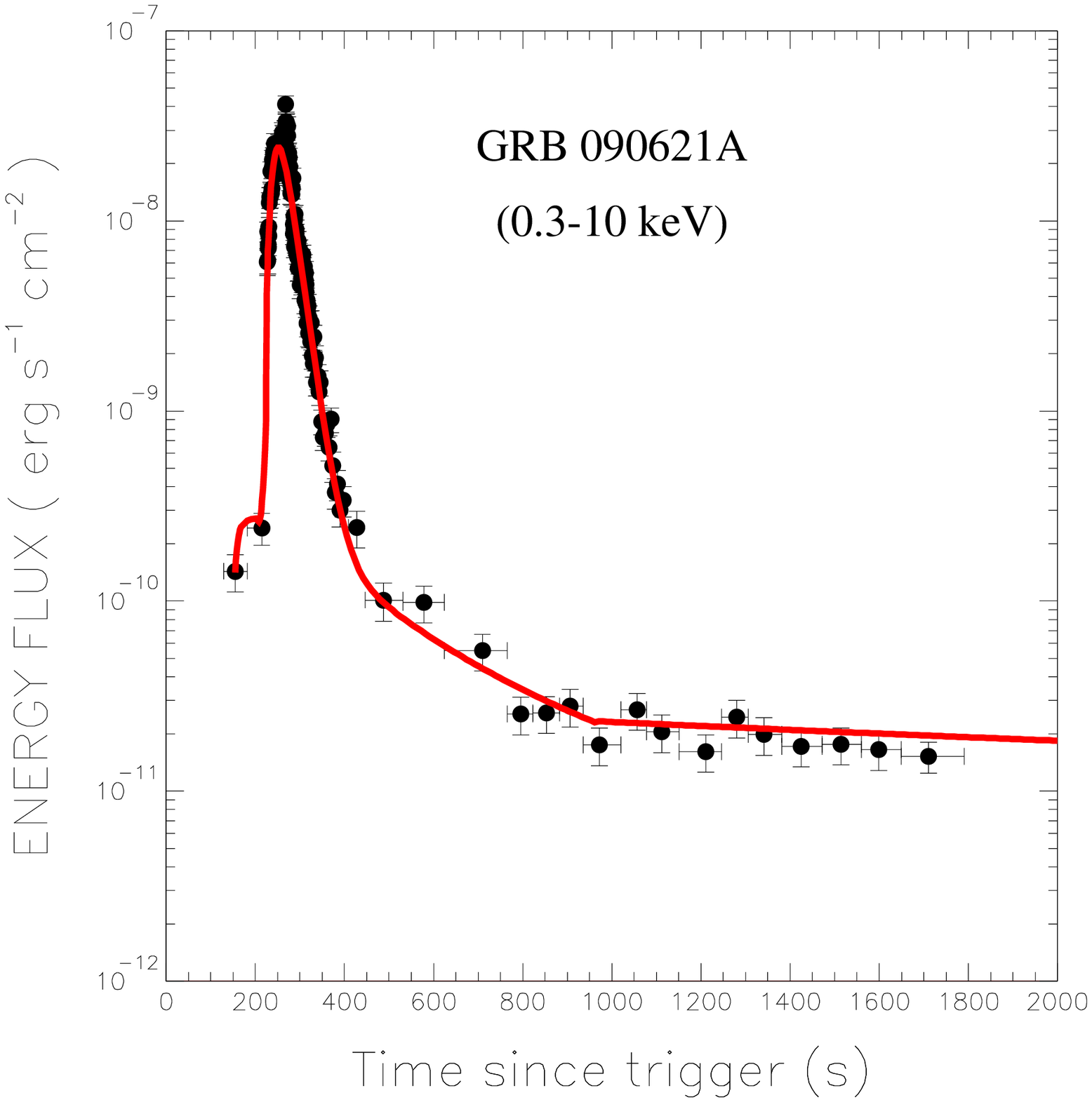,width=7.0cm,height=6cm}
}}
\caption{Comparison between the 0.3-10 KeV X-ray lightcurves
of GRBs with prominent flares in their X-ray afterglow that were
measured by the Swift XRT and reported in the Swift/XRT lightcurve
repository \citep{Evans2009}
and their CB model description with the parameters listed in
Table 1.
{\bf Top left (a):}  GRB 080906.
{\bf Top right (b):}  Zoom on the 
                      prominent X-ray flare
                      in the early-time AG of GRB 080906.
{\bf Middle left (c):} GRB 090417B.
{\bf Middle right (d):} Zoom on 
                       the prominent X-ray flare
                        in the early-time AG of GRB 090417B.
{\bf Bottom left (e):} GRB 090621A.
{\bf Bottom right (f):} Zoom on
                        the prominent X-ray flare
                        in the early-time AG of GRB 090621A.
}
\label{fig5}
\end{figure}

\newpage
\begin{figure}[]
\centering
\vspace{-1cm}
\vbox{
\hbox{
\epsfig{file=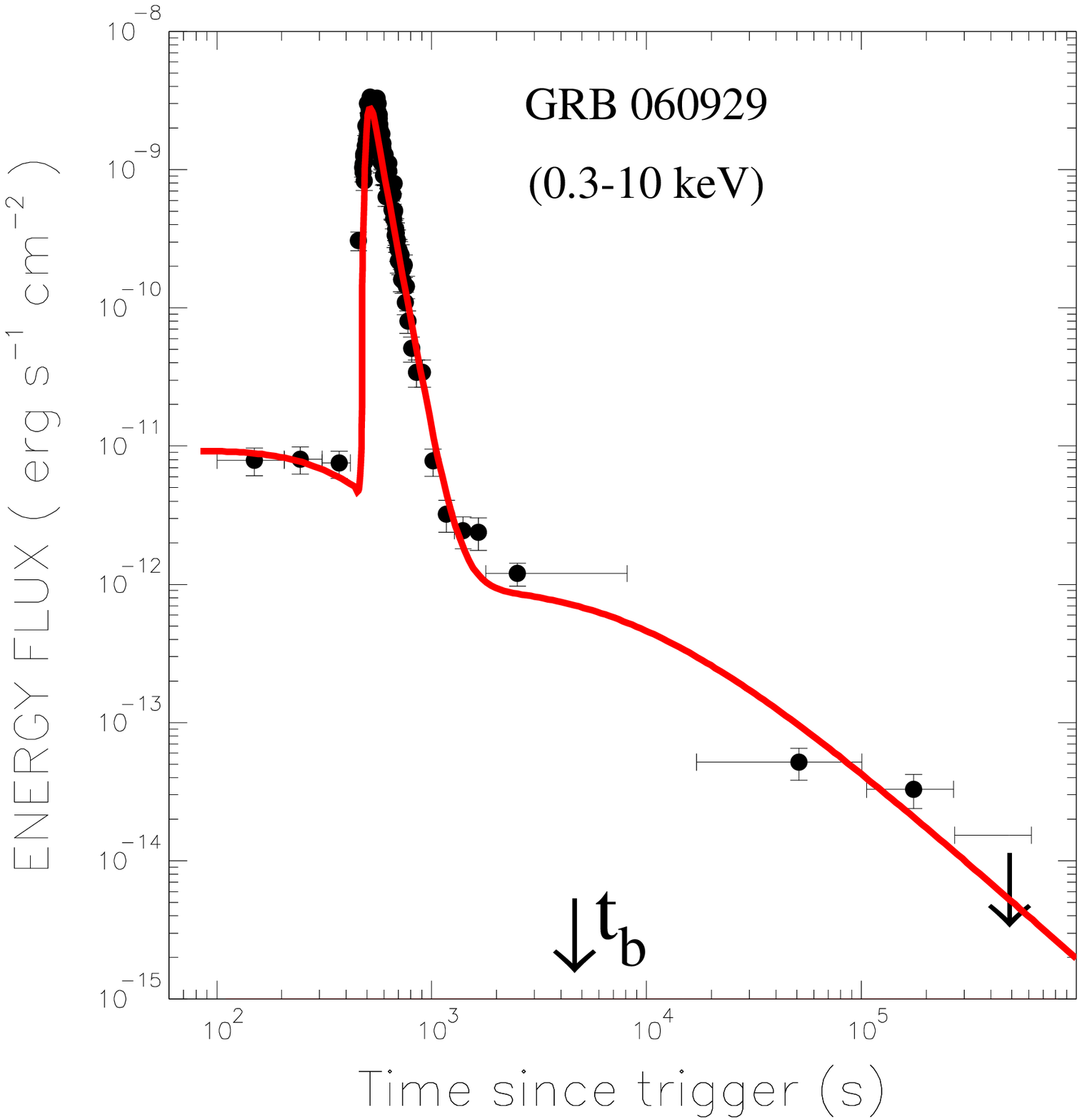,width=8.0cm,height=6.0cm}
\epsfig{file=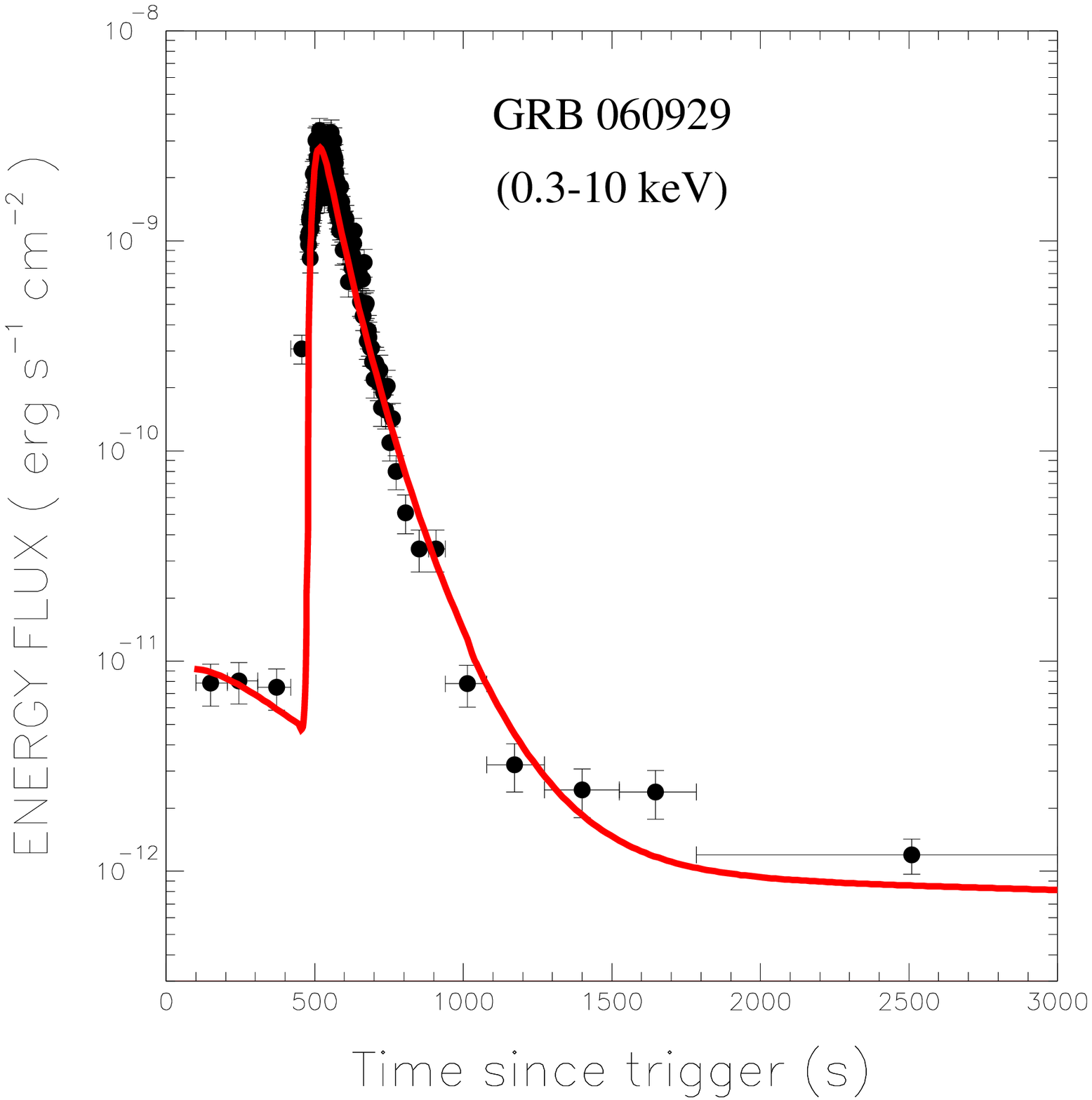,width=8.0cm,height=6.0cm}
}}
\vbox{
\hbox{
\epsfig{file=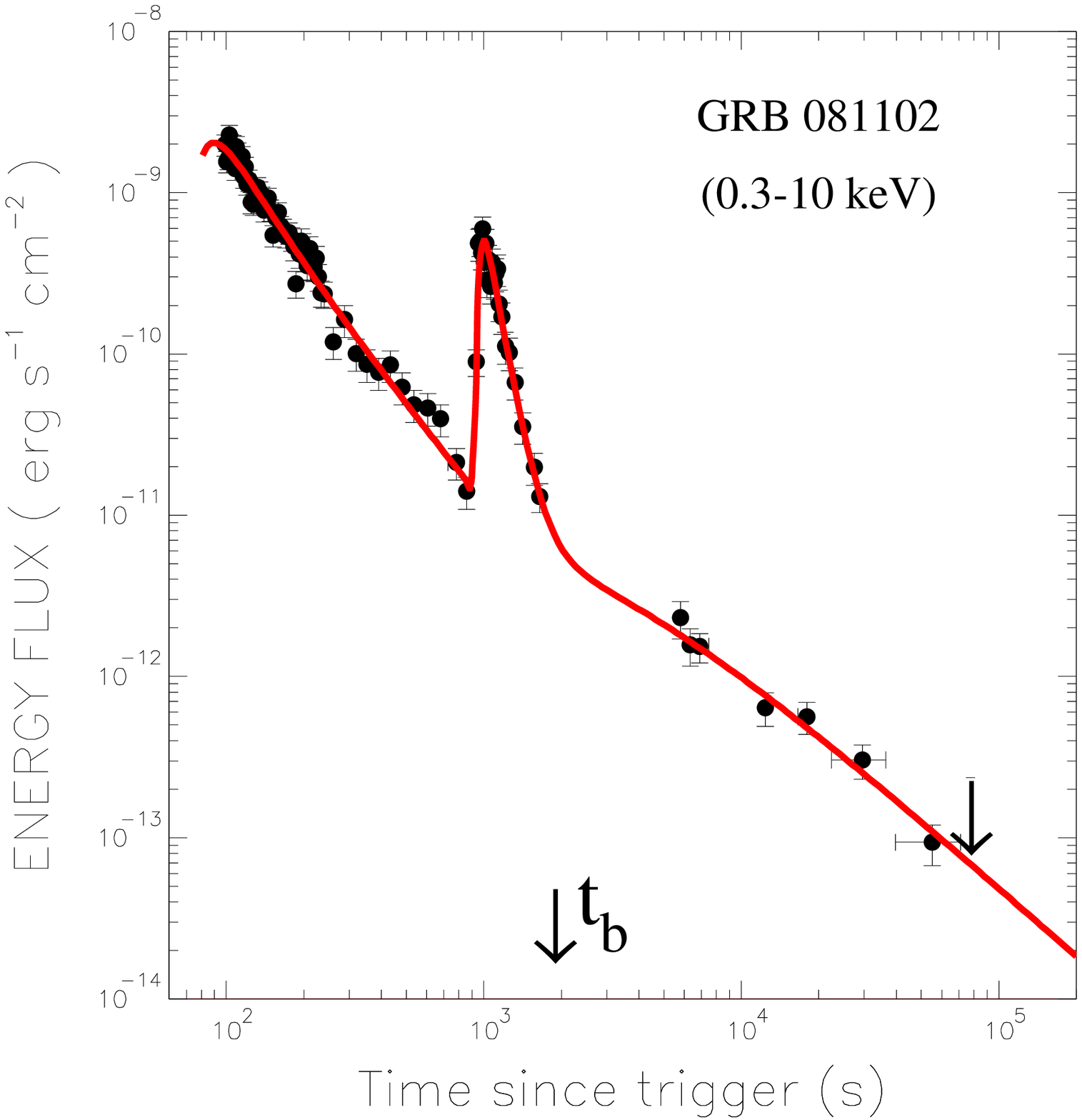,width=8.0cm,height=6.0cm}
\epsfig{file=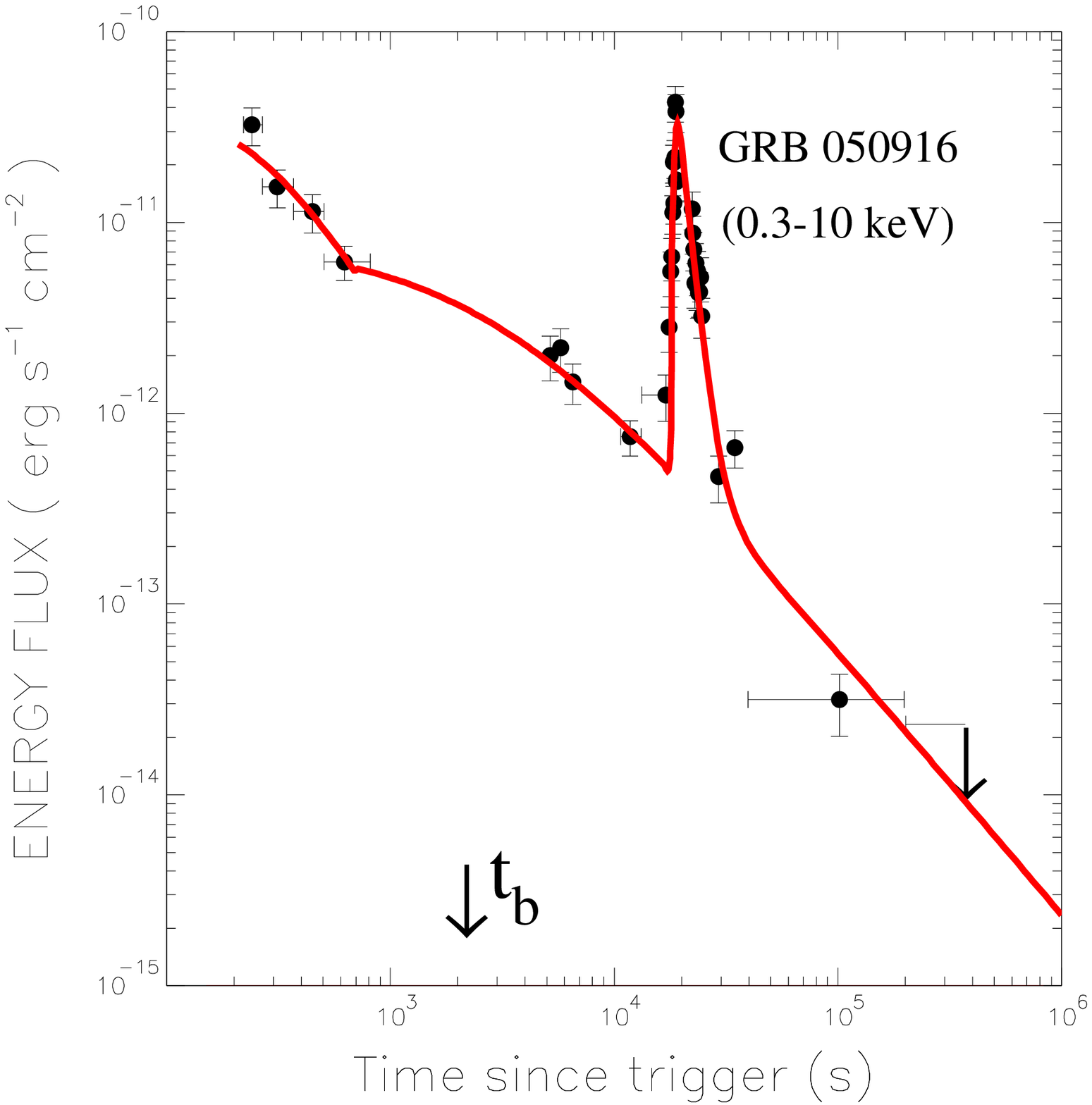,width=8.0cm,height=6.0cm}
}}
\vbox{
\hbox{
\epsfig{file=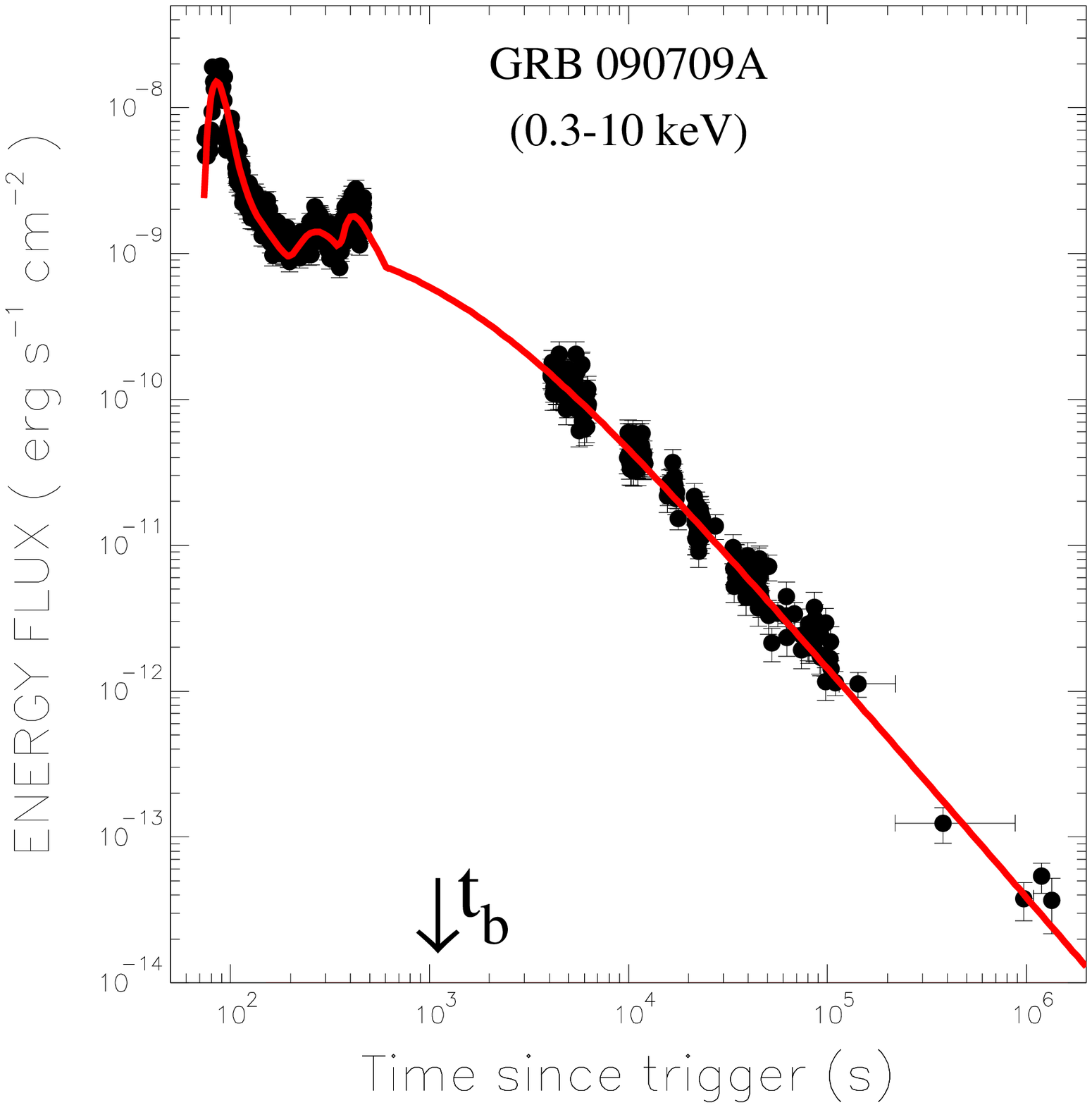,width=8.0cm,height=6cm }
\epsfig{file=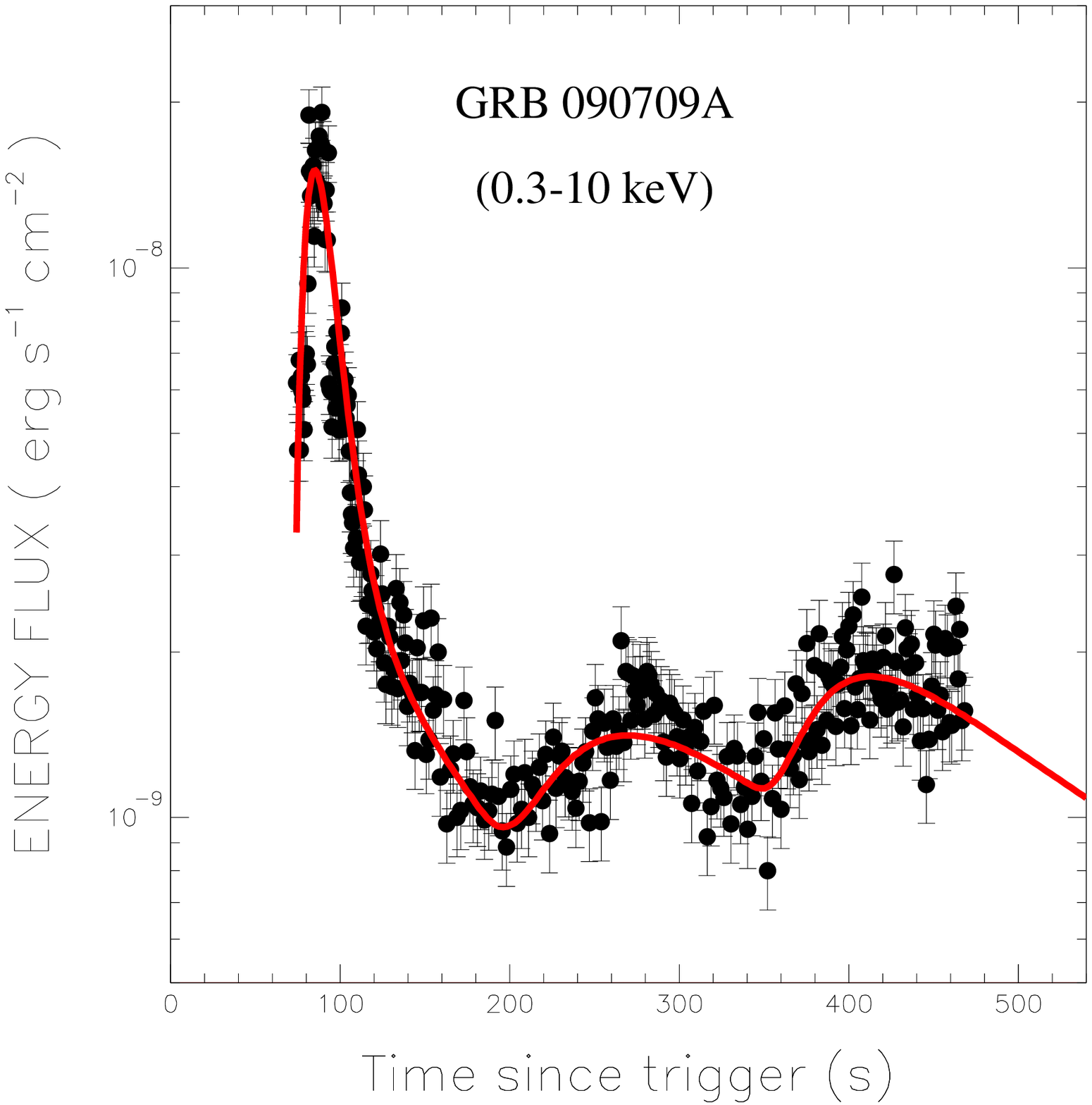,width=8.0cm,height=6cm}
}}

\caption{Comparison between the 0.3-10 KeV X-ray lightcurves
of GRBs with prominent flares in their X-ray afterglow that were
measured by the Swift XRT and reported in the Swift/XRT lightcurve
repository \citep{Evans2009} and their CB model description with the 
parameters listed in Table 1.
{\bf Top left (a):} GRB 060929.
{\bf Top right (b):} Zoom on
the flare in the AG of GRB 060929.
 {\bf Middle left (c):} GRB 081102.
{\bf Middle right (d):} GRB 050916.
{\bf Bottom left (e):} GRB  090709A.
{\bf Bottom right (f):} Zoom on the early-time X-ray flares in the AG of 
GRB 090709A.
 }                       
\label{fig6}
\end{figure}

\newpage
\begin{figure}[]
\centering
\vspace{-1cm}
\vbox{
\hbox{
\epsfig{file=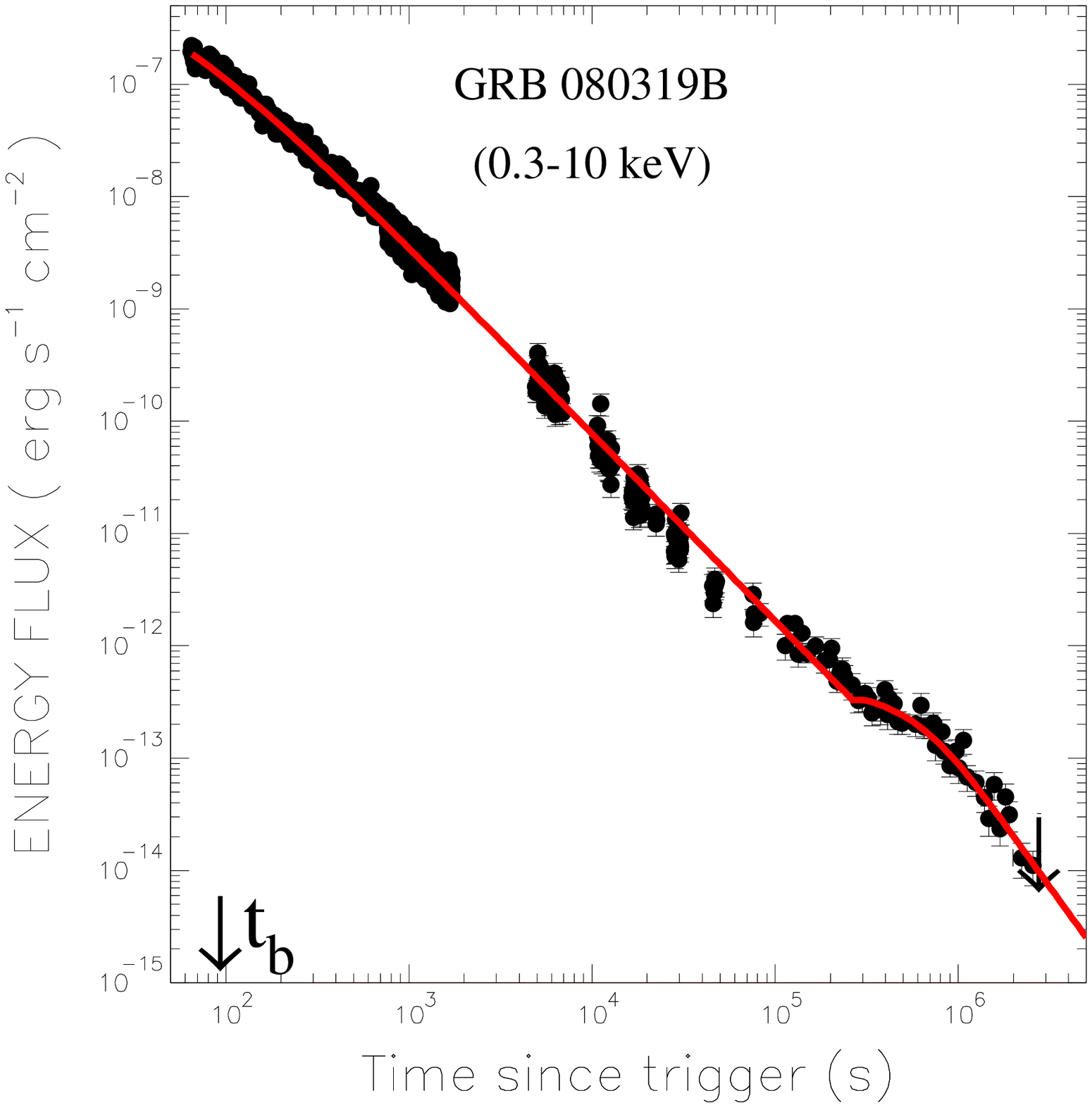,width=8.0cm,height=6.0cm}
\epsfig{file=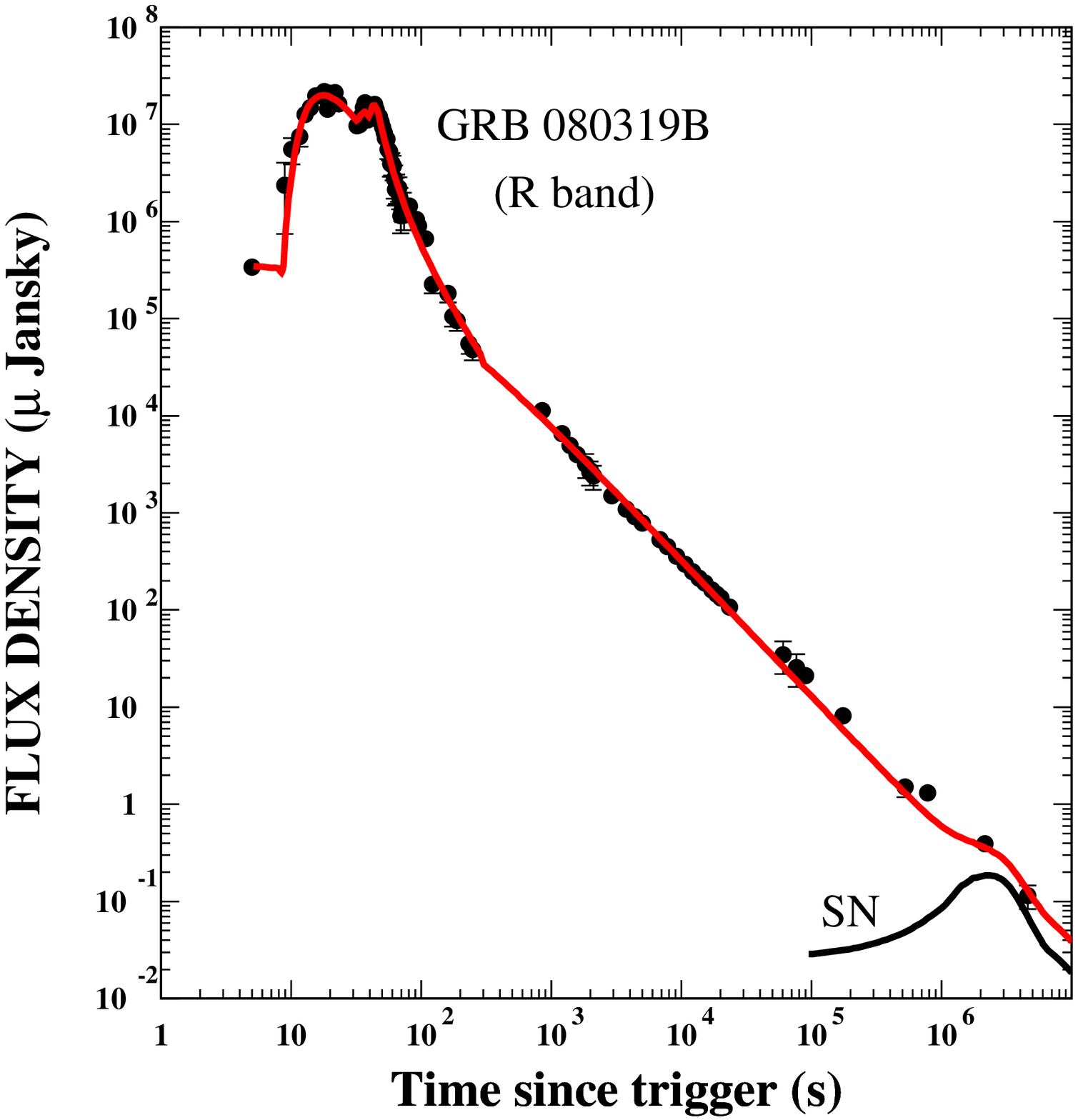,width=8.0cm,height=6.0cm}
}}
\vbox{
\hbox{
\epsfig{file=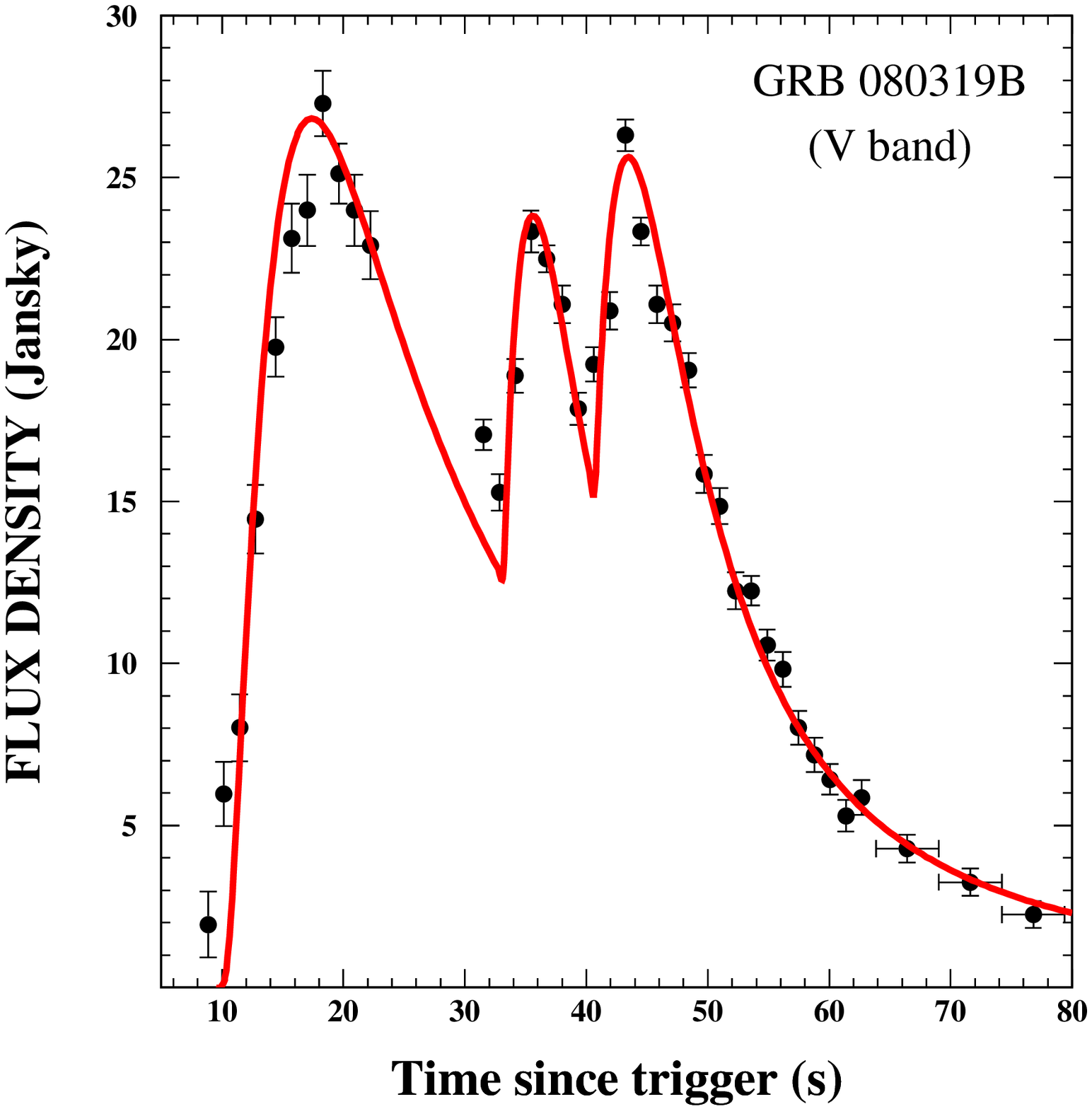,width=8.0cm,height=6.0cm}
\epsfig{file=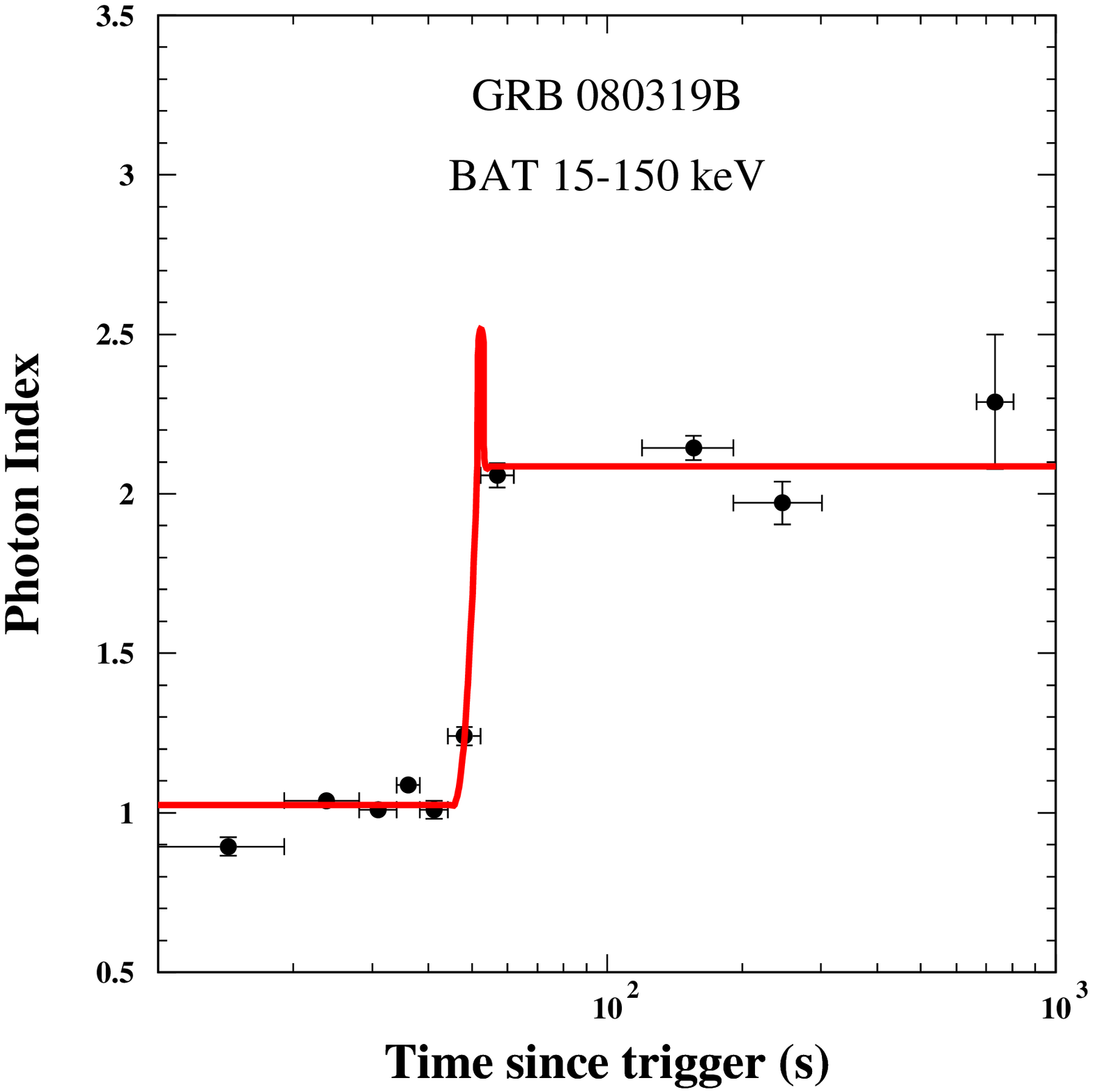,width=8.0cm,height=6.0cm}
}}
\vbox{
\hbox{
\epsfig{file=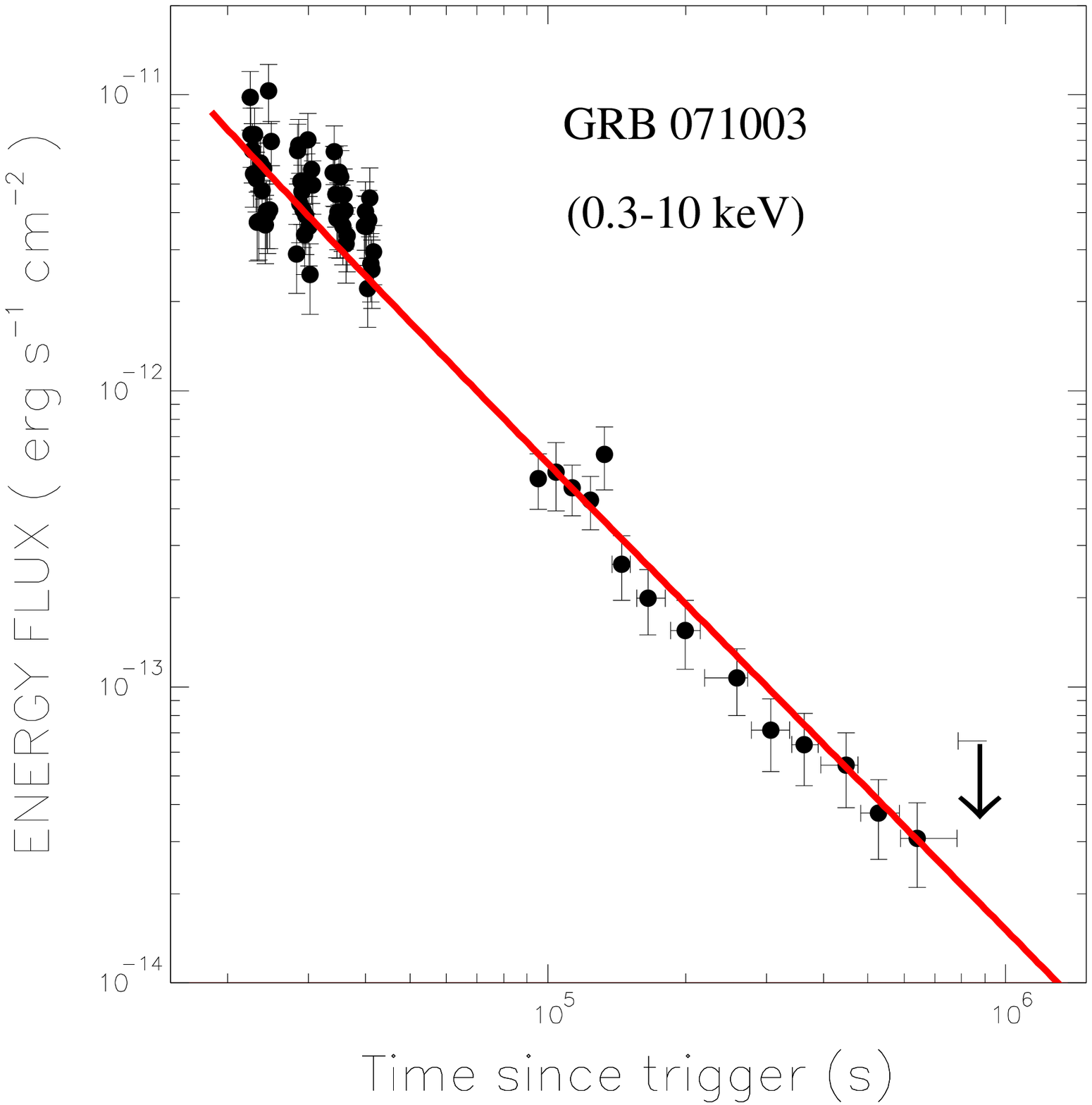,width=8.0cm,height=6.0cm}
\epsfig{file=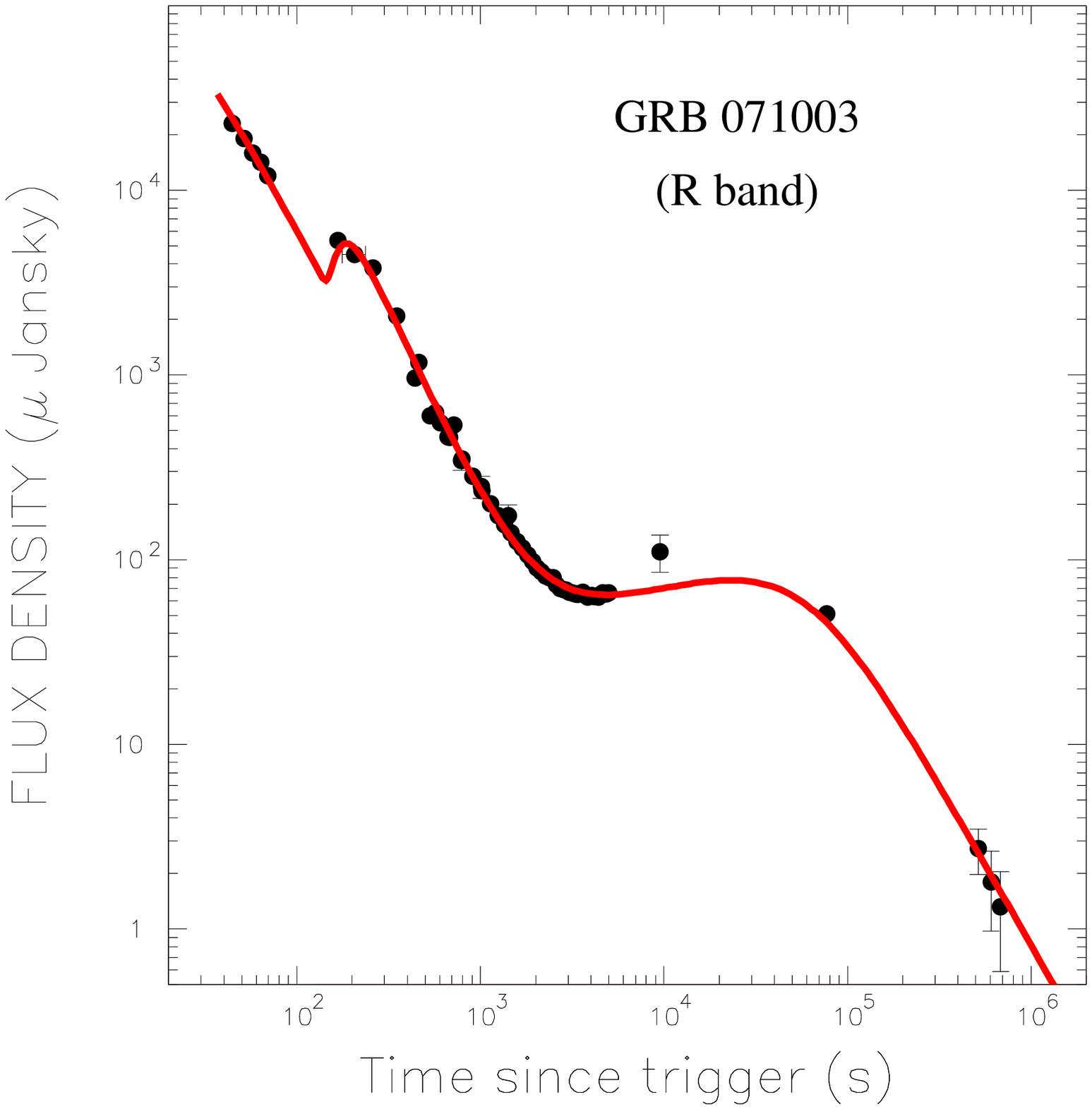,width=8.0cm,height=6.0cm}
}}
\caption{Comparison between the observed early-time X-ray and 
optical lightcurves of very bright GRBs and their CB model descriptions.
GRB 080319B -
{\bf Top left (a):} 0.3-10 keV XRT lightcurve. 
{\bf Top right (b):} R-band lightcurve.
{\bf Middle left (c):} The early-time optical flares.  
{\bf Middle right (d):} The spectral index lightcurve.
GRB 071003 -
{\bf Bottom left (e):} The 0.3-10 keV XRT lightcurve.
{\bf Bottom right (f):} The entire R-band lightcurve.
}
\label{fig7}
\end{figure}

\newpage
\begin{figure}[]
\centering
\vspace{-1cm}
\vbox{
\hbox{
\epsfig{file=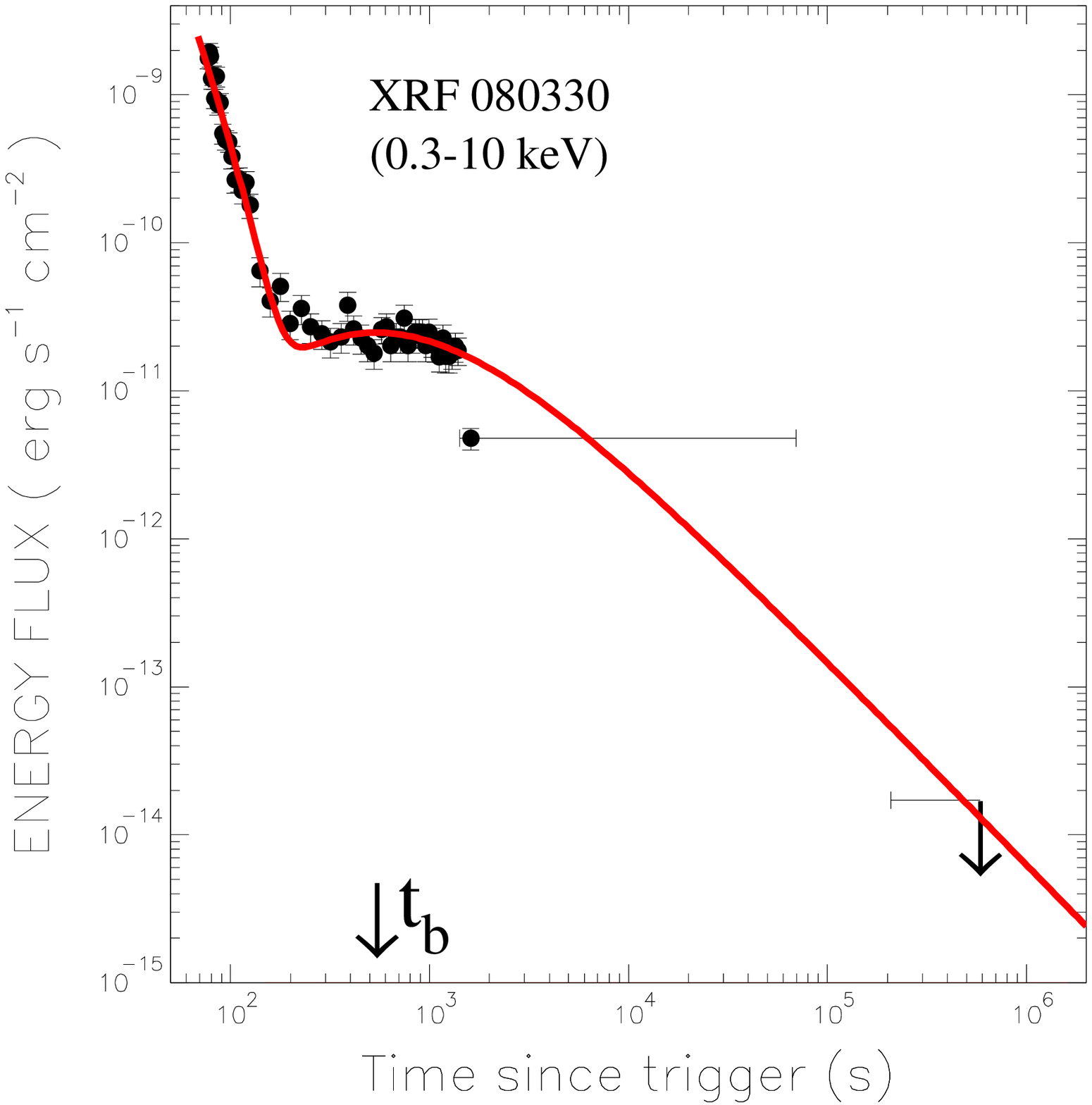,width=8.0cm,height=6.0cm}
\epsfig{file=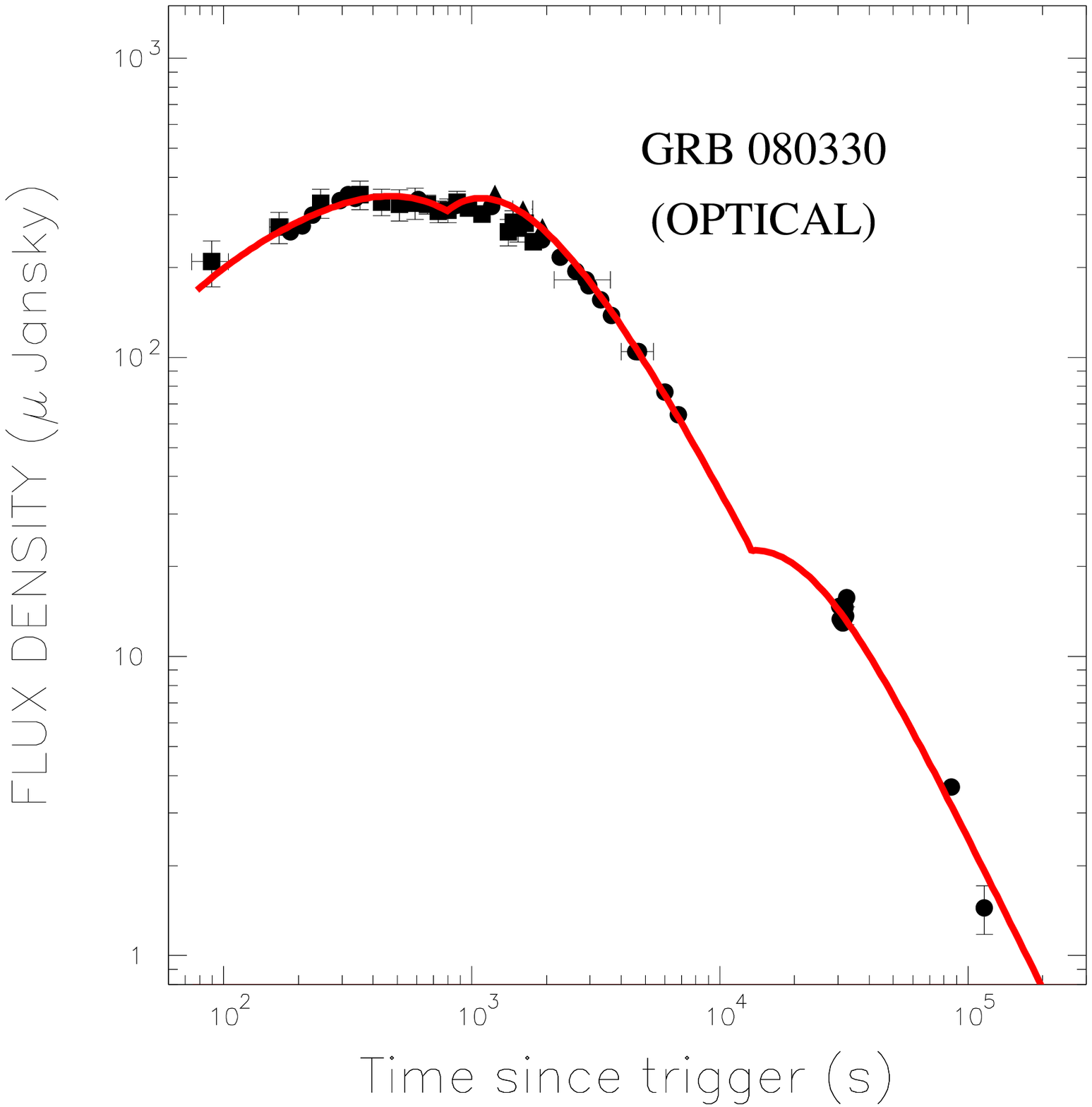,width=8.0cm,height=6.0cm}
}}
\vbox{
\hbox{
\epsfig{file=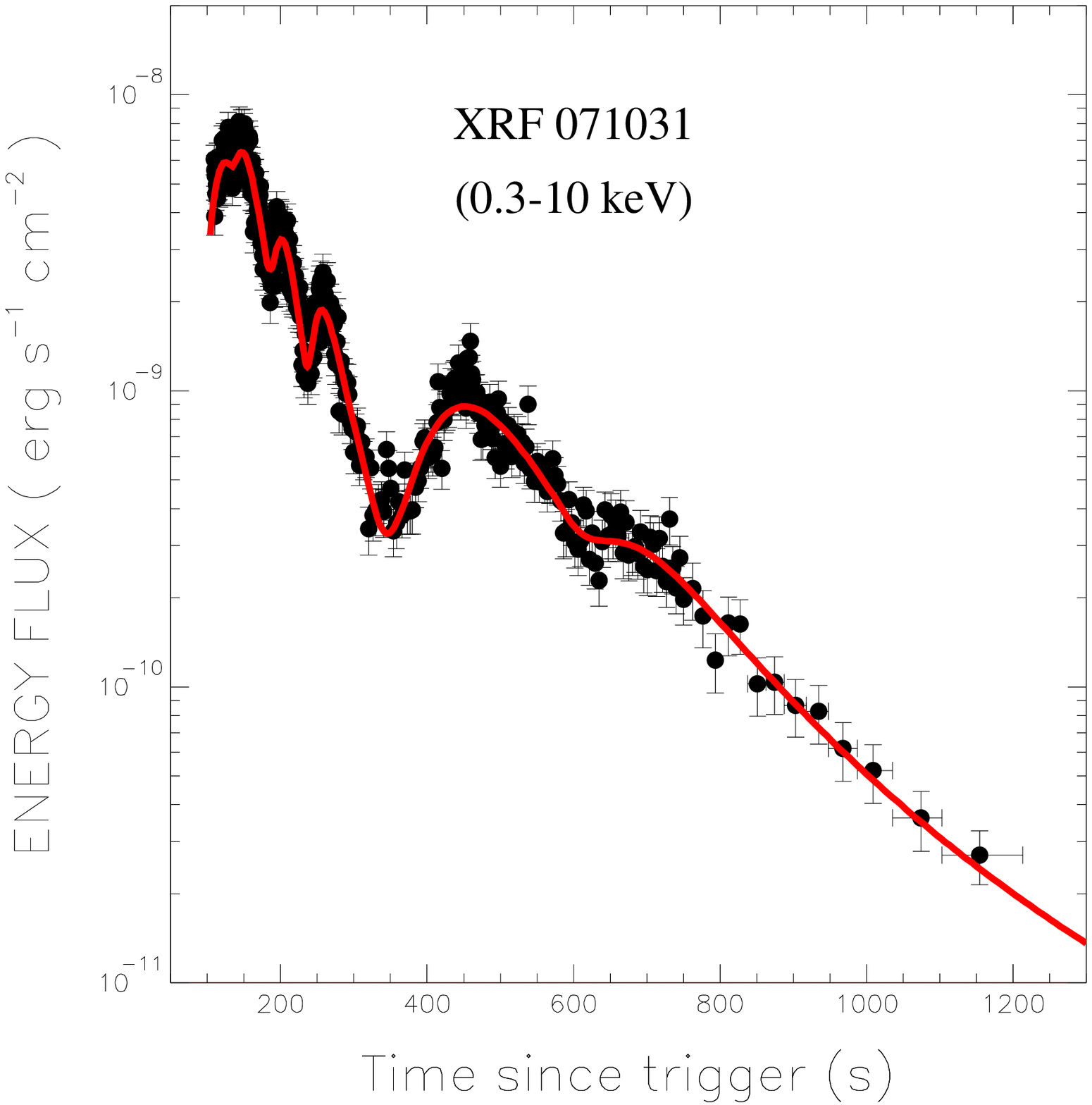,width=8.0cm,height=6.0cm}
\epsfig{file=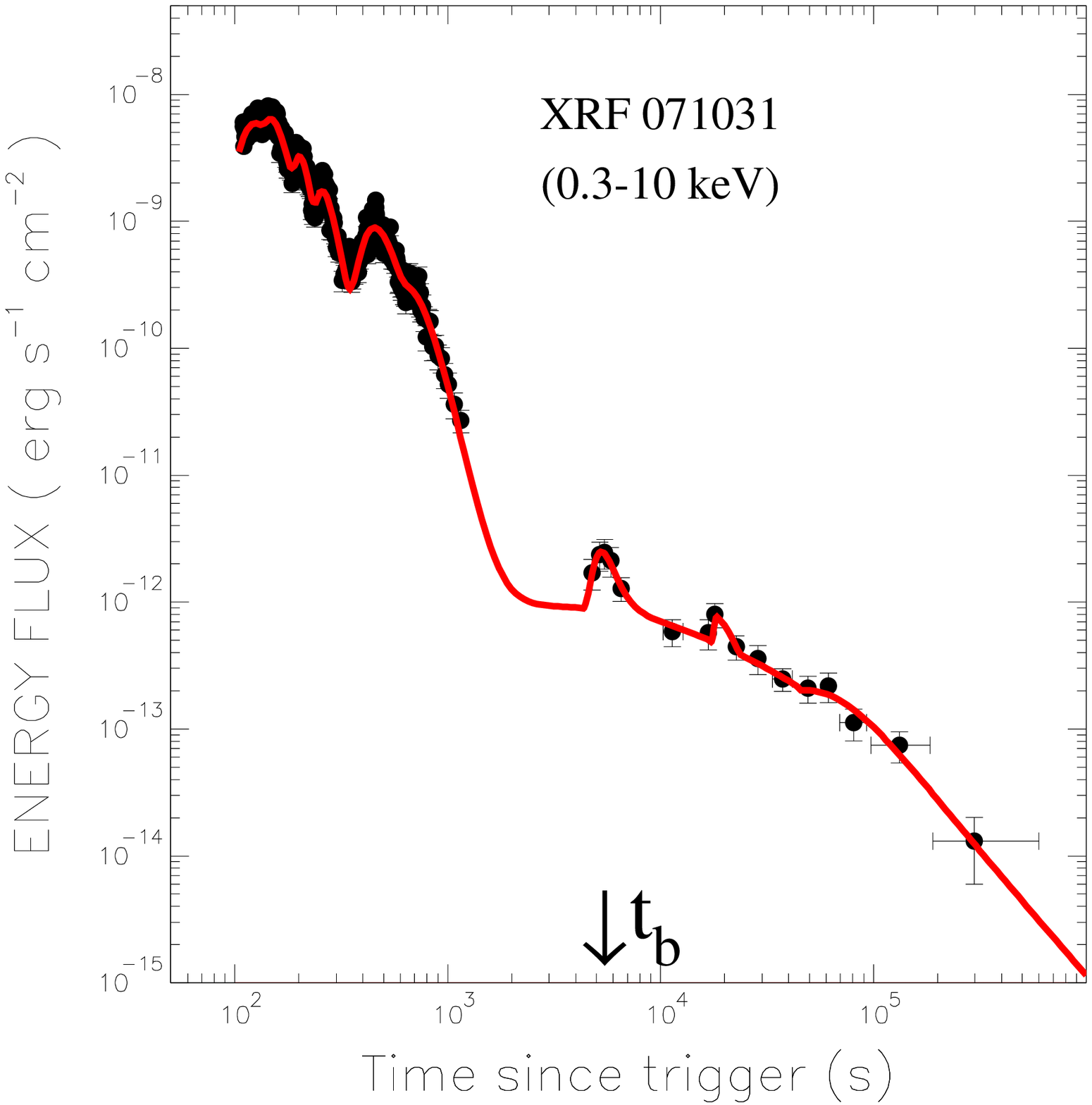,width=8.0cm,height=6.0cm}
}}
\vbox{
\hbox{
\epsfig{file=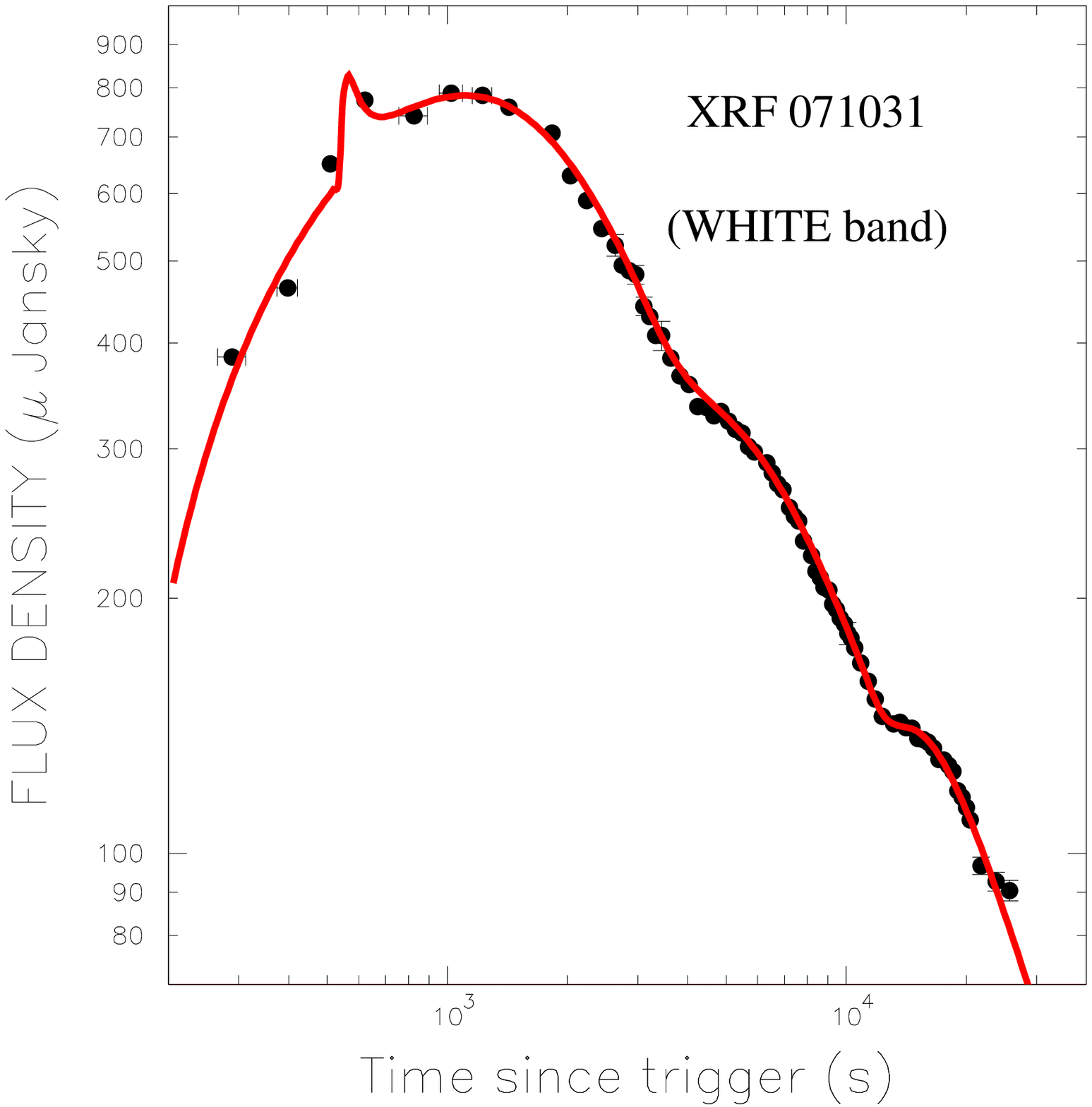,width=8.0cm,height=6.0cm}
\epsfig{file=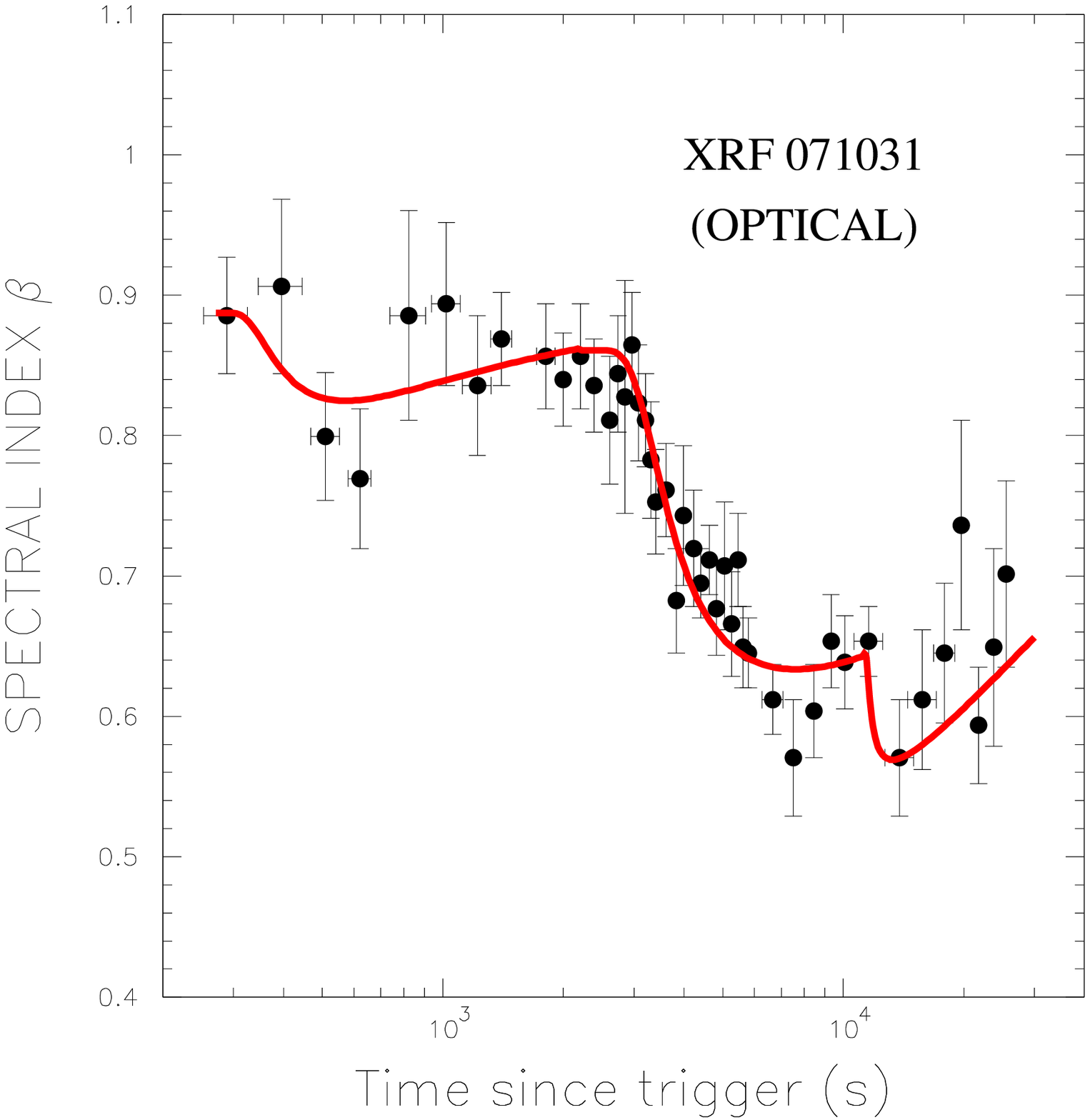,width=8.0cm,height=6.0cm}
}}
\caption{Comparison between the observed early-time X-ray and 
optical lightcurves of XRFs and their CB 
model description. Unlike the X-ray flares, the early-time optical 
flares are barely resolvable to separate  optical flares.
XRF 080330 -
{\bf Top left (a):} The 0.3-10 keV XRT lightcurve. 
{\bf Top right (b):} The optical (white) lightcurve. 
XRF 071031-
{\bf Middle left (c):} The 0.3-10 keV XRT lightcurve.
{\bf Middle right (d):} Zoom on the early time XRT lightcurve.
{\bf Bottom left (e):}  The e
arly-time optical (white) lightcurve.
{\bf Bottom right (f):} Evolution of the optical spectral index. 
}
\label{fig8}
\end{figure}

\newpage
\begin{figure}[]
\centering
\vspace{-1cm}
\vbox{
\hbox{
\epsfig{file=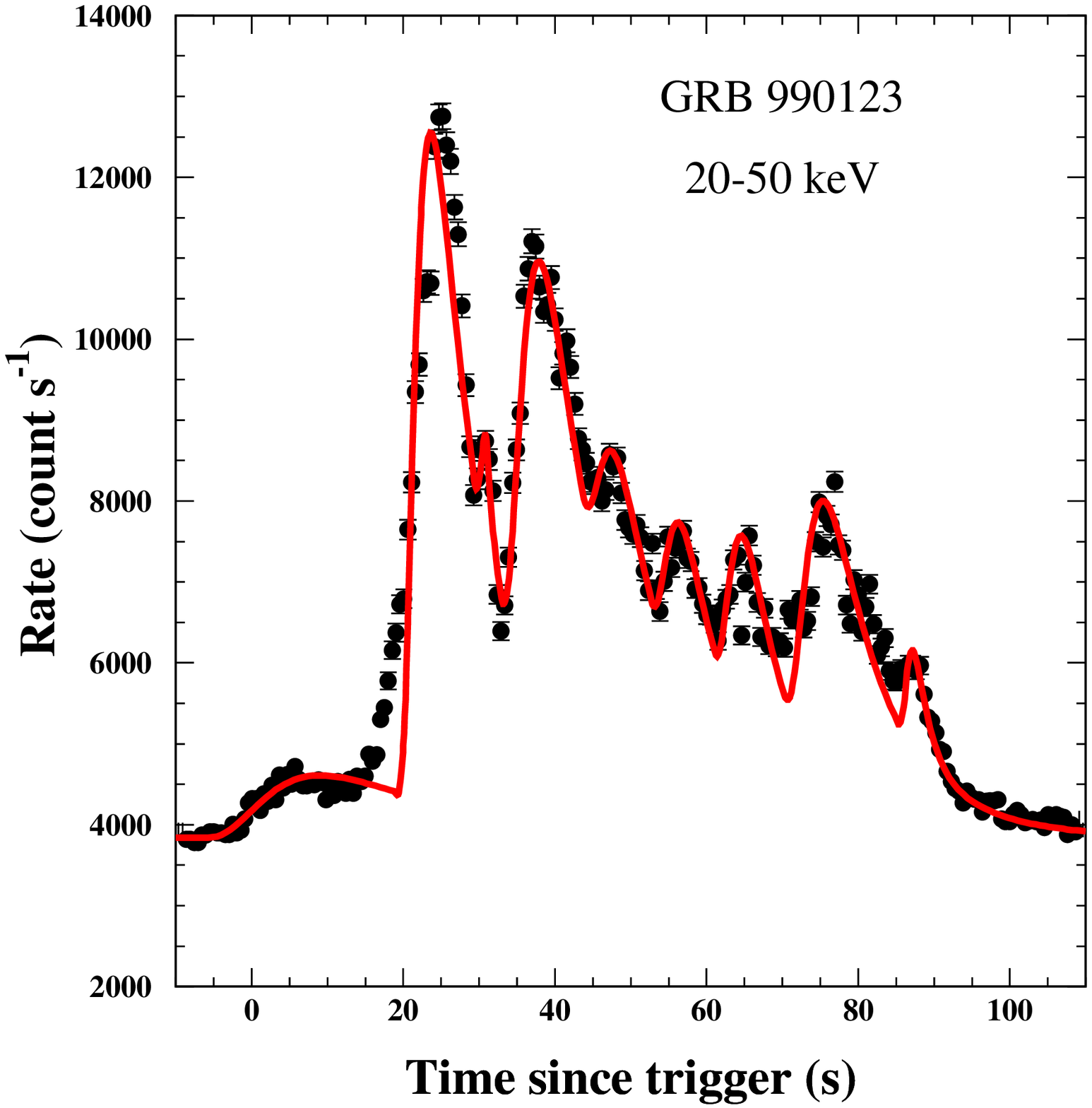,width=8.0cm,height=6.0cm}
\epsfig{file=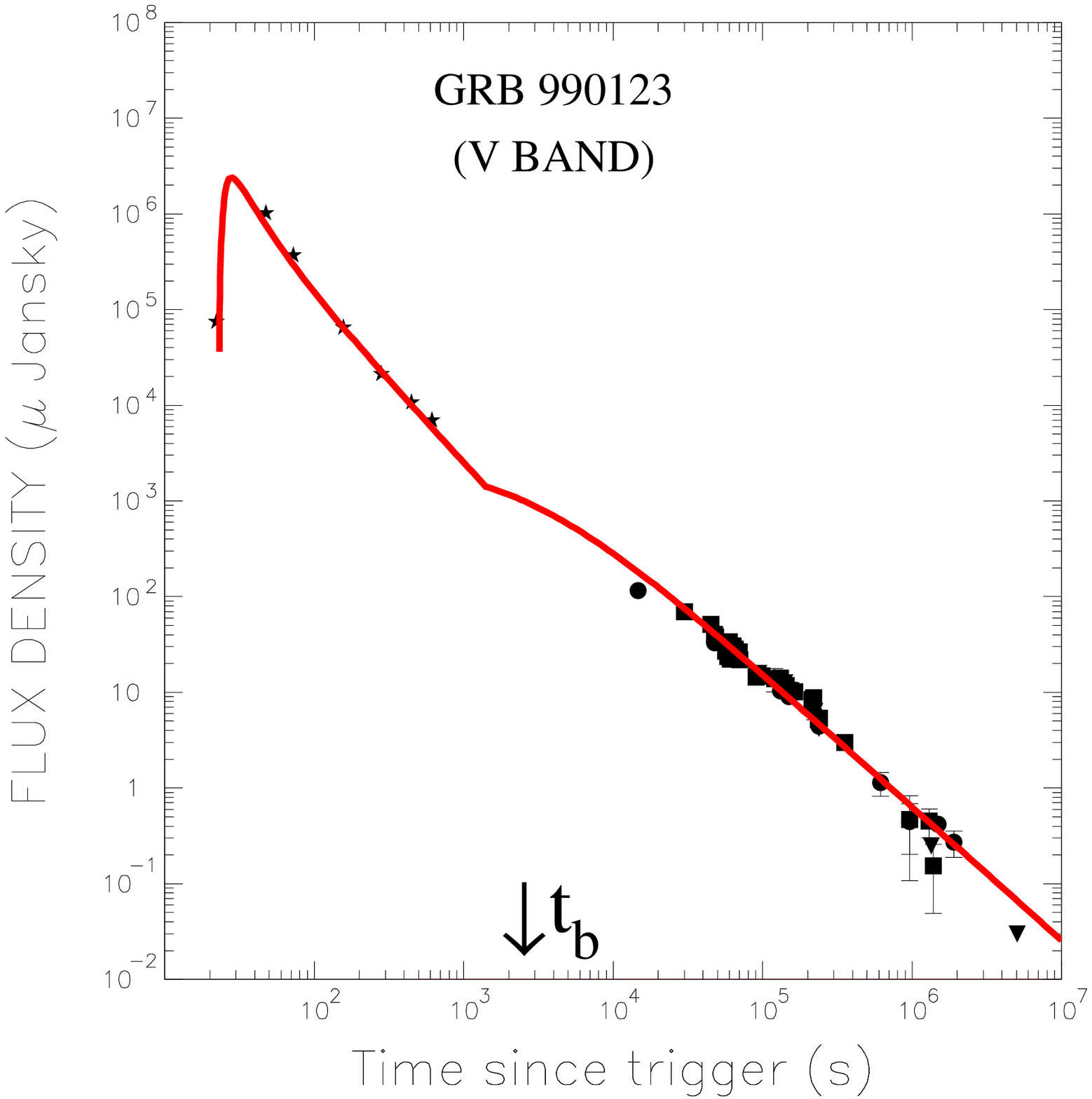,width=8.0cm,height=6.0cm}
}}
\vbox{
\hbox{
\epsfig{file=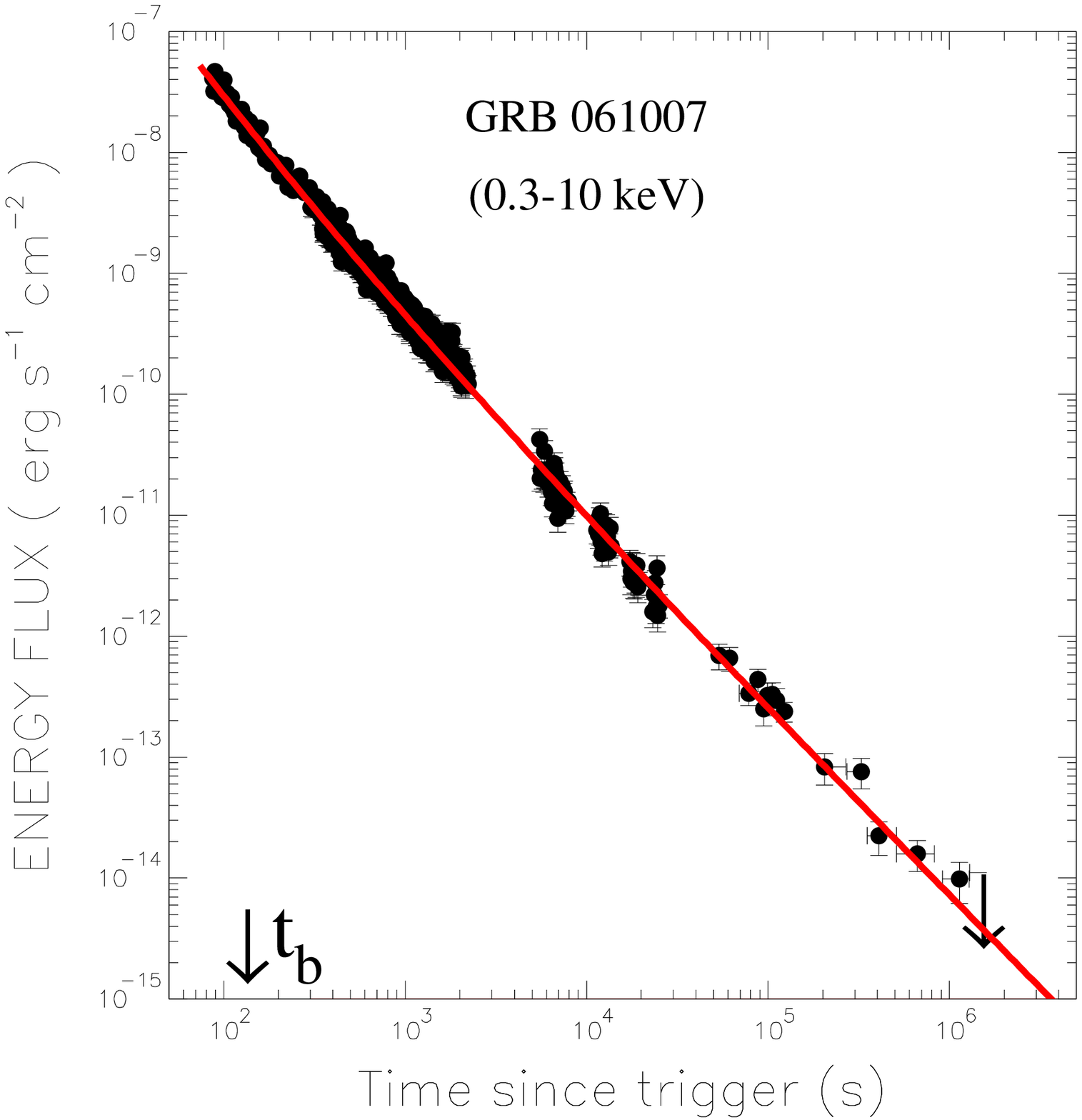,width=8.0cm,height=6.0cm}
\epsfig{file=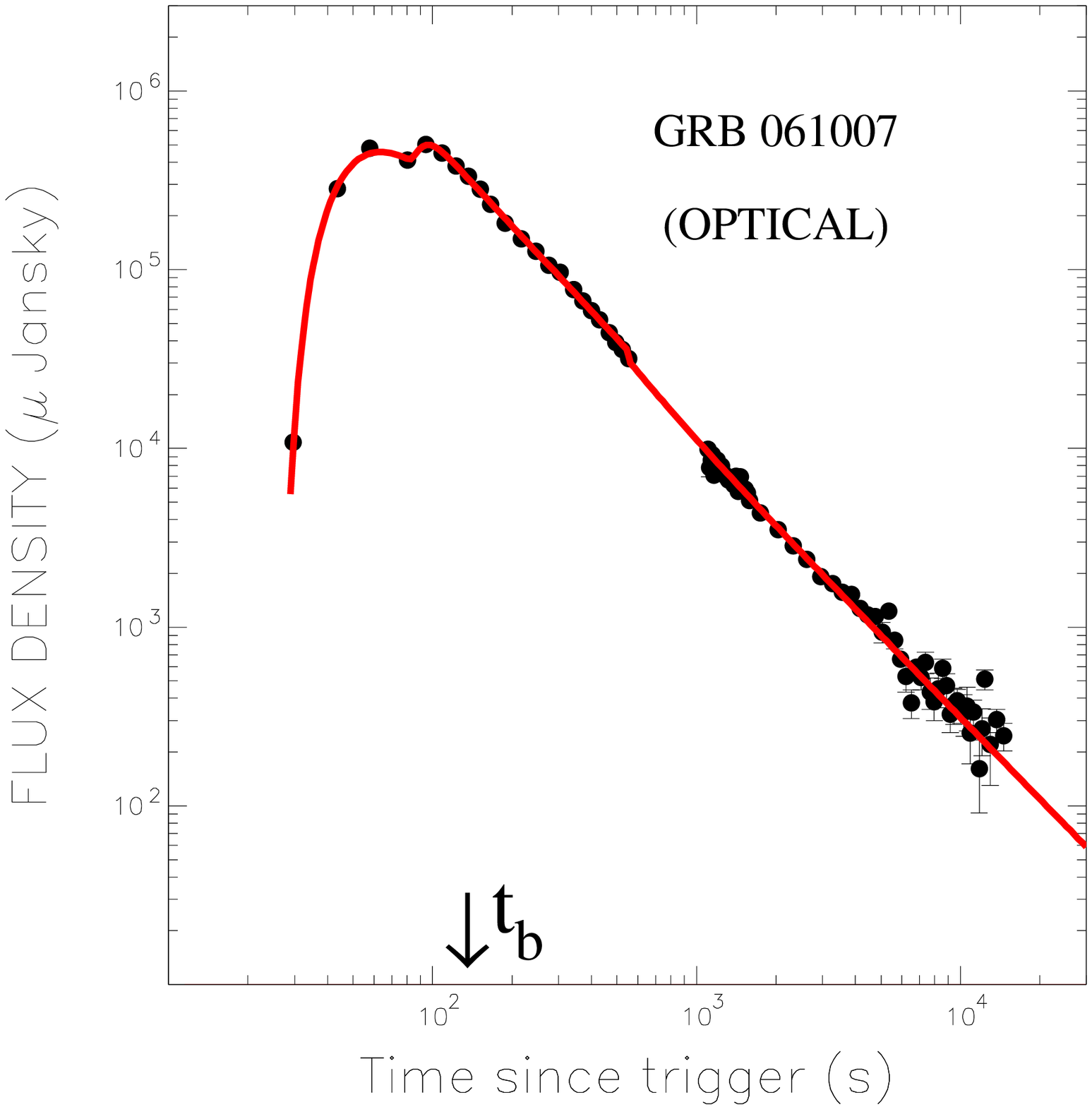,width=8.0cm,height=6.0cm}
}}
\vbox{
\hbox{
\epsfig{file=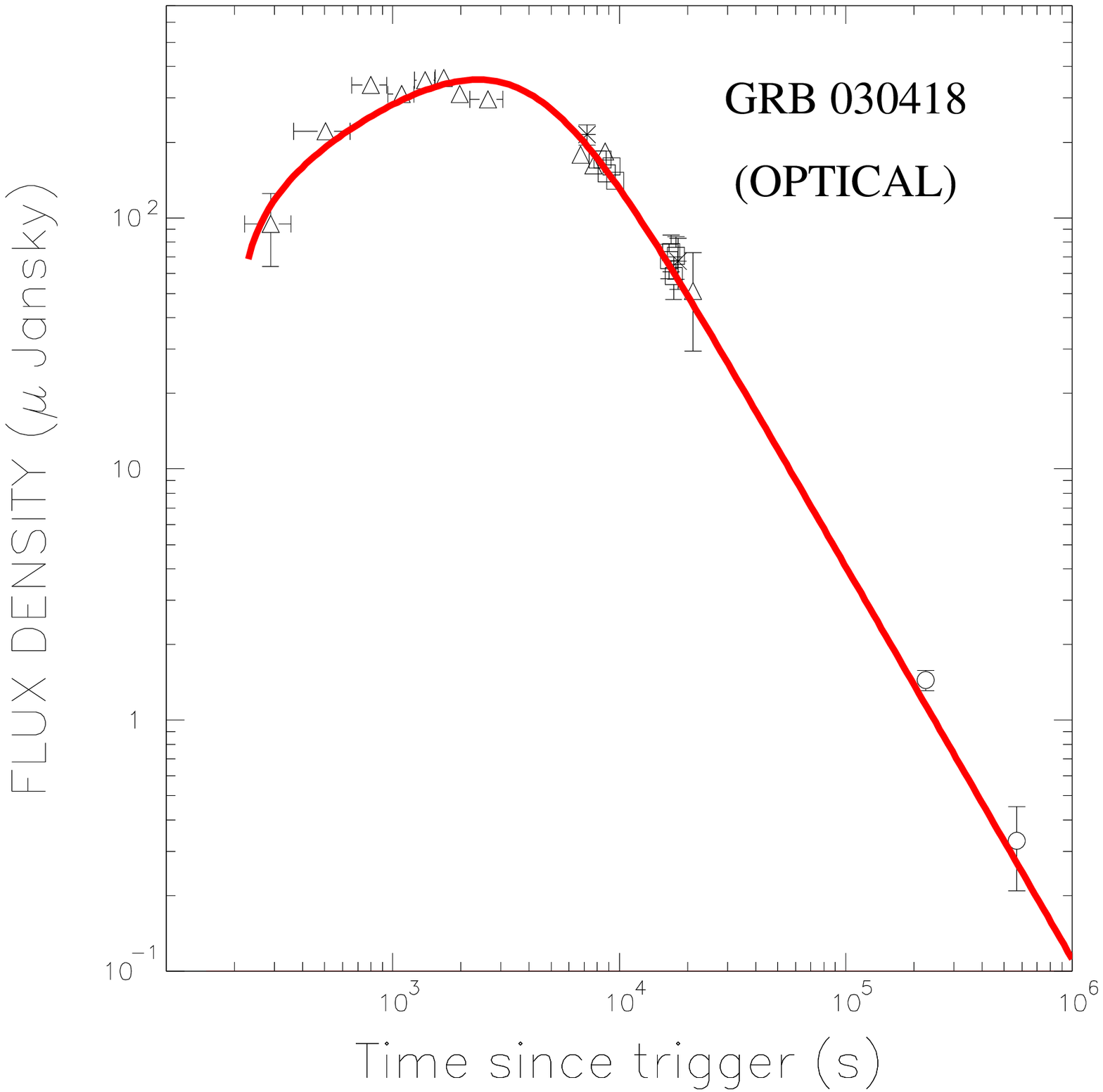,width=8.0cm,height=6.0cm}
\epsfig{file=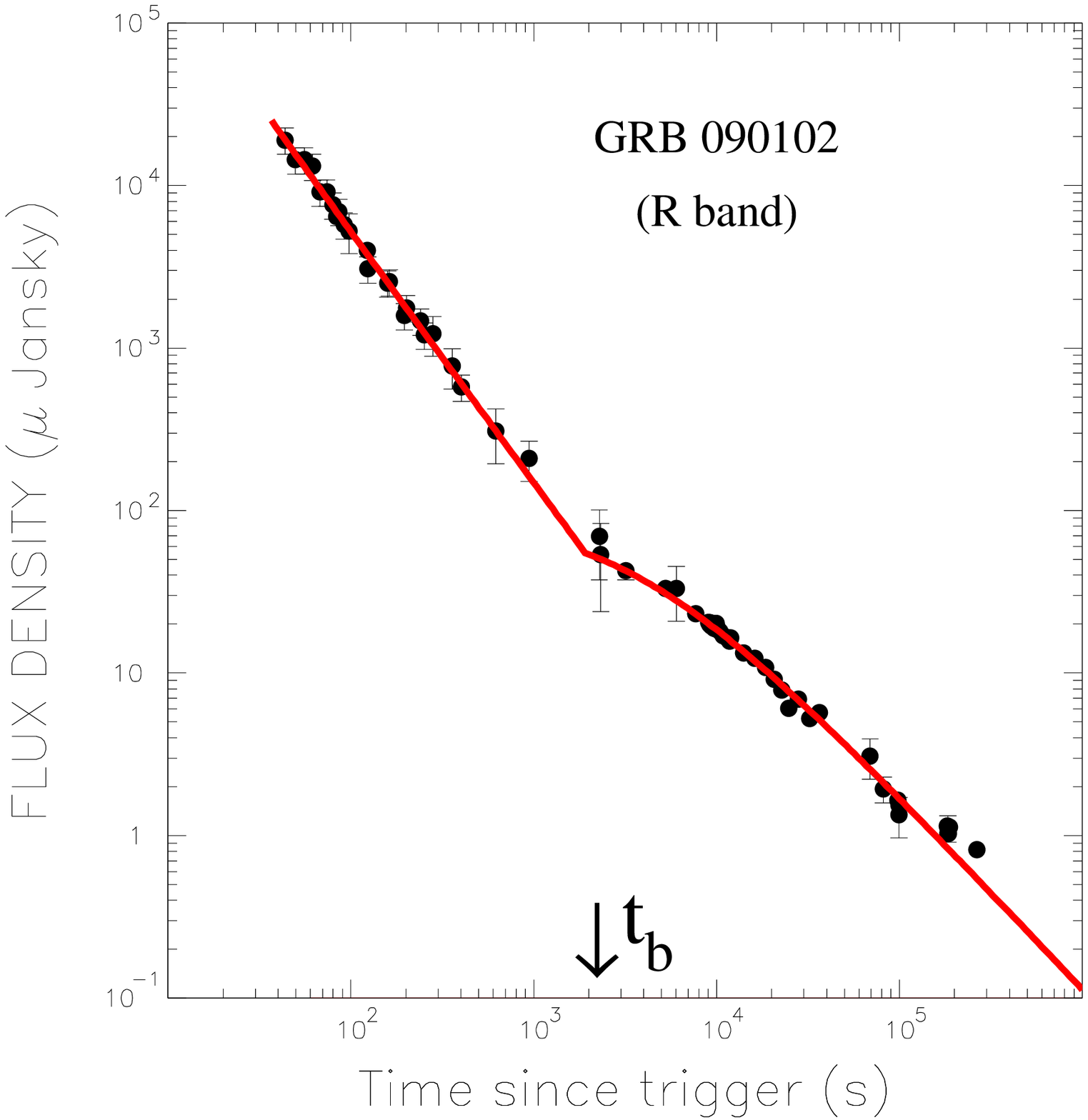,width=8.0cm,height=6.0cm}
}}
\caption{Comparison between observed and CB
model description of lightcurves. 
GRB 990123 -
{\bf Top left (a):} Comparison between the 20-50 keV BATSE lightcurve
\citep{Briggs1999}
and its CB model description, Eq.~(\ref{ICSlc}),
in terms of 9 ICS peaks + a constant background of 3850 counts $\rm s^{-1}$.
{\bf Top right (b):}
Comparison between the $V$ band lightcurve of GRB 990123
and its CB model description assuming a single CB moving
in circumstellar density profile
$\propto\! 1/r^2$  overtaken  by a constant ISM
density around an observer time $t\!=\!1000$ s. The prompt flare was not 
resolved into separate flares. 
GRB 061007 -
{\bf Middle left (c):} The 0.3-10 keV XRT lightcurve.
{\bf Middle right (d):} The optical lightcurve 
with evidence for at least two early time overlapping flares.  
{\bf Bottom left (e):}  The optical (white) lightcurve of GRB 030418 - 
{\bf Bottom right (f):} The R-band lightcurve of GRB 090102
with evidence for a tail of a prompt optical flare. 
}
\label{fig9}
\end{figure}

\newpage
\begin{figure}[]
\centering
\vspace{-1cm}
\vbox{
\hbox{
\epsfig{file=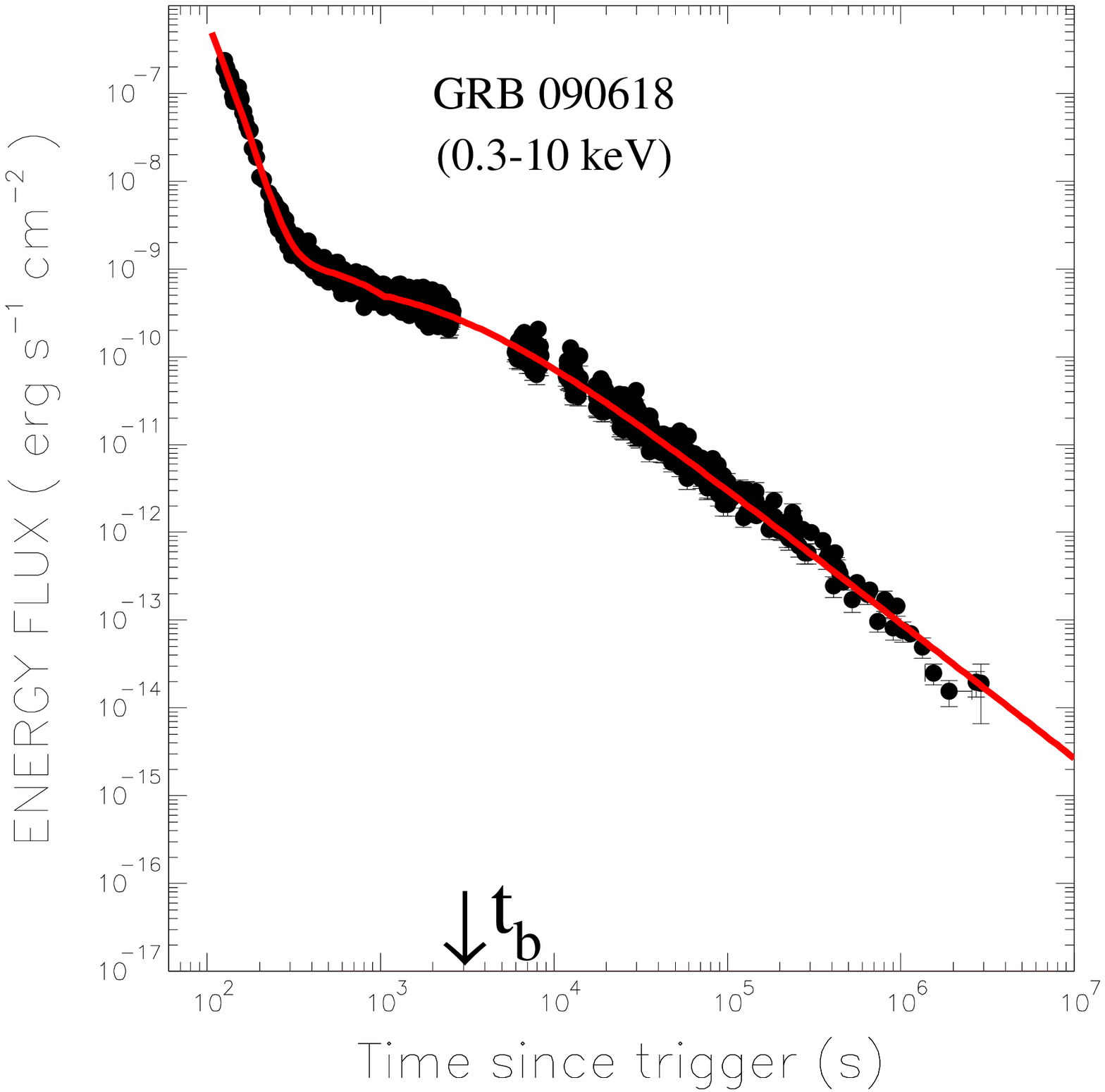,width=8.0cm,height=6.0cm}
\epsfig{file=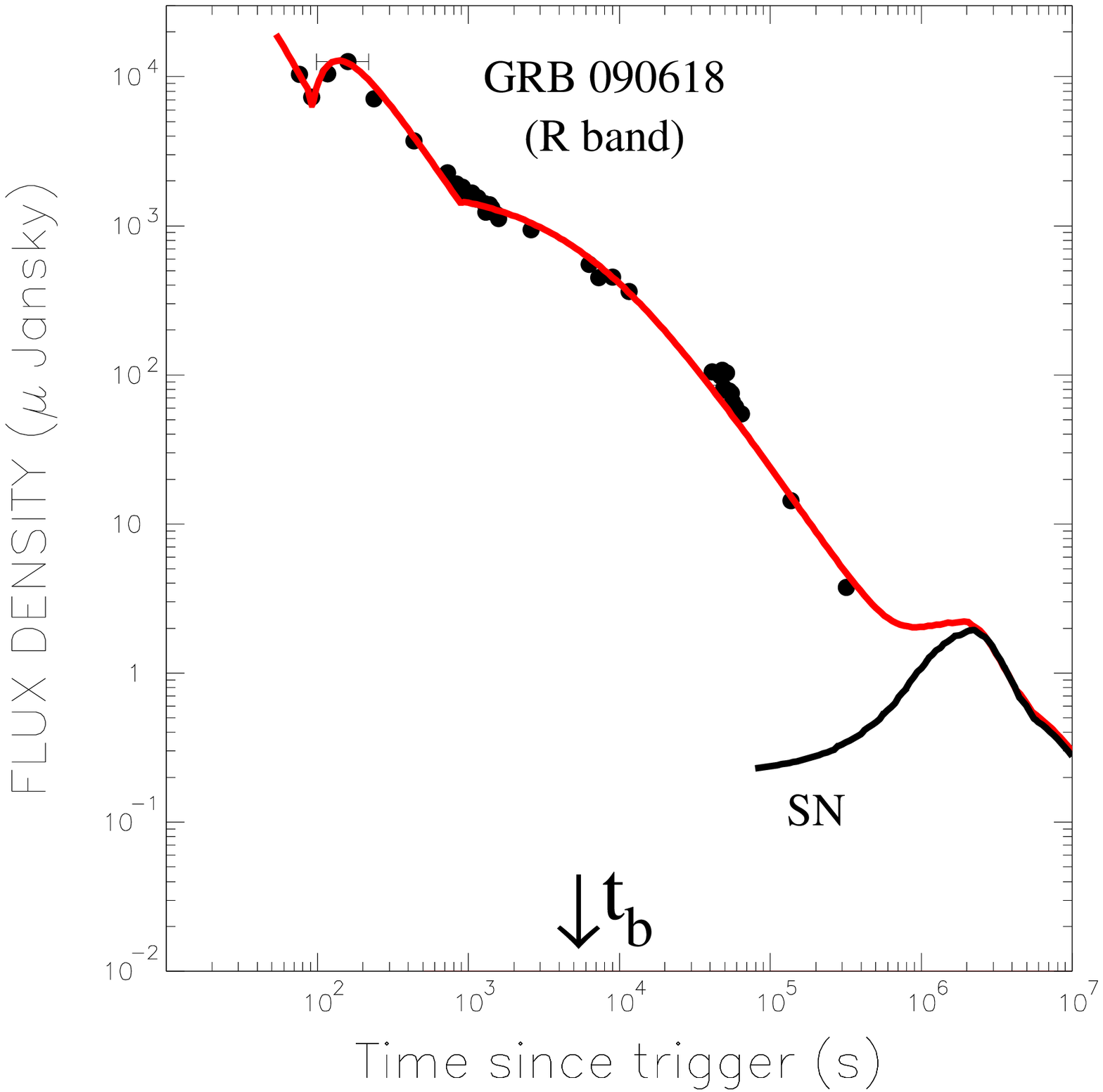,width=8.0cm,height=6.0cm}
}}
\vbox{
\hbox{
\epsfig{file=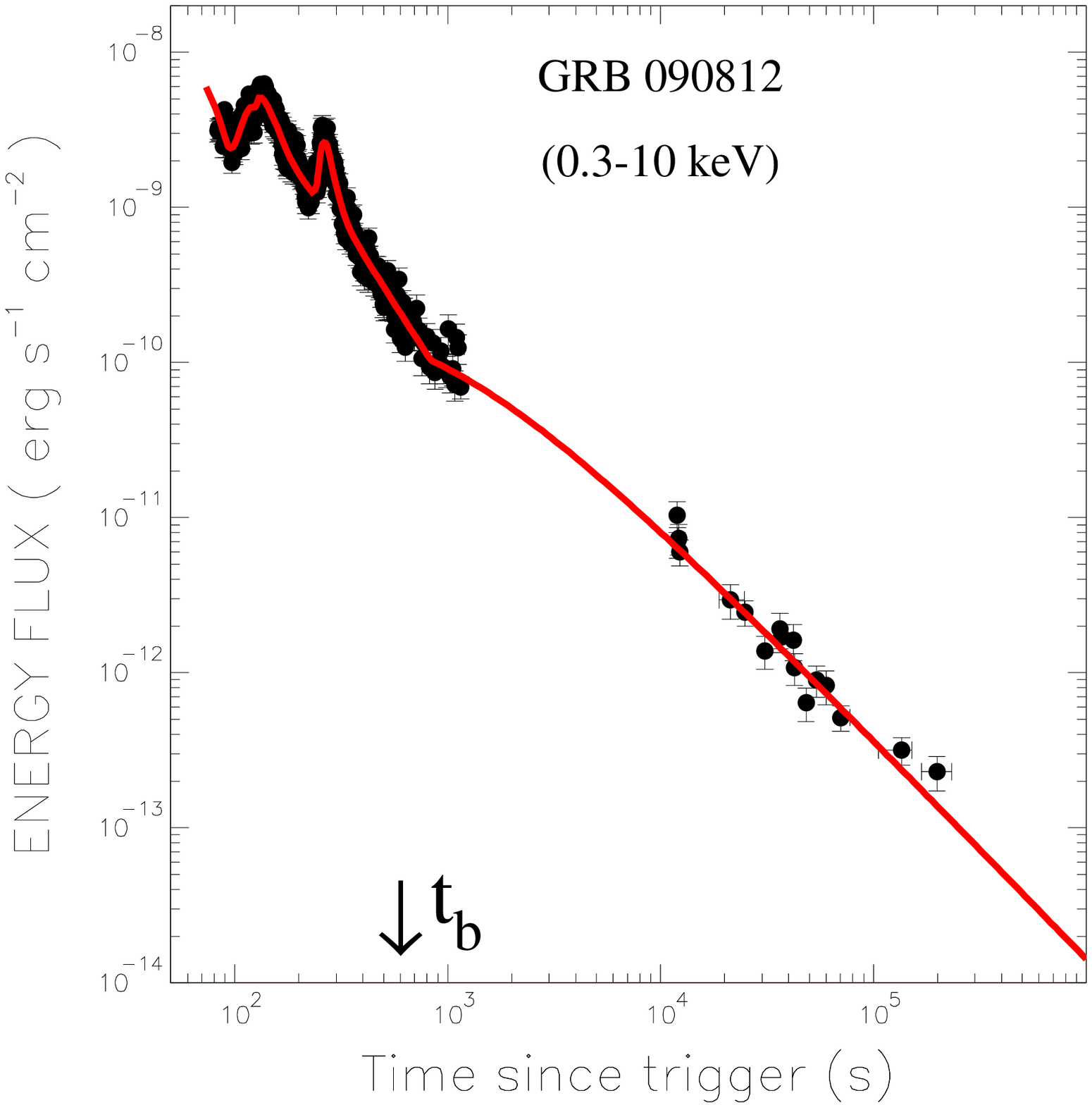,width=8.0cm,height=6.0cm}
\epsfig{file=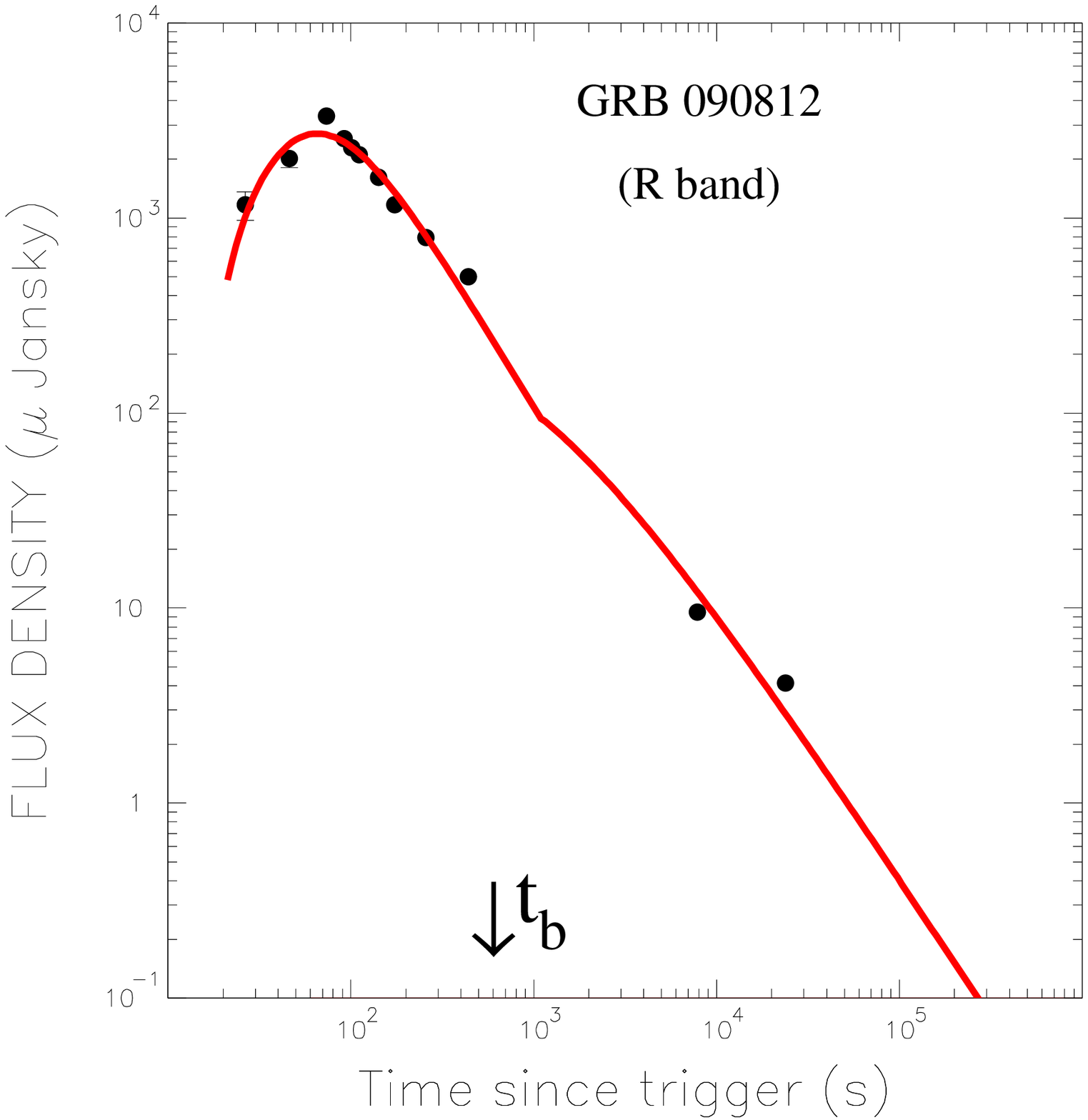,width=8.0cm,height=6.0cm}
}}
\vbox{
\hbox{
\epsfig{file=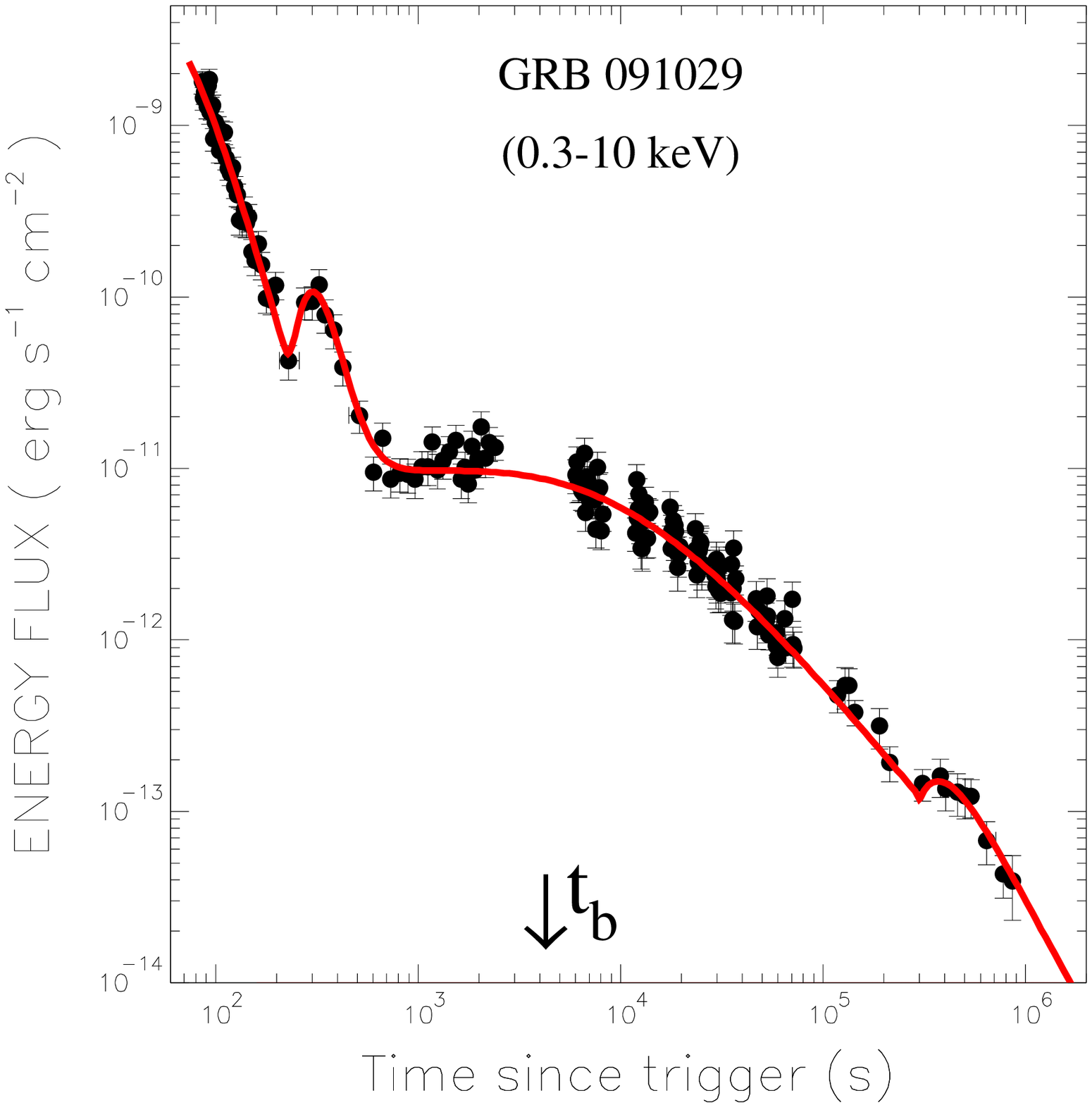,width=8.0cm,height=6.0cm}
\epsfig{file=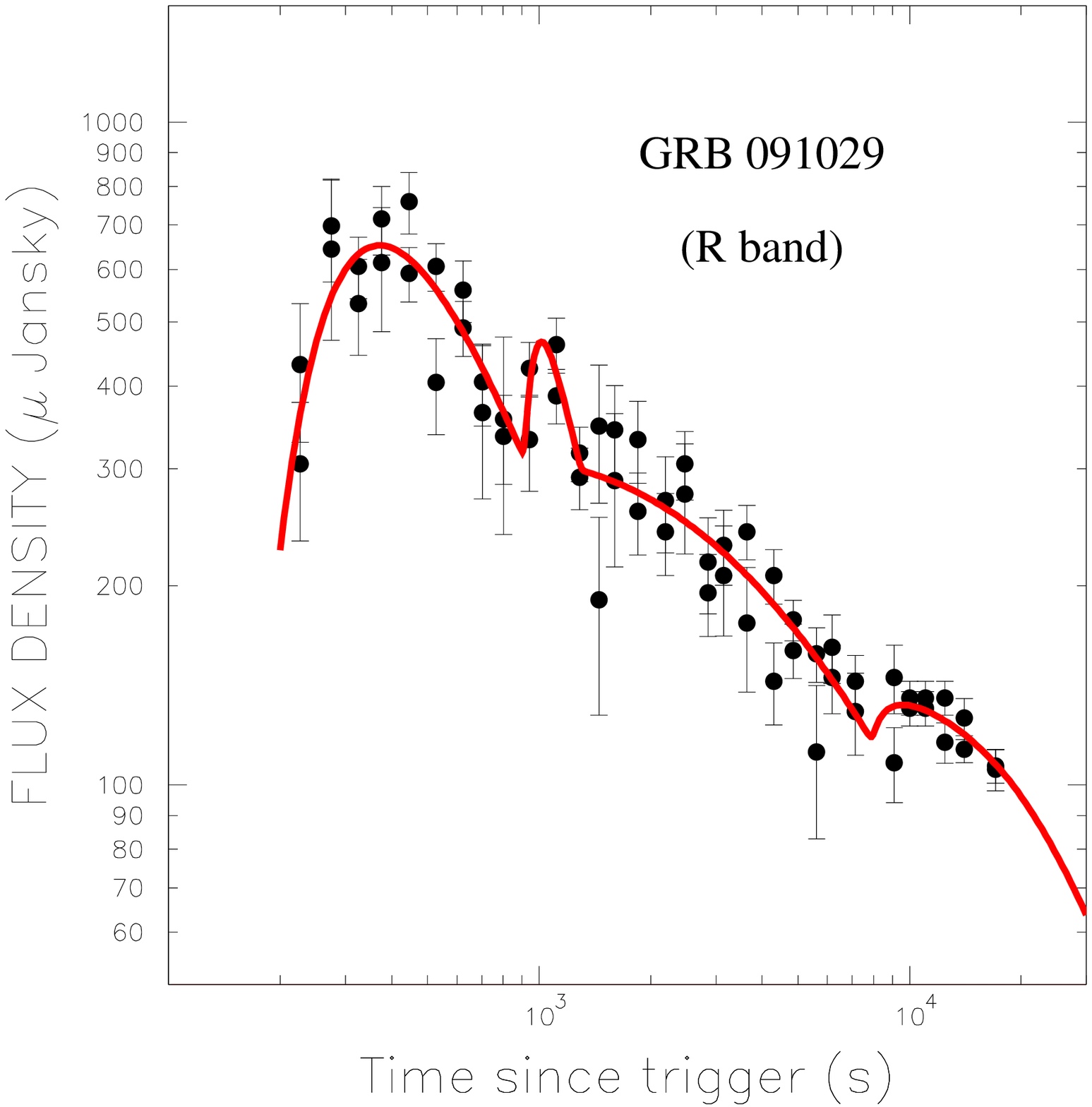,width=8.0cm,height=6.0cm}
}}
\caption{Comparison between the observed early-time X-ray and
optical lightcurve of GRBs and their CB
model description. The X-ray and optical emissions do not appear to be 
correlated.
GRB 090618 -
{\bf Top left (a):} The 0.3-10 keV lightcurve.
{\bf Top right (b):} The optical R-band lightcurve.
GRB 090812 -
{\bf Middle left (c):} The 0.3-10 keV XRT lightcurve.
{\bf Middle right (d):} The optical R-band lightcurve
with evidence for at least two early time overlapping flares.
GRB 091029 -
{\bf Bottom left (e):}  The 0.3-10 keV  X-ray lightcurve.
{\bf Bottom right (f):} The R-band lightcurve
with possible evidence for unresolved optical flares.
}
\label{fig10}
\end{figure}

\newpage
\begin{figure}[]
\centering
\vspace{-1cm}
\vbox{
\hbox{
\epsfig{file=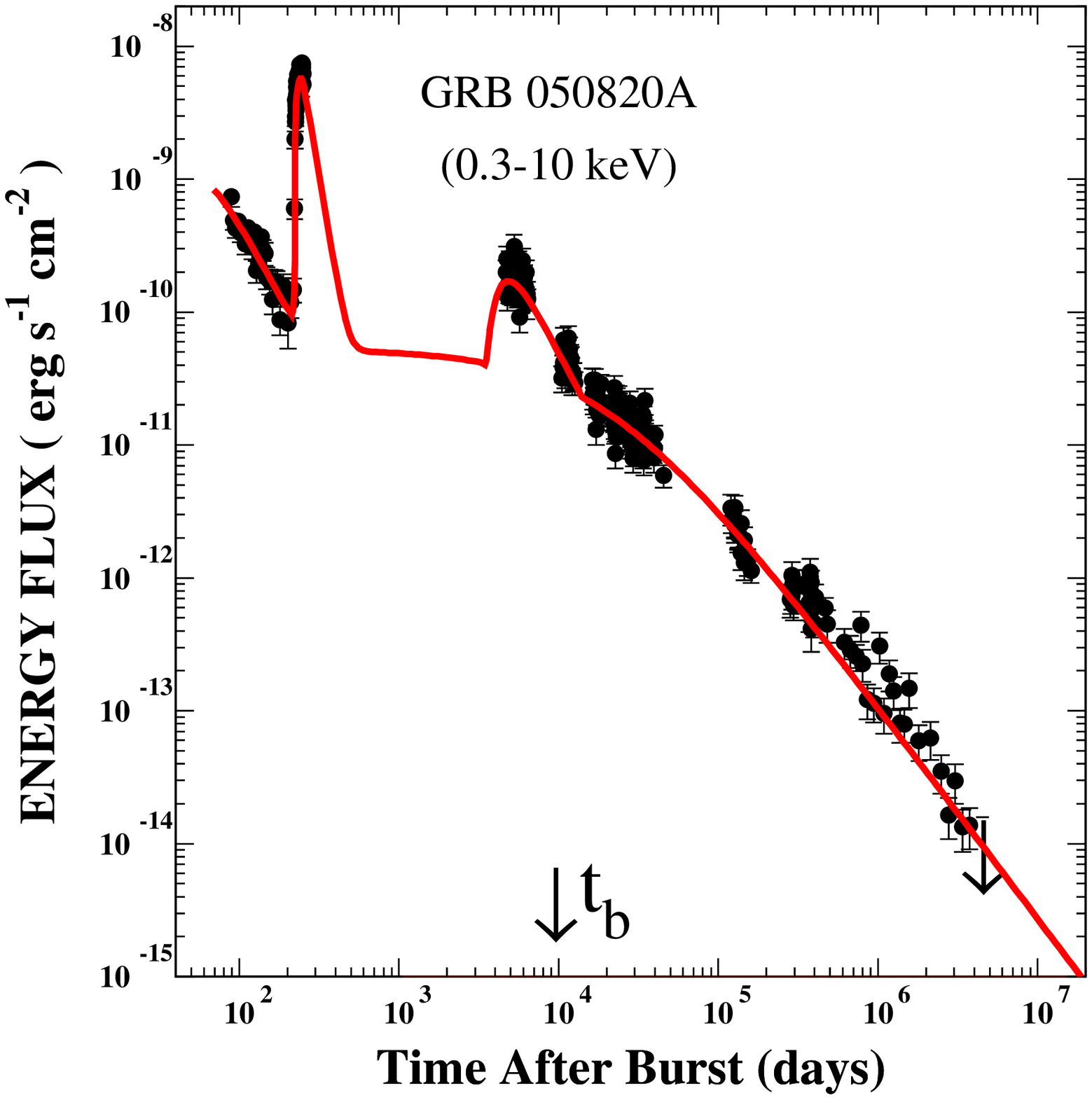,width=8.0cm,height=6.0cm}
\epsfig{file=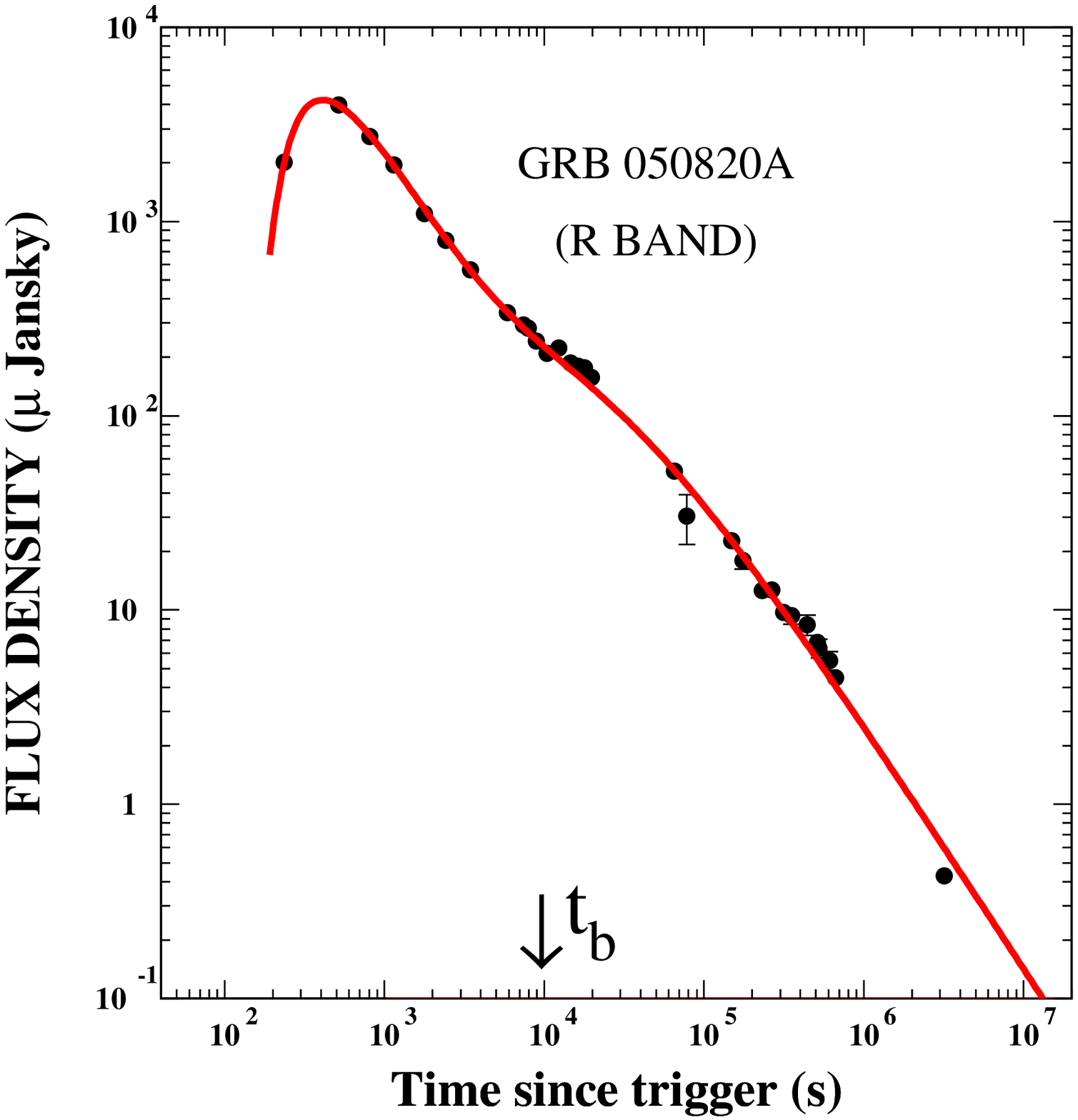,width=8.0cm,height=6.0cm}
}}
\vbox{
\hbox{
\epsfig{file=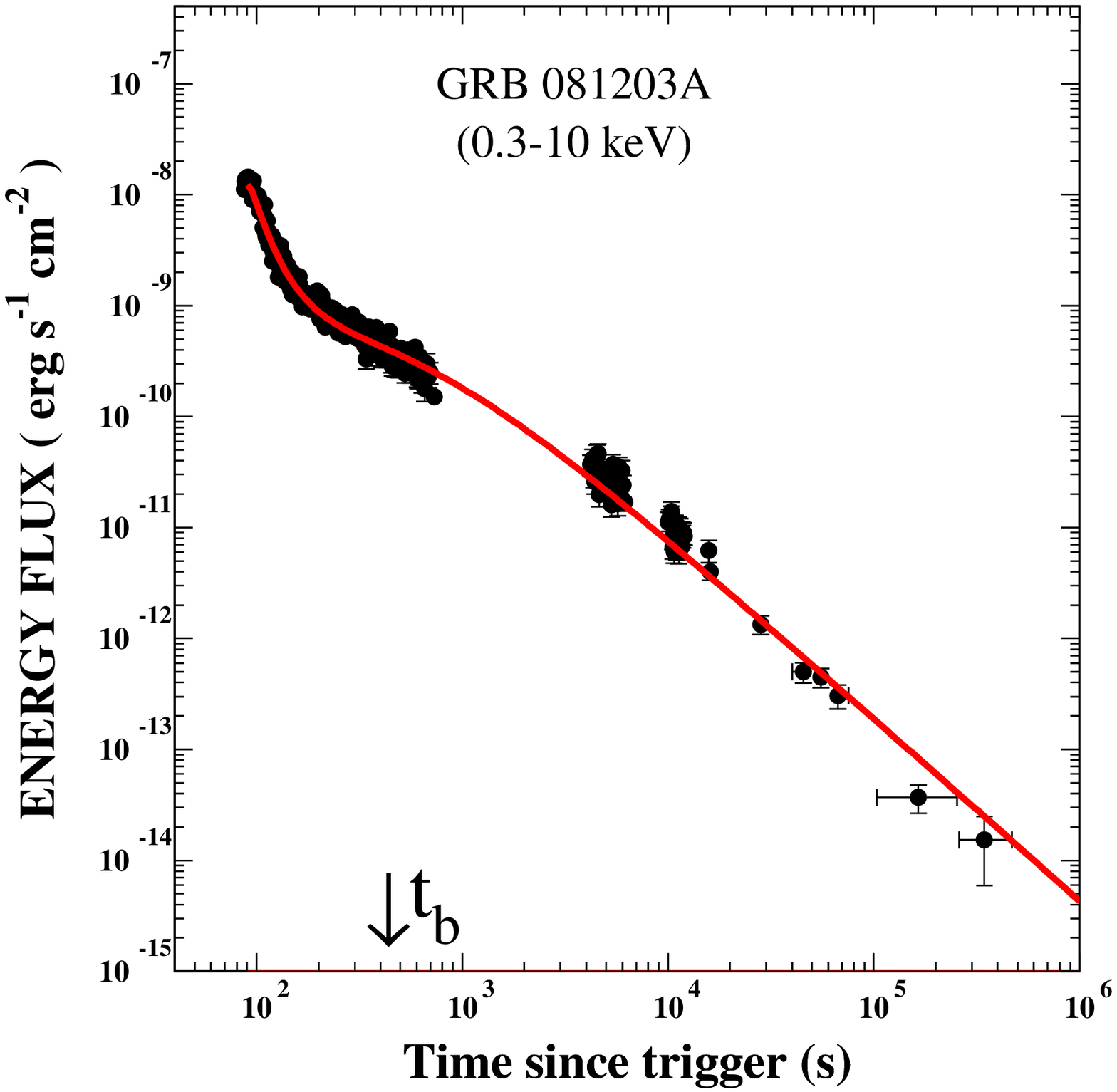,width=8.0cm,height=6.0cm}
\epsfig{file=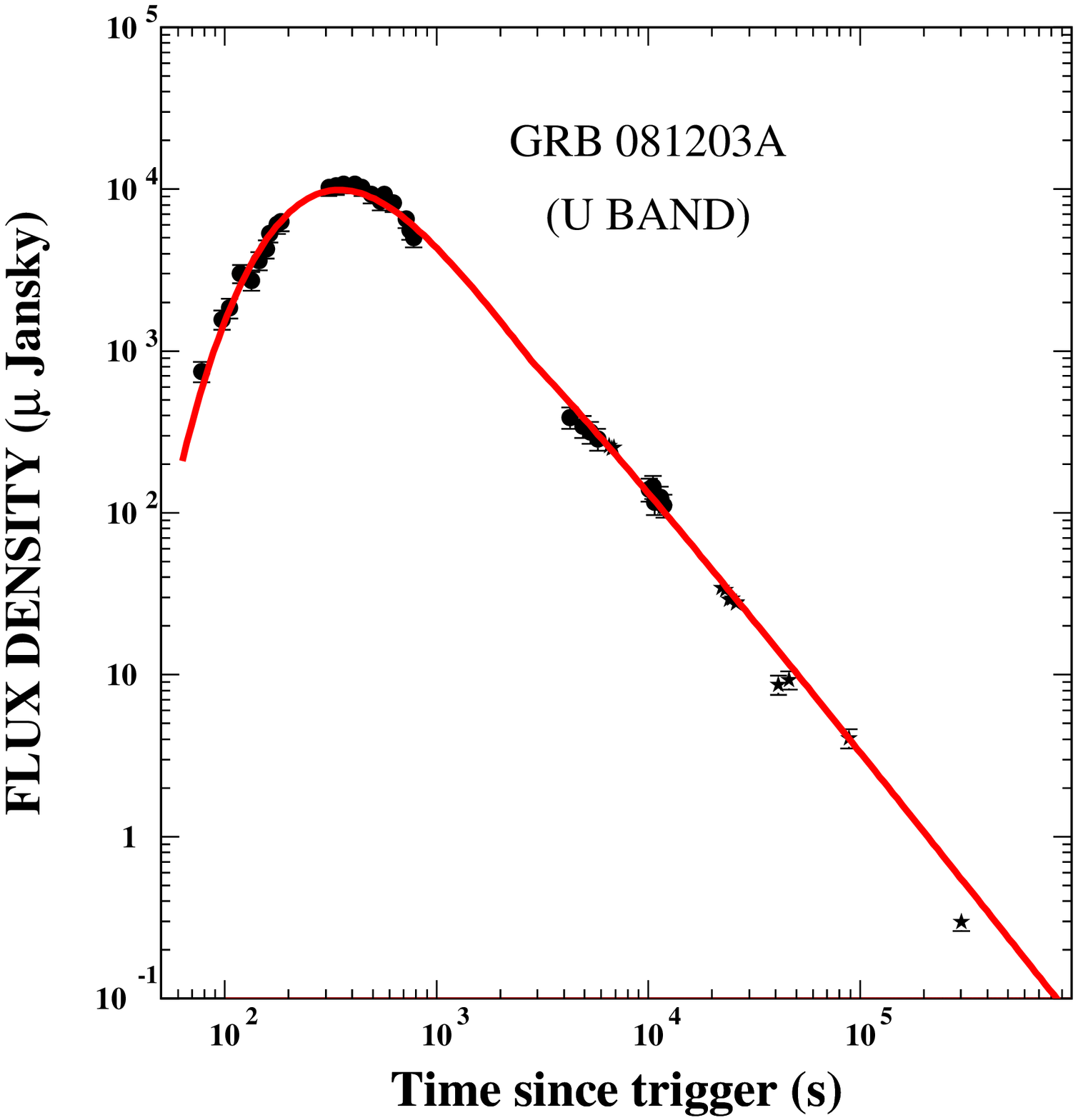,width=8.0cm,height=6.0cm}
}}
\vbox{
\hbox{
\epsfig{file=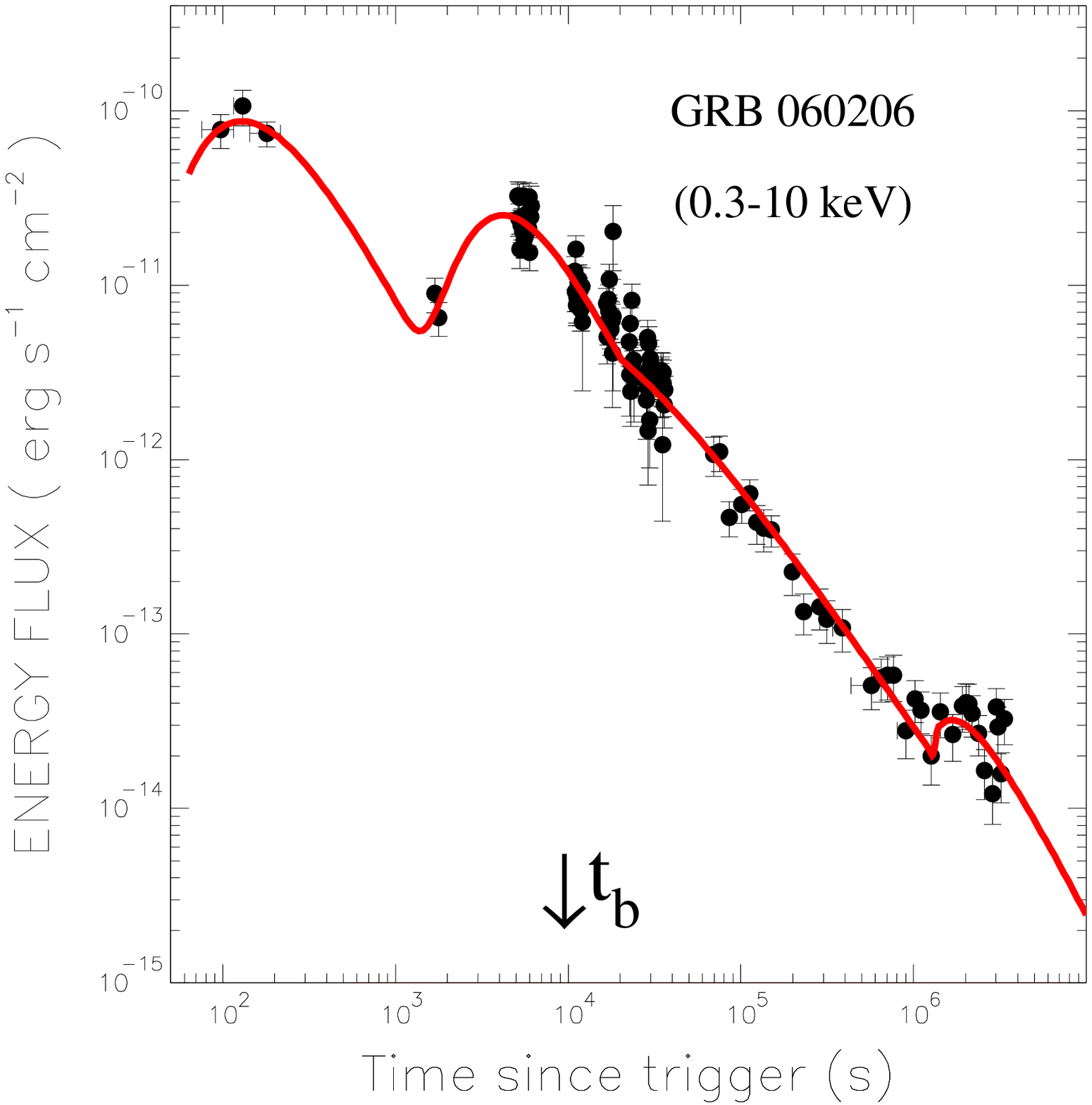,width=8.0cm,height=6.0cm}
\epsfig{file=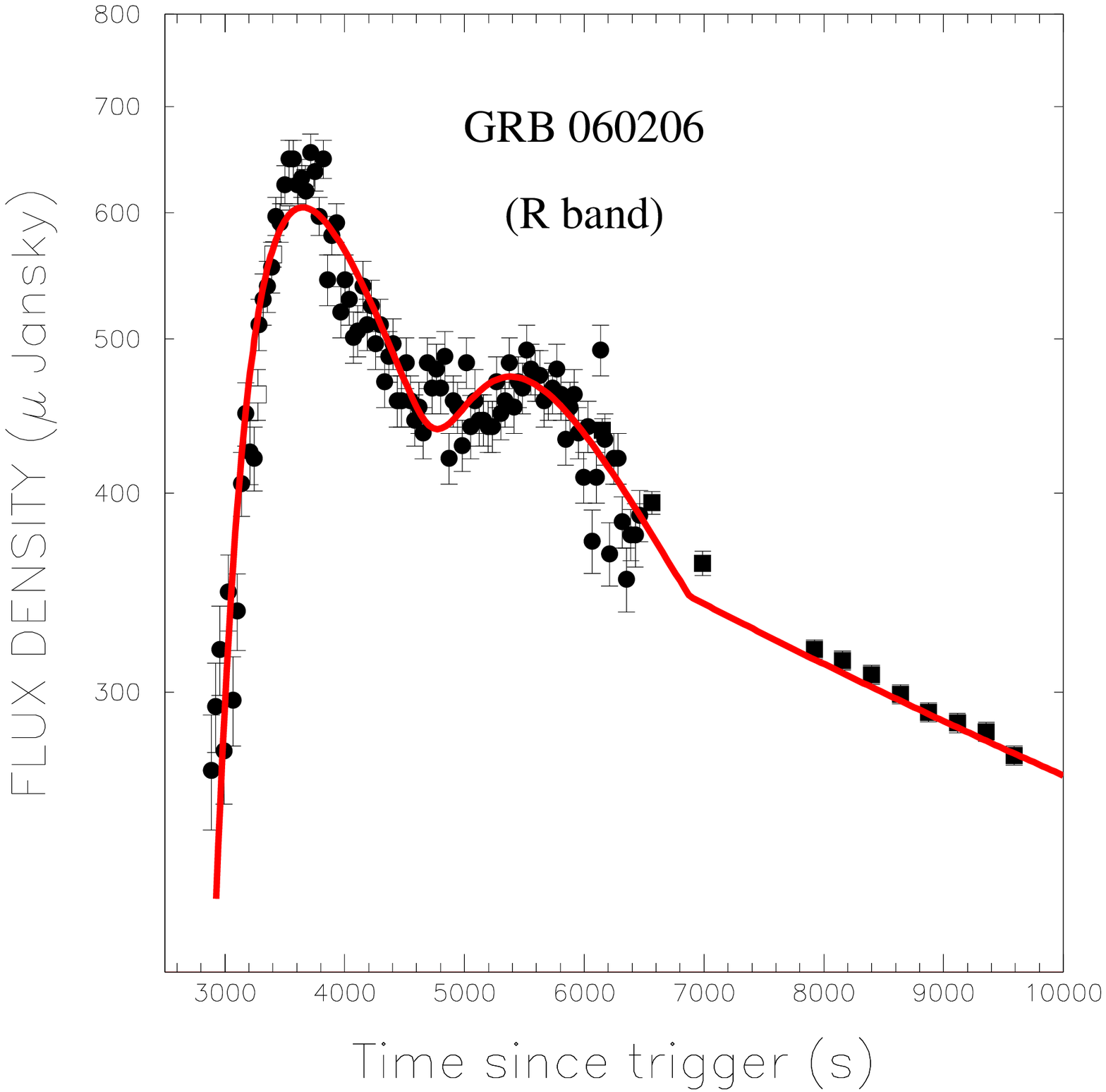,width=8.0cm,height=6.0cm}
}}
\caption{Comparison between the observed X-ray and 
optical lightcurves of XRFs and their CB 
model description. The early-time optical lightcurves
are approximated by a single (unresolved) SRF. 
GRB 050820A -
{\bf Top left (a):} The 0.3-10 keV XRT lightcurve. 
{\bf Top right (b):} The optical R-band lightcurve.
GRB 081203A
{\bf Middle left (c):} The 0.3-10 keV XRT lightcurve.
{\bf Middle right (d):} The optical U-band lightcurve.
GRB 060206
{\bf Bottom left (e):}  The 0.3-10 keV lightcurve with poorly resolved 
late-time flares.
{\bf Bottom right (f):} The R-band lightcurve with 
well resolved late-time SRFs.
}
\label{fig11}
\end{figure}

\end{document}